\documentclass[aip,apl,reprint,a4paper,floatfix,amsmath,amssymb,amsfonts,noshowpacs,citeautoscript]{revtex4-2}
\usepackage{newtxtext,newtxmath}
\usepackage[T1]{fontenc}
\usepackage{graphicx}
\usepackage[dvipsnames]{xcolor}
\usepackage{soul} 
\usepackage{floatrow}
\usepackage[
,textwidth=17.5cm
,textheight=23.5cm
,verbose
,pdftex
]{geometry}
\usepackage{xspace}

\usepackage[pdftex]{hyperref}
\hypersetup{
  pdftitle={Excitonic and deep-level emission from N- and Al-polar homoepitaxial AlN layers grown by molecular beam epitaxy},
	colorlinks,
	citecolor=blue,
	linkcolor=blue
	}

\makeatletter
\newcommand*{\refig}[2]{\hyperref[#1]{\ref*{#1}(#2)}}
\makeatother

\DeclareMathAlphabet{\mathsf}{OT1}{\sfdefault}{m}{n}
\SetMathAlphabet{\mathsf}{bold}{OT1}{\sfdefault}{b}{n}

\begin{document}
\graphicspath{{./figures/}}

\title{Excitonic and deep-level emission from N- and Al-polar homoepitaxial AlN grown by molecular beam epitaxy}

\author{L.~van~Deurzen}
\email[Electronic mail: ]{lhv9@cornell.edu}
\affiliation{School of Applied and Engineering Physics, Cornell University, Ithaca, New York 14853, USA}
\author{J.~Singhal}
\author{J.~Encomendero}
\affiliation{Department of Electrical and Computer Engineering, Cornell University, Ithaca, New York 14853, USA}
\author{N.~Pieczulewski}
\affiliation{Department of Materials Science and Engineering, Cornell University, Ithaca, New York 14853, USA}
\author{C.S.~Chang}
\affiliation{School of Applied and Engineering Physics, Cornell University, Ithaca, New York 14853, USA}
\affiliation{Research Laboratory of Electronics, Massachusetts Institute of Technology, MA 02139, USA}
\author{Y.~Cho}
\affiliation{Department of Electrical and Computer Engineering, Cornell University, Ithaca, New York 14853, USA}
\author{D.A.~Muller}
\affiliation{School of Applied and Engineering Physics, Cornell University, Ithaca, New York 14853, USA}
\affiliation{Kavli Institute at Cornell for Nanoscale Science, Cornell University, Ithaca, New York 14853, USA}
\author{H.G.~Xing}
\affiliation{Department of Electrical and Computer Engineering, Cornell University, Ithaca, New York 14853, USA}
\affiliation{Department of Materials Science and Engineering, Cornell University, Ithaca, New York 14853, USA}
\affiliation{Kavli Institute at Cornell for Nanoscale Science, Cornell University, Ithaca, New York 14853, USA}
\author{D.~Jena}
\affiliation{School of Applied and Engineering Physics, Cornell University, Ithaca, New York 14853, USA}
\affiliation{Department of Electrical and Computer Engineering, Cornell University, Ithaca, New York 14853, USA}
\affiliation{Department of Materials Science and Engineering, Cornell University, Ithaca, New York 14853, USA}
\affiliation{Kavli Institute at Cornell for Nanoscale Science, Cornell University, Ithaca, New York 14853, USA}
\author{O.~Brandt}
\author{J.~L\"{a}hnemann}
\affiliation{Paul-Drude-Institut f\"ur Festk\"orperelektronik, Leibniz-Institut im Forschungsverbund Berlin e.V., 10117 Berlin, Germany}

\begin{abstract}
Using low-temperature cathodoluminescence spectroscopy, we study the properties of N- and Al-polar AlN layers grown by molecular beam epitaxy on bulk AlN{\{0001\}}. Compared to the bulk AlN substrate, layers of both polarities feature a suppression of deep level luminescence, a total absence of the prevalent donor with an exciton binding energy of 28~meV, and a much increased intensity of the emission from free excitons. The dominant donor in these layers is characterized by an associated exciton binding energy of 13~meV. The observation of excited exciton states up to the exciton continuum allows us to directly extract the $\Gamma_{5}$ free exciton binding energy of 57~meV. 
\end{abstract}

\maketitle


\section*{Introduction} 

The ultra-wide gap semiconductor AlN was first synthesized over a century ago,\cite{brieglebUeberStickstoffmagnesiumUnd1862} but it has only recently been recognized  that the unique physical properties of AlN make it of great interest for applications in novel electronic and optoelectronic devices. Apart from its direct band gap of 6.095~eV corresponding to an emission wavelength of 203~nm, most noteworthy are its large piezoelectric coefficients (1.5~C/m$^{2}$ or 5~pm/V),\cite{muralt_Int.J.Microw.Wirel.Technol._2009} its high breakdown-field ($>\!\!10$~MV/cm) \cite{khachariyaRecord10MV2022,hussainHighFigureMerit2023}, and its high thermal conductivity ($>\!\!300$~W/mK) \cite{chengExperimentalObservationHigh2020}. 
Moreover, AlN has an exciton binding energy exceeding twice the value of $kT$ at room temperature,\cite{koppeOverviewBandedgeDefect2016} which is notably higher than in any other III-V semiconductor with the exception of hBN.\cite{doanBandgapExcitonBinding2016}
These properties make AlN an attractive material for applications in high-power and millimeter-wave electronics\cite{baderProspectsWideBandgap2020,hickmanFirstRFPower2021,kimNpolarGaNAlGaN2023} as well as deep-ultraviolet (UV-C) light emission and lasing\cite{amano2020UVEmitter2020,zhang271NmDeepultraviolet2019, zhangKeyTemperaturedependentCharacteristics2022, vandeurzenOpticallyPumpedDeepUV2022}. The large exciton binding energy in AlN is also of high interest for the study of the fundamental science of excitons and their condensation\cite{ginzburg_key_problems_physics_astrophysics}.

This potential of AlN lay dormant until very recently, when researchers succeeded to synthesize large AlN crystals of high structural perfection with low dislocation densities ($<10^{4}$--$10^{6}$~cm$^{-2}$) utilizing either physical vapor transport (PVT)\cite{zhouBandgapPhotoluminescenceAlN2020, fenebergHighexcitationHighresolutionPhotoluminescence2010, chichibuExcitonicEmissionDynamics2013, thonkeOpticalSignaturesSilicon2017} for bulk growth, or hydride vapor phase epitaxy (HVPE)\cite{ishiiStimulatedEmissionMechanism2022} for fabricating thick free-standing layers. The availability of AlN substrates has enabled studies of the fundamental properties of AlN, particularly regarding its spontaneous emission, yielding insight into valence-band ordering, spin-orbit and crystal-field splitting, spin exchange interaction, exciton binding energies, and the detection of impurities and defects. These investigations have been performed by either photo- or cathodoluminescence spectroscopy on PVT-grown bulk crystals,\cite{zhouBandgapPhotoluminescenceAlN2020, fenebergHighexcitationHighresolutionPhotoluminescence2010, chichibuExcitonicEmissionDynamics2013, thonkeOpticalSignaturesSilicon2017} HVPE-grown free-standing substrates,\cite{ishiiStimulatedEmissionMechanism2022} or  homoepitaxial AlN layers grown by metal-organic chemical vapor deposition \cite{funatoHomoepitaxyPhotoluminescenceProperties2012, leutePhotoluminescenceHighlyExcited2009, fenebergSharpBoundFree2011, chichibuMajorImpactsPoint2010, chichibuExcitonicEmissionDynamics2013, bryanExcitonTransitionsOxygen2014}.

Numerous studies have focused on the excitonic near-bandedge emission of AlN, resolving transitions due to lowest-energy free excitons with $\Gamma^c_{7} \otimes \Gamma^v_{7+}$ symmetry (often called A-excitons)\cite{paskovSpinexchangeSplittingExcitons2001} and several bound exciton transitions \cite{fenebergSharpBoundFree2011, neuschlOpticalIdentificationSilicon2012}. At higher excitation densities, bi-exciton emission (M-band), inelastic exciton-exciton scattering bands (P-bands), and electron-hole plasma recombination were observed \cite{leutePhotoluminescenceHighlyExcited2009,fenebergHighexcitationHighresolutionPhotoluminescence2010}. Spectra recorded over a wider spectral range revealed the presence of deep luminescence bands in the 2–4~eV range that frequently dominate over the near-bandedge emission in terms of integrated intensity. The actual origin of these ubiquitous deep luminescence bands is unknown, but is generally believed to be related to complexes of native defects, particularly Al vacancies, with impurities such as Si and O that form DX centers in AlN. However, also other native defects and the impurity C have been suspected to play an important role in the transitions giving rise to these deep bands.\cite{koppeOverviewBandedgeDefect2016, thonkeOpticalSignaturesSilicon2017, zhouBandgapPhotoluminescenceAlN2020}

\begin{figure*}[t]
\includegraphics[width=\textwidth]{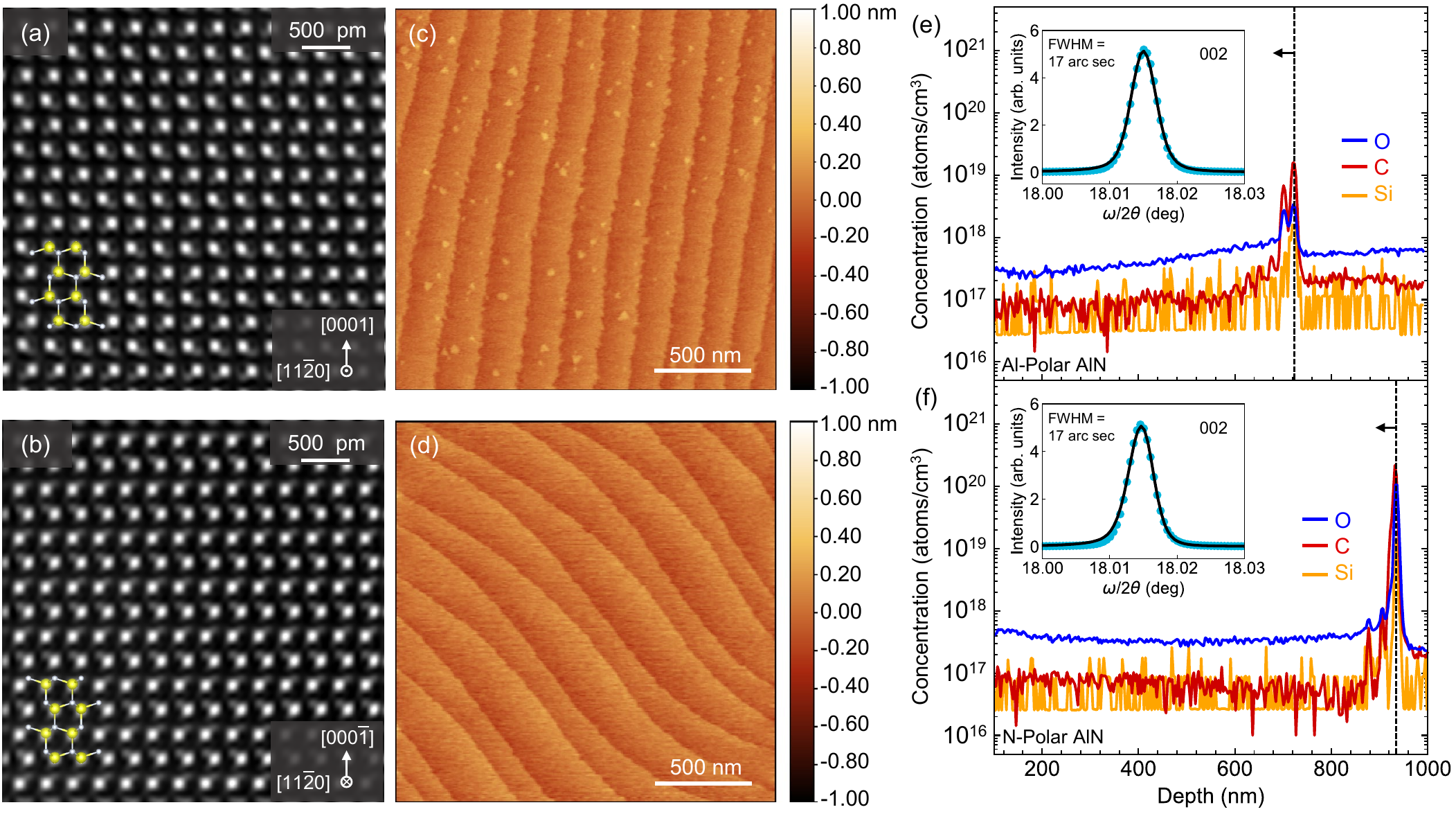}
\caption{[(a) and (b)] HAADF-STEM images with overlayed wurtzite ball-and-stick model, [(c) and (d)] $c$-plane atomic force topographs, and [(e) and (f)] SIMS profiles of samples I (Al-polar AlN, top row) and II (N-polar AlN, bottom row). The insets in figures (e) and (f) show the respective $\omega/2\theta$ x-ray diffraction scans across the 002 reflection of the MBE layers (blue, dotted) and the bare AlN substrate (solid, black). \label{Figure_1}}
\end{figure*}

In general, the incorporation of impurities and the formation of native point defects is related in a complex way to the particular growth method and conditions. Surprisingly, there seem to be no detailed studies of the spontaneous emission of homoepitaxial AlN films grown by molecular beam epitaxy (MBE). Furthermore, the point defect incorporation and formation often strongly depends on the bonding configuration on the growth front, i.\,e., its crystallographic orientation, as reported, for example, for GaN.\cite{tatarczakOpticalPropertiesNpolar2021} To the best of our knowledge, all of the available studies on the excitonic emission of AlN have been performed on films grown along the Al-polar ([0001]), semi-polar, or non-polar ([1$\bar{1}$00] and [1$\bar{2}$10]) directions. However, N-polar ([000$\bar{1}$]) (In,Ga)N/(Al,Ga)N heterostructures are of particular interest because they exhibit internal electrostatic fields opposite to those of their metal-polar counterparts, which is considered favorable for various advanced device concepts.\cite{li_Phys.StatusSolidiA_2011,*akyol_Jpn.J.Appl.Phys._2011,*dong_Appl.Phys.Lett._2012,*han_Jpn.J.Appl.Phys._2012,*wong_Semicond.Sci.Technol._2013,*feng_J.Appl.Phys._2015,*vandeurzenlenEnhancedEfficiencyBottom2021,*choBlueGaLightemitting2019,*bharadwajEnhancedInjectionEfficiency2020,*leeLightemittingDiodesAlN2020,*bharadwajBottomTunnelJunction2020}


In the present work, we use cathodoluminescence (CL) spectroscopy in a scanning electron microscope (SEM) to analyze and compare N- and Al-polar homoepitaxial strain-free AlN films grown by MBE in terms of their near-bandedge (around 6~eV) and deep-level (2--4~eV) light emission. With respect to the bare substrate, we find that both N-polar and Al-polar films feature a suppression of deep-level luminescence, and the total absence of the donor dominating the NBE of the substrate. Furthermore, the intense free exciton emission in the films under investigation allows us to directly measure an exciton binding energy of 57~meV from the $\Gamma_{5}^{n \rightarrow \infty}$ transition.

\section*{Experiments}

The three samples under investigation were grown in an MBE system equipped with a radio-frequency plasma source for generating active N and solid-source effusion cells for Al. As substrates, we used bulk AlN wafers grown by PVT with a dislocation density $<10^{4}$~cm$^{-2}$. The layers were grown at substrate temperatures above 1000\,$^{\circ}$C under Al-stable conditions. Details on the surface preparation and growth parameters for the MBE growth of Al- and N-polar homoepitaxial AlN layers are given in Refs.~\citenum{choMolecularBeamHomoepitaxy2020, leeSurfaceControlMBE2020, singhalMolecularBeamHomoepitaxy2022,zhangMolecularBeamHomoepitaxy2022}. Samples I and II consist of a 700~nm-thick Al-polar and a 900~nm-thick N-polar layer, respectively. Sample III consists of a 1-$\upmu$m-thick N-polar AlN layer that was capped with 6.7~nm heteroepitaxial GaN. This GaN cap layer serves as a passivation layer for the AlN surface and also acts as a charge-spreading layer during electron beam excitation. As reference and for comparison with other studies, a freestanding AlN substrate was used. 

\begin{figure*}[t]
\floatbox[{\capbeside\thisfloatsetup{capbesideposition={right,center},capbesidewidth=4cm}}]{figure}[\FBwidth]
{\caption{(a) Low-temperature, high-resolution near-bandedge and (b) deep-level CL spectra for the bare PVT-AlN substrate, as well as samples I and II. Light was collected with $\mathbf{k} \| \mathbf{c}$. The narrow feature just above 3~eV in (b) is the second order of the near-bandedge emission.}\label{Figure_2}}
{\includegraphics[width=12cm]{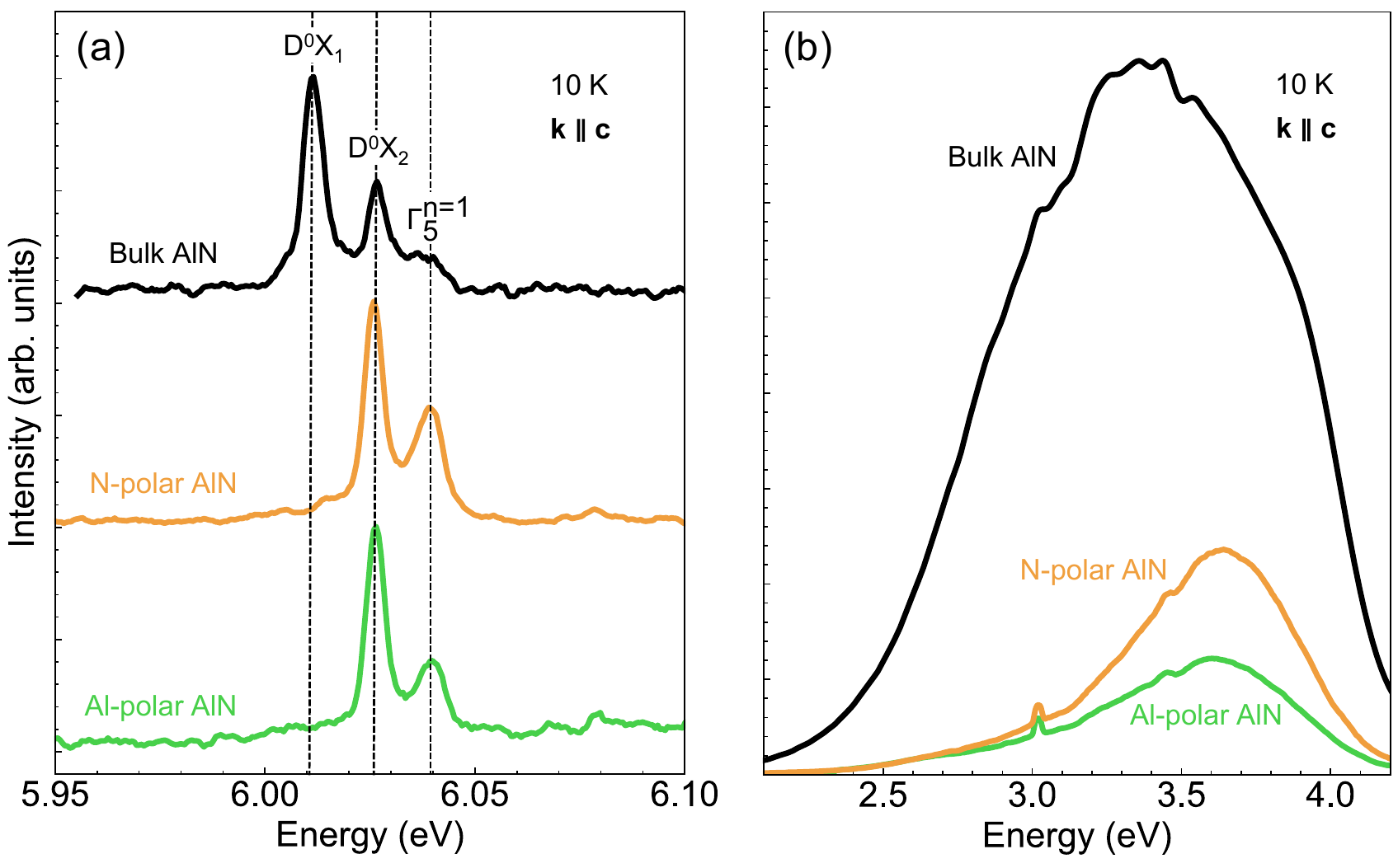}}
\end{figure*}

The samples were investigated by scanning transmission electron microscopy (STEM) on an aberration-corrected ThermoFisher Spectra 300 CFEG operated at 300~keV to confirm their polarity. STEM samples were prepared using a ThermoFisher Helios G4 UX Focused Ion Beam system. Carbon and platinum protective layers were deposited and a final milling step at 5~keV was used to minimize ion-beam damage. Additionally, atomic force microscopy (AFM) was performed on an Asylum Cypher AFM microscope in tapping mode for sample areas of 2~×~2~\textmu m$^2$ and 20~×~20~\textmu m$^2$. Furthermore, the concentrations of O, C, and Si were measured by time-of-flight secondary ion mass spectrometry (ToF-SIMS) performed by Evans Analytical Group. Finally, triple-axis $\omega/2\theta$ x-ray diffraction (XRD) scans across the symmetrical 002 wurtzite reflection were measured using a Panalytical Empyrean x-ray diffractometer equipped with a PIXcel$^{3\mathrm{D}}$ detector and Xe proportional detector. The monochromator consists of a hybrid two-bounce Ge220 crystal utilizing Cu~${\mathrm{K}\alpha_{1}}$ radiation.

To analyze and compare the optical properties of these samples, low-temperature emission spectra were recorded by CL spectroscopy in a Zeiss Ultra55 SEM equipped with a He-cryo-stage allowing sample temperatures down to 10~K and a Gatan MonoCL4 detection system. The spectrometer is operated with two gratings, one blazed at 300~nm with 2400~grooves/mm (Fig.~\ref{Figure_2}) and the other blazed at 250~nm with 1200~grooves/mm (Fig.~\ref{Figure_3}), resulting in a spectral resolution of approximately 0.0035 and 0.007~eV for a slit width of 0.1~mm, respectively. Note that the line widths of the near-bandedge transitions are limited by this spectral resolution. The monochromator was calibrated using a Hg(Ar) spectral calibration lamp. The error of this calibration amounts to $\pm2$~\AA\ because of the spectral separation between the highest energy line of the lamp and the AlN near-bandedge emission. The samples were aligned with respect to the detection setup such that mostly light propagating parallel to the $c$-axis, $\mathbf{k} \| \mathbf{c}$, and electric field polarization ($\mathbf{E}$) orthogonal to the $c$-axis, $\mathbf{E} \perp \mathbf{c}$, was collected. The light was detected monochromatically with a photomultipier tube (PMT) while stepping through the spectral range and the spectra were converted to energy scale by a Jacobian transformation, taking into account the refractive index dispersion of air \cite{LumiSpy2023}. For the CL measurements, all of the AlN samples were sputter coated with 3~nm of Ti to reduce charging effects in the SEM.

\section*{Results and Discussion}

Figure \ref{Figure_1} summarizes data pertaining to the structural and morphological properties of our AlN layers as well as to their purity. The polarity of the layers is examined by high-angle annular dark-field STEM (HAADF-STEM) as shown in Fig.~\ref{Figure_1}(a) and \ref{Figure_1}(b), respectively. The yellow spheres in the overlayed ball-and-stick model of the wurtzite crystal structure represent Al atoms, and the white spheres represent N atoms, confirming the nominal polarity of the layers (growth direction pointing upwards). Figures ~\ref{Figure_1}(c) and \ref{Figure_1}(d) depict $2 \times 2~\upmu$m$^2$ atomic force topographs of the $c$-plane surface of the as-grown layers. The trains of well-resolved monolayer steps are characteristic for step-flow growth, resulting in a root-mean-square roughness as low as 80 and 160~pm, respectively. Figures \ref{Figure_1}(e) and \ref{Figure_1}(f) display ToF-SIMS data for the respective samples revealing comparatively low impurity concentrations for layers of either polarity. The level of $[\mathrm{Si}]<5\times 10^{16}$~cm$^{-3}$ is at, and the levels of O and C are close to the detection limit (given in parentheses): $[\mathrm{O}] \approx 3-4 \times 10^{17}$~cm$^{-3}$ ($1 \times 10^{17}$~cm$^{-3}$) and $[\mathrm{C}] \approx 7-8 \times 10^{16}$~cm$^{-3}$ ($5 \times 10^{16}$~cm$^{-3}$). Hydrogen (not shown) has a concentration below the detection limit for both samples, $[\mathrm{H}]< 3 \times 10^{17}$~cm$^{-3}$. Note that the spike at the substrate/MBE interface is less pronounced for the Al-polar sample (near 700~nm depth) than for the N-polar sample (near 900~nm depth). This is likely due to a greater adsorption efficiency of impurities on the N-polar surface \cite{yeAsymmetryAdsorptionOxygen2008, miaoEffectsSurfaceReconstructions2010}, as well as the less extensive ex-situ chemical treatment for the N-polar face, as it is more reactive \cite{choMolecularBeamHomoepitaxy2020, singhalMolecularBeamHomoepitaxy2022, zhangMolecularBeamHomoepitaxy2022}. Finally, the insets in Figs.~\ref{Figure_1}(e) and \ref{Figure_1}(f) depict $\omega/2\theta$ x-ray diffraction scans across the 002 reflection for the Al- and N-polar AlN samples (blue, dotted) compared with the bare AlN substrates they were grown on (solid, black). The line widths of approximately 17~arcsec confirm that the MBE-grown AlN layers are free of strain. The absence of strain is further confirmed by reciprocal space maps around the asymmetric $\bar{1}05$ reflection for Samples I and II (see Fig.~S1 in the Supplementary Material). Symmetric and asymmetric $\omega$ scans recorded for Samples I and II exhibit widths between 20 and 30~arcsec, comparable or below to that measured on the corresponding bulk AlN substrates (see Fig.~S2 in the Supplementary Material). For the structural characterization and SIMS data of sample III, we refer to Figs.~S3, S4 and S5 in the Supplementary Material.

\begin{figure*}[t]
\includegraphics[width=\textwidth]{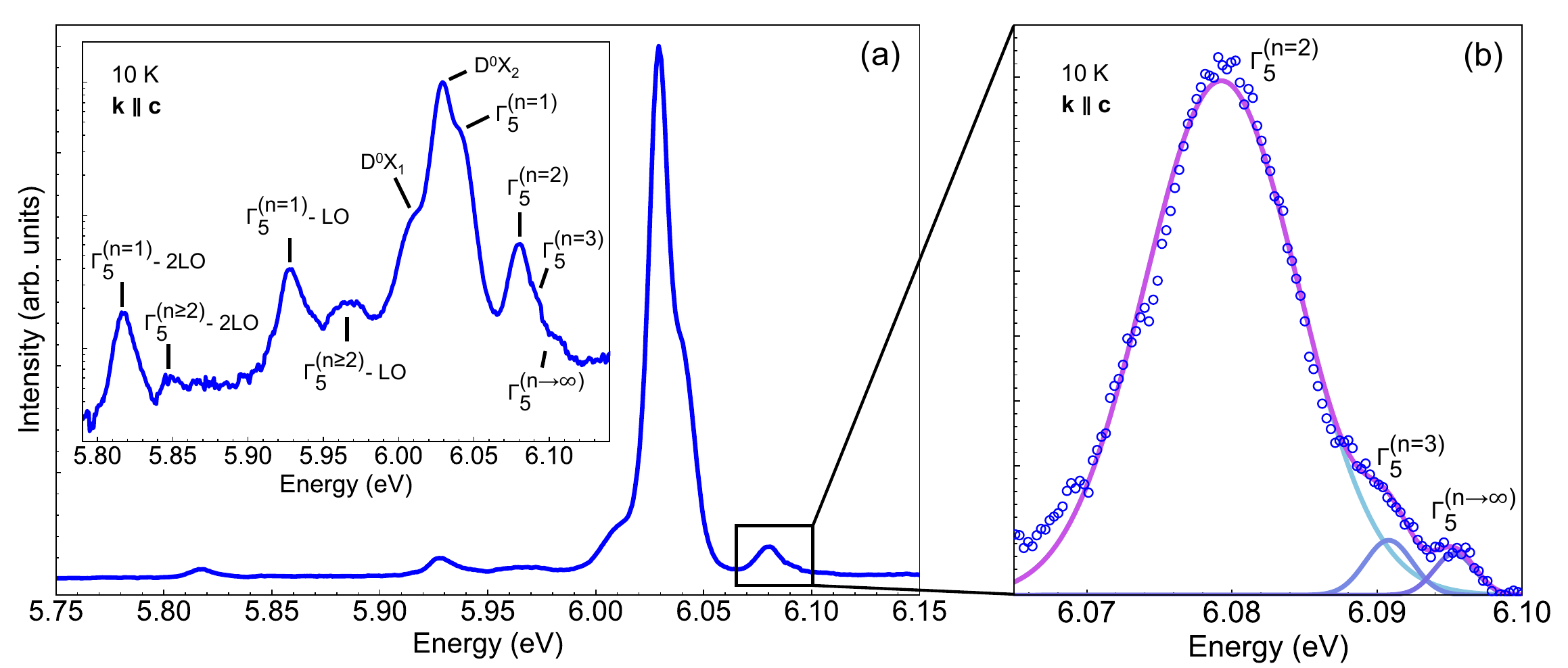}
\caption{(a) Low-temperature, high-resolution near-bandedge CL spectrum for sample III on a linear and logarithmic (inset) intensity scale. The integration time was chosen to be $10\times$ longer than for the spectra in Fig.~\ref{Figure_2}. (b) Expanded view of the excited free exciton transitions, with $\Gamma_{5}^{n=2}$, $\Gamma_{5}^{n=3}$, and $\Gamma_{5}^{n \rightarrow \infty}$ resolved. The line is a fit of the data with three Gaussians (also shown).}
\label{Figure_3}
\end{figure*}

CL spectra of samples I and II are compared with the one of the bare PVT-AlN substrate (N-polar face) in Fig.~\ref{Figure_2}. An acceleration voltage of 7~keV was used to excite the samples, ensuring that electron-hole pairs are generated only within the homoepitaxial layers (first few hundred nm from the AlN/air interface, see Fig.~S7 in the Supplementary Material). Figure \ref{Figure_2}(a) shows the low-temperature, high-resolution near-bandedge emission spectra normalized to their peak intensities. For all three samples, a distinct high-energy line is observed at $(6.038 \pm 0.005)$~eV, with the error due to the uncertainty of the wavelength calibration. This transition energy is close to the average energy (6.041~eV) reported for the exciton spin triplet with $\Gamma_{5}$ symmetry in bulk (strain-free) AlN.\cite{funatoHomoepitaxyPhotoluminescenceProperties2012,fenebergHighexcitationHighresolutionPhotoluminescence2010,leutePhotoluminescenceHighlyExcited2009,bryanExcitonTransitionsOxygen2014,fenebergSharpBoundFree2011,neuschlOpticalIdentificationSilicon2012} 
 
For AlN, the negative crystal-field splitting causes the $\Gamma^v_{7+}$ band to be the uppermost valence band,\cite{liBandStructureFundamental2003} resulting in exciton states of $\Gamma^c_{7} \otimes \Gamma^v_{7+}$ symmetry (often called A-excitons).\cite{choUnifiedTheorySymmetrybreaking1976} Spin-exchange interaction splits these states into excitons with the irreducible representation $\Gamma_{1}$~$\oplus$~$\Gamma_{2}$~$\oplus$~$\Gamma_{5}$, with the $\Gamma_{5}$ spin triplet being the only optically active state that satisfies $\mathbf{E} \perp \mathbf{c}$ and $\mathbf{k} \parallel \mathbf{c}$, i.\,e., our measurement geometry. It should be noted that there currently is no consensus in the literature on the assignment of the experimentally observed emission lines and thus the ordering of the $\Gamma_{1}^{n=1}$ and $\Gamma_{5}^{n=1}$ states. In fact, researchers have reported both negative\cite{feneberg_Appl.Phys.Lett._2013, leutePhotoluminescenceHighlyExcited2009,bryanExcitonTransitionsOxygen2014,fenebergHighexcitationHighresolutionPhotoluminescence2010, fenebergSharpBoundFree2011} and positive\cite{chichibuExcitonicEmissionDynamics2013,ishiiStimulatedEmissionMechanism2022,ishiiLongrangeElectronholeExchange2020,chichibuInplaneOpticalPolarization2019,ishiiHugeElectronholeExchange2013} values for the spin-exchange splitting $j$. A possible explanation of this apparent discrepancy is the fact that $j$ depends sensitively on strain.\cite{koda_Phys.Rev.Lett._1968,akimoto_Phys.Rev.Lett._1968,paskovSpinexchangeSplittingExcitons2001} For the case of GaN, ~ \citet{paskovSpinexchangeSplittingExcitons2001} calculated the dependence of the $\Gamma_{1}^{n=1}-\Gamma_{5}^{n=1}$ splitting (which is given by $2j$) on biaxial strain, and observed that it changes sign for a biaxial compressive strain as small as $2 \times 10^{-4}$. The scatter of $a$ lattice constants of bulk AlN amount to a variation of $\pm 2.5 \times 10^{-4}$ with respect to the strain-free value,\cite{nilsson_J.Phys.Appl.Phys._2016} which, in analogy to GaN, may already be sufficient to result in a significant change of the spin-exchange splitting in AlN. An explicit calculation as done by ~ \citet{paskovSpinexchangeSplittingExcitons2001} for GaN is required to see if this hypothesis can account for the conflicting results in the literature.

The data in the present work, which are obtained on strain-free AlN, support a negative value of $j$ and thus a lower energy for the $\Gamma_{1}$ as compared to the $\Gamma_{5}$ exciton. The geometry ($\mathbf{k \parallel c}$) of the measurements depicted in Figs.~\ref{Figure_2} and \ref{Figure_3} is compatible with the detection of the $\Gamma_{5}$, but not the $\Gamma_{1}$ exciton. This fact is confirmed by a comparison to measurements taken on the cross-section of sample I (see Fig.~S6 in the Supplementary Material). In the emission normal to the cross-section ($\mathbf{k \perp c}$), we observe a superposition of the $\Gamma_{5}^{n=1}$ and $\Gamma_{1}^{n=1}$ excitons as expected from the selection rules. The intense emission of the $\Gamma_{1}^{n=1}$ exciton is situated at 6.031~eV, in between the $\Gamma_{5}^{n=1}$ and $\mathrm{D^{0}X_{2}}$ lines.
The additional peak observed at higher energy (6.08~eV) for the homoepitaxial layers is attributed to the first excited state of the free $\Gamma_{5}$ exciton as discussed in more detail for sample III. 

For the PVT substrate, two intense and narrow lines are observed 28 and 13~meV below the $\Gamma_{5}^{n=1}$ free exciton ground state. The former of these lines is absent for both samples I and II. These lines have been frequently observed in previous work, and are attributed to donor-bound exciton transitions. $\mathrm{D^{0}X_{1}}$, with an exciton binding energy of 28~meV, has been tentatively assigned to either O \cite{fenebergSharpBoundFree2011} or Si \cite{neuschlOpticalIdentificationSilicon2012} as the shallow donor, while the $\mathrm{D^{0}X_{2}}$ with a binding energy of 13~meV has been speculated to be related to a native shallow donor.\cite{fenebergSharpBoundFree2011} 
The dominance of the $\mathrm{D^{0}X_{1}}$ line in the spectra of the substrate, and the total absence of this transition in our epitaxial layers, implies a significantly lower concentration of the responsible shallow donor in the layers compared to the substrate. Accordingly, it is unlikely that this line is related to O, the concentration of which is even slightly higher in the layers than in the substrate [cf.\ Figs.~\ref{Figure_1}(e) and \ref{Figure_1}(f)]. Si, in contrast, is below the detection limit for all samples, and it is thus possible that the actual concentration in the layers is much below that in the substrate. Our experiments are thus consistent with Si being the donor related to the $\mathrm{D^{0}X_{1}}$ line, adding support of the assignments made in Refs.~\citenum{neuschlOpticalIdentificationSilicon2012,neuschlDirectDeterminationSilicon2013,ishiiStimulatedEmissionMechanism2022}.

To probe radiative deep levels, we performed CL measurements with lower resolution in the range of 2--4.2~eV as shown in Fig.~\ref{Figure_2}(b). All spectra in this figure are normalized to their respective near-bandedge peak intensities. The integrated intensity of the broad luminescence bands observed in this spectral range exceeds the near-bandedge intensity by a factor 2.5 for the bulk AlN substrate, but is lower by factors of 0.5 and 0.8 for samples I and II, respectively. Apart from the higher intensity, the CL band from the substrate is much broader and seems to consist of two individual contributions, with the one at higher energy matching the energy of the band in samples I and II. The lineshape of samples I and II are almost identical, suggesting that similar types of defects are incorporated for both polarities.

The overall similarity between the radiative properties of the Al- and N-polar AlN samples is remarkable. In contrast, studies on N-polar GaN layers, regardless of whether being grown by MBE or MOCVD, have shown that it is difficult to obtain layers with a purity and emission characteristics equivalent to their Ga-polar counterparts. A higher incorporation efficiency of impurities as well as a lower formation energy for N vacancies in N-polar GaN growth are discussed as possible reasons for this difference in material quality \cite{zywietzAdsorptionOxygenGaN1999}. Consequently, N-rich growth seems to be a prerequisite to suppress the formation of N vacancies for GaN($000\bar{1}$) \cite{tatarczakOpticalPropertiesNpolar2021}. For AlN, our data show that the situation is much more favorable, suggesting that we may use the alignment of the polarization fields as an additional degree of freedom for the design of AlN-based light-emitters. Note, however, that interface recombination may become important in group-III nitride heterostructures, and may strongly depend on their polarity.\cite{fernandez-garrido_Phys.Rev.Appl._2016,cheze_Appl.Phys.Lett._2018,auzelle_Phys.Rev.Appl._2022} 

To further elucidate the details of the near-bandedge emission, we turn to sample III, which exhibits the highest emission intensity of all samples. Figure \ref{Figure_3}(a) shows its low-temperature CL spectrum, collected at an acceleration voltage of 11~keV (still exciting mostly the homoepitaxial layer, see Fig.~S7 in the Supplementary Material). The higher beam energy ensures a high signal intensity, facilitating the reliable detection of weak transitions. As seen best in the inset, ten transitions are resolved, the energies of which are compiled in Table~\ref{table:1}. The highest energy transitions associated to excited states of the free exciton are better visible in the magnified view displayed in Fig.~\ref{Figure_3}(b). Where available, our values are in good agreement with previous studies. In particular, the $E(\Gamma_{5}^{n=1})$ exciton and its excited states are resolved at 6.038, 6.079, 6.091, and 6.095~eV, together with their longitudinal optical (LO) phonon replicas at lower energies. Again, the $\mathrm{D^{0}X_{2}}$ transition and a weak $\mathrm{D^{0}X_{1}}$ line (originating from the substrate) contribute to the spectra as well. The prominent transition at 6.079~eV can be readily assigned to the $E(\Gamma_{5}^{n=2})$ transition, for which energy separations of 37--43~meV with respect to the $E(\Gamma_{5}^{n=1})$ transition have been reported by other groups,\cite{murotaniTemperatureDependenceExcitonic2009,fenebergHighexcitationHighresolutionPhotoluminescence2010,funatoHomoepitaxyPhotoluminescenceProperties2012,chichibuExcitonicEmissionDynamics2013,leutePhotoluminescenceHighlyExcited2009,fenebergSharpBoundFree2011,neuschlOpticalIdentificationSilicon2012} while no values are available for the higher-order transitions. 


To identify the origin of the transitions at 6.091 and 6.095~eV, we first consider the hydrogenic effective-mass approximation, in which the energies of the excited states as well as the binding energy of the free exciton can be determined based on the 1S-2S splitting, i.\,e., the energy difference between the first excited state and the ground state $E(\Gamma_{5}^{n=2})-E(\Gamma^{n=1}_{5})$. Specifically, the energy of the second excited state is predicted to be
\begin{equation}
E(\Gamma_{5}^{n = 3}) = E(\Gamma_{5}^{n = 1}) + \frac{32}{27}\left[E(\Gamma_{5}^{n=2})-E(\Gamma^{n=1}_{5})\right].
\label{Second_Excited_State}
\end{equation}
Similarly, the energy of the $n \rightarrow \infty$ transition is given by
\begin{equation}
    \begin{aligned}
    E(\Gamma_{5}^{n \rightarrow \infty}) & = E(\Gamma_{5}^{n = 1}) + \frac{4}{3}\left[E(\Gamma_{5}^{n=2})-E(\Gamma^{n=1}_{5})\right].
\end{aligned}
\label{Free_Electron_Hole}
\end{equation}
These equations yield values of 6.087 and 6.093~eV for the $\Gamma_{5}^{n=3}$ and $\Gamma_{5}^{n \rightarrow \infty}$ transitions, respectively, slightly lower than the transition energies observed. 

\begin{table}[t]
\begin{tabular}{cccc}
\hline Transition & $E$ & $\Delta E(\Gamma^{n=1}_{5})$ & \textbf{$\Delta E_\mathrm{lit}(\Gamma^{n=1}_{5})$}\\
\text{} & $(\mathrm{eV})$ & $(\mathrm{meV})$ & $(\mathrm{meV})$ \\
\hline $\boldsymbol{\Gamma}^{\mathbf{n}=\mathbf{1}}_{\mathbf{5}}-\mathbf{2} \mathbf{L} \mathbf{O}$ & 5.815 & $-223$ & $-225$, \cite{leutePhotoluminescenceHighlyExcited2009} $-231$ \cite{ishiiStimulatedEmissionMechanism2022} \\
$\boldsymbol{\Gamma}^{\mathbf{n} \geq \mathbf{2}}_{\mathbf{5}}-\mathbf{2} \mathbf{L} \mathbf{O}$ & $5.849$ & $-189$ & $-188$\cite{leutePhotoluminescenceHighlyExcited2009} \\
$\boldsymbol{\Gamma}^{\mathbf{n}=\mathbf{1}}_{\mathbf{5}}-\mathbf{L} \mathbf{O}$ & $5.926$ & $-112$ & $-109$, \cite{leutePhotoluminescenceHighlyExcited2009} $-117$ \cite{ishiiStimulatedEmissionMechanism2022} \\
$\boldsymbol{\Gamma}^{\mathbf{n} \geq \mathbf{2}}_{\mathbf{5}}-\mathbf{L O}$ & $5.965$ & $-73$ & $-73$\cite{leutePhotoluminescenceHighlyExcited2009} \\
$\mathrm{D^{0}X_{\alpha}}$ & {} & {} & $-37$\cite{bryanExcitonTransitionsOxygen2014} \\
$\mathbf{D^{\mathbf{0}} \mathbf{X}_{\mathbf{1}} }$ & $6.010$ & $-28$ & $-28$ \cite{bryanExcitonTransitionsOxygen2014}, $-28.5$\cite{funatoHomoepitaxyPhotoluminescenceProperties2012}, $-28.5$\cite{neuschlOpticalIdentificationSilicon2012} \\
$\mathrm{D^{0}X_{\beta}}$ & {} & {} & $-25$\cite{bryanExcitonTransitionsOxygen2014} \\
$\mathrm{D^{0}X_{\gamma}}$ & {} & {} & $-22.1$ \cite{fenebergHighexcitationHighresolutionPhotoluminescence2010}, $-22.4$ \cite{fenebergSharpBoundFree2011}, $-22.5$\cite{neuschlOpticalIdentificationSilicon2012} \\
$\mathrm{D^{0}X_{\delta}}$ & {} & {} &  $-19$ \cite{bryanExcitonTransitionsOxygen2014}, $-19.0$ \cite{neuschlOpticalIdentificationSilicon2012} \\
$\mathbf{D}^{\mathbf{0}} \mathbf{X}_{\mathbf{2}}$ & $6.025$ & $-13$ & $-13$ \cite{bryanExcitonTransitionsOxygen2014}, $-13.3$\cite{neuschlOpticalIdentificationSilicon2012}, $-13.6$ \cite{funatoHomoepitaxyPhotoluminescenceProperties2012} \\
$\mathrm{D^{0}X_{\epsilon}}$ & {} & {} & $-9.5$\cite{funatoHomoepitaxyPhotoluminescenceProperties2012}, $-9.5$\cite{neuschlOpticalIdentificationSilicon2012} \\
$\boldsymbol{\Gamma^{\mathbf{n}=\mathbf{1}}_{1}}$ & 6.031 & {$-7$} & $-8$\cite{bryanExcitonTransitionsOxygen2014} \\
$\boldsymbol{\Gamma}^{\mathbf{n}=\mathbf{1}}_{\mathbf{5}}$ & $6.038$ & {0} & {0}\\
$\Gamma^{\mathrm{n=2}}_{1}$ & {} & {} & $+31$\cite{bryanExcitonTransitionsOxygen2014} \\
$\boldsymbol{\Gamma}^{\mathbf{n}=\mathbf{2}}_{\mathbf{5}}$ & $6.079$ & $+41$ & $+39.4$\cite{funatoHomoepitaxyPhotoluminescenceProperties2012}, $+39.4$\cite{neuschlOpticalIdentificationSilicon2012}, $+41$\cite{bryanExcitonTransitionsOxygen2014} \\
$\boldsymbol{\Gamma}^{\mathbf{n}=\mathbf{3}}_{\mathbf{5}}$ & $6.091$ & $+53$ & \\
$\boldsymbol{\Gamma}^{\mathbf{n} \rightarrow \infty}_{\mathbf{5}}$ & $6.095$ & $+57$ & \\
\\
\hline
\end{tabular}
\caption{Low-temperature (10~K) near-bandedge transitions in AlN. Transitions resolved in the homoepitaxial, N-polar AlN layers measured in the present work are highlighted in \textbf{bold}. All of these values have an experimental uncertainty of $\pm 1$~meV. We also provide the shift with respect to the $\Delta E(\Gamma^{n=1}_{5})$ transitions and, where available, corresponding literature values. An extended version of this table is provided in the Supplementary Material.}
\label{table:1}
\end{table}

The simple hydrogenic approximation is not expected to be in perfect agreement with the actual transition energies, since AlN exhibits a notable anisotropy regarding both the reduced exciton mass $\mu$ and the relative static permittivity $\varepsilon$. This anisotropy is commonly summarized in the form of the anisotropy parameter $\gamma =\varepsilon_{\perp} \mu_{\perp}/(\varepsilon_{\parallel} \mu_{\parallel})$ with the directions relative to that of the $c$ axis.\cite{baldereschi_IlNuovoCimentoBSer.10_1970,muljarov_J.Math.Phys._2000,gil_Phys.StatusSolidiB_2012,ishiiEffectsStrongElectron2014} For values of $\gamma$ other than 1 or 0, exact analytical expressions for the excitonic levels do not exist. Gil \emph{et al.}\cite{gil_Phys.StatusSolidiB_2012} have provided numerical values for the $\Gamma_{5}^{n=2}$, $\Gamma_{5}^{n=3}$, and $\Gamma_{5}^{n \rightarrow \infty}$ transitions, placing them at 6.076, 6.083, and 6.086~eV for $\gamma \approx 1.8$, the value obtained for AlN using the material parameters given in Ref.~\onlinecite{ishiiEffectsStrongElectron2014}. These energies are systematically lower than those obtained by the simple hydrogen model because $\gamma > 1$, which results in a \emph{reduction} of the exciton binding energy. In fact, while the effective Rydberg energy $\text{Ry}^* = \text{Ry(H)} \,\mu_{\perp}/(\varepsilon_{\parallel} \varepsilon_{\perp})$ amounts to 59~meV, where $\text{Ry(H)} = 13.6$~eV is the ionization energy of the hydrogen atom, the binding energy obtained within the framework of an anisotropic exciton is only 48~meV, as reported previously by Funato \emph{et al.}\cite{funatoHomoepitaxyPhotoluminescenceProperties2012} Even the experimentally observed 1S-2S splitting of 41~meV only allows for an exciton binding energy of 52~meV, still inconsistent with the highest energy feature observed in Fig.~\ref{Figure_3}(b). 

Indeed, for an accurate determination of the exciton binding energy in AlN, Ishii \emph{et al.}\cite{ishiiEffectsStrongElectron2014} have pointed out that it is crucial to not only consider crystal anisotropy, but exchange and electron-phonon interactions as well. They showed the latter to result in a significantly increased exciton binding energy for the $\Gamma_5$ triplet, namely, 64~meV, accompanied by a 1S-2S splitting of 51~meV. Based on our $\Gamma_5$ transition energy of 6.038~eV, these values would result in transition energies for the $\Gamma_{5}^{n=2}$ and $\Gamma_{5}^{n \rightarrow \infty}$ excitons of 6.089 and 6.102~eV, too large compared to the experimental values. Clearly, theory has still to be refined to accurately reproduce the excitonic transition energies observed experimentally. Currently, we share the observation in the literature that the hydrogenic approximation still seems to reproduce experimental data best.\cite{feneberg_Appl.Phys.Lett._2013} 
Consequently, we assign the emission lines observed at 6.091 and 6.095~eV to the $n=3$ and $n \rightarrow \infty$ states of the $\Gamma_{5}$ free exciton. The energy of the $\Gamma^{n \rightarrow \infty}_{5}$ transition corresponds to the exciton continuum and thus the bandgap of AlN. Therefore, the energy difference between $\Gamma^{n=1}_{5}$ and $\Gamma^{n \rightarrow \infty}_{5}$ directly provides the binding energy of the $\Gamma_{5}$ exciton, which thus amounts to 57~meV.

Other experimental values for the free exciton binding energy reported in the literature range from 47 to 67~meV. Those derived by relying on the 1S-2S spitting \cite{murotaniTemperatureDependenceExcitonic2009,fenebergHighexcitationHighresolutionPhotoluminescence2010,funatoHomoepitaxyPhotoluminescenceProperties2012,chichibuExcitonicEmissionDynamics2013,leutePhotoluminescenceHighlyExcited2009,fenebergSharpBoundFree2011,neuschlOpticalIdentificationSilicon2012} or the energy difference between the $n=1$ free exciton transition and the exciton-exciton scattering band $\mathrm{P}^{\infty}$ \cite{fenebergHighexcitationHighresolutionPhotoluminescence2010,ishiiStimulatedEmissionMechanism2022} tend to be close to our value, with the average being about 54~meV. The large overall variation may be due to strain effects, sample heating (especially for high-excitation densities), and the large energy splitting between the $\Gamma_{1}$ and $\Gamma_{5}$ excitons \cite{ishiiEffectsStrongElectron2014}.

Besides the zero-phonon exciton emission, we resolve two longitudinal optical (LO) phonon replicas each for the free exciton state and for the excited states from this sample. In Fig.~\ref{Figure_3}(a), the LO phonon replicas are denoted as $\Gamma^{(n=1)}_{5}\mathrm{-}m\mathrm{LO}$ and $\Gamma^{(n \geq 2)}_{5}\mathrm{-}m\mathrm{LO}$ with $m \in \{1,2\}$. The energy spacing between the phonon replicas and the main peaks is a multiple of 112~meV, in close agreement with the quantized LO phonon energy at the $\Gamma$-point in AlN \cite{davydovPhononDispersionRaman1998}.

Table~\ref{table:1} summarizes all near-bandedge transition energies measured for our MBE films. Column $E$ lists the absolute transition energy, and column $\Delta E(\Gamma^{n=1}_{5})$ the energy shift with respect to the ground state exciton energy $\Gamma^{n=1}_{5}$. Table~\ref{table:1} also gives values for $\Delta E_\mathrm{lit}(\Gamma^{n=1}_{5})$ of other transitions observed in studies in the literature but which are not observed in our films. Note that the actual origin of the $\mathrm{D^{0}X_{\alpha}}$, $\mathrm{D^{0}X_{\beta}}$, $\mathrm{D^{0}X_{\gamma}}$, $\mathrm{D^{0}X_{\delta}}$, $\mathrm{D^{0}X_{2}}$  and $\mathrm{D^{0}X_{\epsilon}}$ lines is not yet known. These lines are generally believed to originate from other donor-bound excitons because of their narrow linewidths, but the contribution of acceptor-bound excitons, inelastic scattering processes or many-body effects  cannot be ruled out. \cite{bryanExcitonTransitionsOxygen2014,funatoHomoepitaxyPhotoluminescenceProperties2012,neuschlOpticalIdentificationSilicon2012} An extended table of all near-bandedge transitions that have been reported in literature can be found in the Supplementary Material (Table ST1).

Finally, it is interesting to discuss the lack of emission from the $n\geq3$ free exciton states in any previous study. At the first glance, it may appear surprising to see recombination from free carriers at a nominal temperature of 10~K, considering the strong Coulomb interaction in AlN. The fraction of excitons at a given temperature and electron-hole density is governed by Saha's law\cite{bieker_Phys.Rev.Lett._2015} and is predicted to decrease with increasing carrier temperature and to increase with increasing electron-hole density below the Mott transition. In the majority of previous works, the samples were excited with an ArF laser, which delivers intense ns pulses with a low repetition rate and an energy close to the band edge. Due to the low repetition rate, the lowest excitation densities are typically around 50~kW/cm$^2$, resulting in a carrier density on the order of $5 \times 10^{17}$~cm$^{-3}$. On the other hand, the excess energy delivered per pulse is comparatively low. For the present CL experiments, the situation is opposite. The continuous-wave excitation by the electron beam facilitates a comparatively low excitation density. In the present case, we estimate a carrier density of not more than $4 \times 10^{16}$~cm$^{-3}$ for an acceleration voltage of 11~kV, as detailed in the Supplementary Material.\cite{jahn_Phys.Rev.Appl._2022} The carrier temperature, however, is typically rather high, since the electron beam creates highly energetic secondary electrons and holes. The cooling of these hot carriers proceeds via the emission of LO phonons, thus creating a nonequilibrium population of hot LO phonons \cite{lugli_Appl.Phys.Lett._1987} that in turn heats the carrier distribution by strongly increasing the probability of LO phonon absorption. Indeed, from the high-energy slope of the $\Gamma^{(n=1)}_{5}\mathrm{-}2\mathrm{LO}$ transition,\cite{bieker_Phys.Rev.Lett._2015,bieker_Phys.Rev.B_2015} we deduce a carrier temperature of 120~K. With this carrier temperature and excitation density, Saha's law (See figure S8 in the Supplementary Material) predicts an excitonic fraction of only 0.82, while for the carrier density estimated for ArF excitation, the fraction increases to 0.95 even for the same carrier temperature. These estimates provide a clear physical explanation for the prominence of excited exciton states in our CL experiments.


\section*{Conclusion}

In summary and conclusion, we have investigated the properties of AlN layers grown by plasma-assisted MBE on both the N- and Al-polar faces of bulk AlN substrates. Regardless of polarity, the layers exhibit atomically smooth surfaces, high structural perfection and purity, and feature intense free exciton emission and suppressed emission from bound excitons and deep-level defects. These results highlight the potential of MBE for the growth of UV emitters. The ability to grow N-polar samples without compromising the crystal quality and thus to change the orientation of the polarization fields in AlN adds an additional degree of freedom that can be exploited in the design of MBE-grown deep-UV emitters. Finally, we have shown that the unique excitation conditions in CL spectroscopy facilitates the detection of several excited states of the $\Gamma_{5}$ exciton, including the excitonic continuum, directly yielding the exciton binding energy of 57~meV.

\section*{Supplementary Material}

See the Supplementary Material for reciprocal space maps as well as symmetric and asymmetric $\omega$ scans of samples I and II, a detailed characterization of sample III (AFM, XRD, SIMS), CL spectra measured under different geometries (sample surface and cross-section), a discussion of the generation volume and excitation density in CL spectroscopy, the phase diagram of the coupled exciton-carrier system according to Saha's law, as well as an extended table of the near-bandedge transitions observed in AlN.

\acknowledgements

The authors thank Chandrashekhar Savant, Zexuan Zhang, and Shivali Agrawal for helpful discussions. This work was partially supported by the Cornell Center for Materials Research with funding from the NSF MRSEC program (Grant No. DMR-1719875). Further support was granted by the National Science Foundation with grant No. NNCI-2025233, DMR-1539918, RAISE-TAQS 1839196, MRI 1631282, and AFOSR Grant No. FA9550-20-1-0148. Finally, LvD acknowledges the "Deutscher Akademischer Austauschdienst" (DAAD) for funding provided through the Research Internships in Science and Engineering (RISE) Professional Scholarship.

\section*{Author Declarations}
\subsection*{Conflict of Interest}
The authors have no conflicts to disclose.


\section*{Data Availability}
The data that support the findings of this study are available within the article and its Supplementary Material. The raw data may be obtained from the corresponding author upon reasonable request.

\bibliography{ms}

\begin{thebibliography}{70}%
\makeatletter
\providecommand \@ifxundefined [1]{%
 \@ifx{#1\undefined}
}%
\providecommand \@ifnum [1]{%
 \ifnum #1\expandafter \@firstoftwo
 \else \expandafter \@secondoftwo
 \fi
}%
\providecommand \@ifx [1]{%
 \ifx #1\expandafter \@firstoftwo
 \else \expandafter \@secondoftwo
 \fi
}%
\providecommand \natexlab [1]{#1}%
\providecommand \enquote  [1]{``#1''}%
\providecommand \bibnamefont  [1]{#1}%
\providecommand \bibfnamefont [1]{#1}%
\providecommand \citenamefont [1]{#1}%
\providecommand \href@noop [0]{\@secondoftwo}%
\providecommand \href [0]{\begingroup \@sanitize@url \@href}%
\providecommand \@href[1]{\@@startlink{#1}\@@href}%
\providecommand \@@href[1]{\endgroup#1\@@endlink}%
\providecommand \@sanitize@url [0]{\catcode `\\12\catcode `\$12\catcode
  `\&12\catcode `\#12\catcode `\^12\catcode `\_12\catcode `\%12\relax}%
\providecommand \@@startlink[1]{}%
\providecommand \@@endlink[0]{}%
\providecommand \url  [0]{\begingroup\@sanitize@url \@url }%
\providecommand \@url [1]{\endgroup\@href {#1}{\urlprefix }}%
\providecommand \urlprefix  [0]{URL }%
\providecommand \Eprint [0]{\href }%
\providecommand \doibase [0]{https://doi.org/}%
\providecommand \selectlanguage [0]{\@gobble}%
\providecommand \bibinfo  [0]{\@secondoftwo}%
\providecommand \bibfield  [0]{\@secondoftwo}%
\providecommand \translation [1]{[#1]}%
\providecommand \BibitemOpen [0]{}%
\providecommand \bibitemStop [0]{}%
\providecommand \bibitemNoStop [0]{.\EOS\space}%
\providecommand \EOS [0]{\spacefactor3000\relax}%
\providecommand \BibitemShut  [1]{\csname bibitem#1\endcsname}%
\let\auto@bib@innerbib\@empty
\bibitem [{\citenamefont {Briegleb}\ and\ \citenamefont
  {Geuther}(1862)}]{brieglebUeberStickstoffmagnesiumUnd1862}%
  \BibitemOpen
  \bibfield  {author} {\bibinfo {author} {\bibfnamefont {{\relax
  Fr}.}~\bibnamefont {Briegleb}}\ and\ \bibinfo {author} {\bibfnamefont
  {A.}~\bibnamefont {Geuther}},\ }\bibfield  {title} {\enquote {\bibinfo
  {title} {Ueber das {{Stickstoffmagnesium}} und die {{Affinit\"aten}} des
  {{Stickgases}} zu {{Metallen}}},}\ }\href
  {https://doi.org/10.1002/jlac.18621230212} {\bibfield  {journal} {\bibinfo
  {journal} {Justus Liebigs Ann. Chem.}\ }\textbf {\bibinfo {volume} {123}},\
  \bibinfo {pages} {228--241} (\bibinfo {year} {1862})}\BibitemShut {NoStop}%
\bibitem [{\citenamefont {Muralt}\ \emph {et~al.}(2009)\citenamefont {Muralt},
  \citenamefont {Conde}, \citenamefont {Artieda}, \citenamefont {Martin},\ and\
  \citenamefont {Cantoni}}]{muralt_Int.J.Microw.Wirel.Technol._2009}%
  \BibitemOpen
  \bibfield  {author} {\bibinfo {author} {\bibfnamefont {P.}~\bibnamefont
  {Muralt}}, \bibinfo {author} {\bibfnamefont {J.}~\bibnamefont {Conde}},
  \bibinfo {author} {\bibfnamefont {A.}~\bibnamefont {Artieda}}, \bibinfo
  {author} {\bibfnamefont {F.}~\bibnamefont {Martin}},\ and\ \bibinfo {author}
  {\bibfnamefont {M.}~\bibnamefont {Cantoni}},\ }\bibfield  {title} {\enquote
  {\bibinfo {title} {Piezoelectric materials parameters for piezoelectric thin
  films in {{GHz}} applications},}\ }\href
  {https://doi.org/10.1017/S1759078709000038} {\bibfield  {journal} {\bibinfo
  {journal} {Int. J. Microw. Wirel. Technol.}\ }\textbf {\bibinfo {volume}
  {1}},\ \bibinfo {pages} {19--27} (\bibinfo {year} {2009})}\BibitemShut
  {NoStop}%
\bibitem [{\citenamefont {Khachariya}\ \emph {et~al.}(2022)\citenamefont
  {Khachariya}, \citenamefont {Mita}, \citenamefont {Reddy}, \citenamefont
  {Dangi}, \citenamefont {Dycus}, \citenamefont {Bagheri}, \citenamefont
  {Breckenridge}, \citenamefont {Sengupta}, \citenamefont {Rathkanthiwar},
  \citenamefont {Kirste}, \citenamefont {Kohn}, \citenamefont {Sitar},
  \citenamefont {Collazo},\ and\ \citenamefont
  {Pavlidis}}]{khachariyaRecord10MV2022}%
  \BibitemOpen
  \bibfield  {author} {\bibinfo {author} {\bibfnamefont {D.}~\bibnamefont
  {Khachariya}}, \bibinfo {author} {\bibfnamefont {S.}~\bibnamefont {Mita}},
  \bibinfo {author} {\bibfnamefont {P.}~\bibnamefont {Reddy}}, \bibinfo
  {author} {\bibfnamefont {S.}~\bibnamefont {Dangi}}, \bibinfo {author}
  {\bibfnamefont {J.~H.}\ \bibnamefont {Dycus}}, \bibinfo {author}
  {\bibfnamefont {P.}~\bibnamefont {Bagheri}}, \bibinfo {author} {\bibfnamefont
  {M.~H.}\ \bibnamefont {Breckenridge}}, \bibinfo {author} {\bibfnamefont
  {R.}~\bibnamefont {Sengupta}}, \bibinfo {author} {\bibfnamefont
  {S.}~\bibnamefont {Rathkanthiwar}}, \bibinfo {author} {\bibfnamefont
  {R.}~\bibnamefont {Kirste}}, \bibinfo {author} {\bibfnamefont
  {E.}~\bibnamefont {Kohn}}, \bibinfo {author} {\bibfnamefont {Z.}~\bibnamefont
  {Sitar}}, \bibinfo {author} {\bibfnamefont {R.}~\bibnamefont {Collazo}},\
  and\ \bibinfo {author} {\bibfnamefont {S.}~\bibnamefont {Pavlidis}},\
  }\bibfield  {title} {\enquote {\bibinfo {title} {Record {$>$}10\,{{MV}}/cm
  mesa breakdown fields in {{Al$_{0.85}$Ga$_{0.15}$N/Al$_{0.6}$Ga$_{0.4}$N}}
  high electron mobility transistors on native {{AlN}} substrates},}\ }\href
  {https://doi.org/10.1063/5.0083966} {\bibfield  {journal} {\bibinfo
  {journal} {Appl. Phys. Lett.}\ }\textbf {\bibinfo {volume} {120}},\ \bibinfo
  {pages} {172106} (\bibinfo {year} {2022})}\BibitemShut {NoStop}%
\bibitem [{\citenamefont {Hussain}\ \emph {et~al.}(2023)\citenamefont
  {Hussain}, \citenamefont {Mamun}, \citenamefont {Floyd}, \citenamefont
  {Alam}, \citenamefont {Liao}, \citenamefont {Huynh}, \citenamefont {Wang},
  \citenamefont {Goorsky}, \citenamefont {Chandrashekhar}, \citenamefont
  {Simin},\ and\ \citenamefont {Khan}}]{hussainHighFigureMerit2023}%
  \BibitemOpen
  \bibfield  {author} {\bibinfo {author} {\bibfnamefont {K.}~\bibnamefont
  {Hussain}}, \bibinfo {author} {\bibfnamefont {A.}~\bibnamefont {Mamun}},
  \bibinfo {author} {\bibfnamefont {R.}~\bibnamefont {Floyd}}, \bibinfo
  {author} {\bibfnamefont {M.~D.}\ \bibnamefont {Alam}}, \bibinfo {author}
  {\bibfnamefont {M.~E.}\ \bibnamefont {Liao}}, \bibinfo {author}
  {\bibfnamefont {K.}~\bibnamefont {Huynh}}, \bibinfo {author} {\bibfnamefont
  {Y.}~\bibnamefont {Wang}}, \bibinfo {author} {\bibfnamefont {M.~S.}\
  \bibnamefont {Goorsky}}, \bibinfo {author} {\bibfnamefont {{\relax
  MVS}.}~\bibnamefont {Chandrashekhar}}, \bibinfo {author} {\bibfnamefont
  {G.}~\bibnamefont {Simin}},\ and\ \bibinfo {author} {\bibfnamefont
  {A.}~\bibnamefont {Khan}},\ }\bibfield  {title} {\enquote {\bibinfo {title}
  {High figure of merit extreme bandgap
  {{Al0}}.{{87Ga0}}.{{13N-Al0}}.{{64Ga0}}.{{36N}} heterostructures over bulk
  {{AlN}} substrates},}\ }\href {https://doi.org/10.35848/1882-0786/acb487}
  {\bibfield  {journal} {\bibinfo  {journal} {Appl. Phys. Express}\ } (\bibinfo
  {year} {2023}),\ 10.35848/1882-0786/acb487}\BibitemShut {NoStop}%
\bibitem [{\citenamefont {Cheng}\ \emph {et~al.}(2020)\citenamefont {Cheng},
  \citenamefont {Koh}, \citenamefont {Mamun}, \citenamefont {Shi},
  \citenamefont {Bai}, \citenamefont {Huynh}, \citenamefont {Yates},
  \citenamefont {Liu}, \citenamefont {Li}, \citenamefont {Lee}, \citenamefont
  {Liao}, \citenamefont {Wang}, \citenamefont {Yu}, \citenamefont {Kushimoto},
  \citenamefont {Luo}, \citenamefont {Goorsky}, \citenamefont {Hopkins},
  \citenamefont {Amano}, \citenamefont {Khan},\ and\ \citenamefont
  {Graham}}]{chengExperimentalObservationHigh2020}%
  \BibitemOpen
  \bibfield  {author} {\bibinfo {author} {\bibfnamefont {Z.}~\bibnamefont
  {Cheng}}, \bibinfo {author} {\bibfnamefont {Y.~R.}\ \bibnamefont {Koh}},
  \bibinfo {author} {\bibfnamefont {A.}~\bibnamefont {Mamun}}, \bibinfo
  {author} {\bibfnamefont {J.}~\bibnamefont {Shi}}, \bibinfo {author}
  {\bibfnamefont {T.}~\bibnamefont {Bai}}, \bibinfo {author} {\bibfnamefont
  {K.}~\bibnamefont {Huynh}}, \bibinfo {author} {\bibfnamefont
  {L.}~\bibnamefont {Yates}}, \bibinfo {author} {\bibfnamefont
  {Z.}~\bibnamefont {Liu}}, \bibinfo {author} {\bibfnamefont {R.}~\bibnamefont
  {Li}}, \bibinfo {author} {\bibfnamefont {E.}~\bibnamefont {Lee}}, \bibinfo
  {author} {\bibfnamefont {M.~E.}\ \bibnamefont {Liao}}, \bibinfo {author}
  {\bibfnamefont {Y.}~\bibnamefont {Wang}}, \bibinfo {author} {\bibfnamefont
  {H.~M.}\ \bibnamefont {Yu}}, \bibinfo {author} {\bibfnamefont
  {M.}~\bibnamefont {Kushimoto}}, \bibinfo {author} {\bibfnamefont
  {T.}~\bibnamefont {Luo}}, \bibinfo {author} {\bibfnamefont {M.~S.}\
  \bibnamefont {Goorsky}}, \bibinfo {author} {\bibfnamefont {P.~E.}\
  \bibnamefont {Hopkins}}, \bibinfo {author} {\bibfnamefont {H.}~\bibnamefont
  {Amano}}, \bibinfo {author} {\bibfnamefont {A.}~\bibnamefont {Khan}},\ and\
  \bibinfo {author} {\bibfnamefont {S.}~\bibnamefont {Graham}},\ }\bibfield
  {title} {\enquote {\bibinfo {title} {Experimental observation of high
  intrinsic thermal conductivity of {{AlN}}},}\ }\href
  {https://doi.org/10.1103/PhysRevMaterials.4.044602} {\bibfield  {journal}
  {\bibinfo  {journal} {Phys. Rev. Mater.}\ }\textbf {\bibinfo {volume} {4}},\
  \bibinfo {pages} {044602} (\bibinfo {year} {2020})}\BibitemShut {NoStop}%
\bibitem [{\citenamefont {Koppe}, \citenamefont {Hofs{\"a}ss},\ and\
  \citenamefont {Vetter}(2016)}]{koppeOverviewBandedgeDefect2016}%
  \BibitemOpen
  \bibfield  {author} {\bibinfo {author} {\bibfnamefont {T.}~\bibnamefont
  {Koppe}}, \bibinfo {author} {\bibfnamefont {H.}~\bibnamefont {Hofs{\"a}ss}},\
  and\ \bibinfo {author} {\bibfnamefont {U.}~\bibnamefont {Vetter}},\
  }\bibfield  {title} {\enquote {\bibinfo {title} {Overview of band-edge and
  defect related luminescence in aluminum nitride},}\ }\href
  {https://doi.org/10.1016/j.jlumin.2016.05.055} {\bibfield  {journal}
  {\bibinfo  {journal} {J. Lumin.}\ }\textbf {\bibinfo {volume} {178}},\
  \bibinfo {pages} {267--281} (\bibinfo {year} {2016})}\BibitemShut {NoStop}%
\bibitem [{\citenamefont {Doan}\ \emph {et~al.}(2016)\citenamefont {Doan},
  \citenamefont {Li}, \citenamefont {Lin},\ and\ \citenamefont
  {Jiang}}]{doanBandgapExcitonBinding2016}%
  \BibitemOpen
  \bibfield  {author} {\bibinfo {author} {\bibfnamefont {T.~C.}\ \bibnamefont
  {Doan}}, \bibinfo {author} {\bibfnamefont {J.}~\bibnamefont {Li}}, \bibinfo
  {author} {\bibfnamefont {J.~Y.}\ \bibnamefont {Lin}},\ and\ \bibinfo {author}
  {\bibfnamefont {H.~X.}\ \bibnamefont {Jiang}},\ }\bibfield  {title} {\enquote
  {\bibinfo {title} {Bandgap and exciton binding energies of hexagonal boron
  nitride probed by photocurrent excitation spectroscopy},}\ }\href
  {https://doi.org/10.1063/1.4963128} {\bibfield  {journal} {\bibinfo
  {journal} {Appl. Phys. Lett.}\ }\textbf {\bibinfo {volume} {109}},\ \bibinfo
  {pages} {122101} (\bibinfo {year} {2016})}\BibitemShut {NoStop}%
\bibitem [{\citenamefont {Bader}\ \emph {et~al.}(2020)\citenamefont {Bader},
  \citenamefont {Lee}, \citenamefont {Chaudhuri}, \citenamefont {Huang},
  \citenamefont {Hickman}, \citenamefont {Molnar}, \citenamefont {Xing},
  \citenamefont {Jena}, \citenamefont {Then}, \citenamefont {Chowdhury},\ and\
  \citenamefont {Palacios}}]{baderProspectsWideBandgap2020}%
  \BibitemOpen
  \bibfield  {author} {\bibinfo {author} {\bibfnamefont {S.~J.}\ \bibnamefont
  {Bader}}, \bibinfo {author} {\bibfnamefont {H.}~\bibnamefont {Lee}}, \bibinfo
  {author} {\bibfnamefont {R.}~\bibnamefont {Chaudhuri}}, \bibinfo {author}
  {\bibfnamefont {S.}~\bibnamefont {Huang}}, \bibinfo {author} {\bibfnamefont
  {A.}~\bibnamefont {Hickman}}, \bibinfo {author} {\bibfnamefont
  {A.}~\bibnamefont {Molnar}}, \bibinfo {author} {\bibfnamefont {H.~G.}\
  \bibnamefont {Xing}}, \bibinfo {author} {\bibfnamefont {D.}~\bibnamefont
  {Jena}}, \bibinfo {author} {\bibfnamefont {H.~W.}\ \bibnamefont {Then}},
  \bibinfo {author} {\bibfnamefont {N.}~\bibnamefont {Chowdhury}},\ and\
  \bibinfo {author} {\bibfnamefont {T.}~\bibnamefont {Palacios}},\ }\bibfield
  {title} {\enquote {\bibinfo {title} {Prospects for {{Wide Bandgap}} and
  {{Ultrawide Bandgap CMOS Devices}}},}\ }\href
  {https://doi.org/10.1109/TED.2020.3010471} {\bibfield  {journal} {\bibinfo
  {journal} {IEEE Trans. Electron Devices}\ }\textbf {\bibinfo {volume} {67}},\
  \bibinfo {pages} {4010--4020} (\bibinfo {year} {2020})}\BibitemShut {NoStop}%
\bibitem [{\citenamefont {Hickman}\ \emph {et~al.}(2021)\citenamefont
  {Hickman}, \citenamefont {Chaudhuri}, \citenamefont {Li}, \citenamefont
  {Nomoto}, \citenamefont {Bader}, \citenamefont {Hwang}, \citenamefont
  {Xing},\ and\ \citenamefont {Jena}}]{hickmanFirstRFPower2021}%
  \BibitemOpen
  \bibfield  {author} {\bibinfo {author} {\bibfnamefont {A.}~\bibnamefont
  {Hickman}}, \bibinfo {author} {\bibfnamefont {R.}~\bibnamefont {Chaudhuri}},
  \bibinfo {author} {\bibfnamefont {L.}~\bibnamefont {Li}}, \bibinfo {author}
  {\bibfnamefont {K.}~\bibnamefont {Nomoto}}, \bibinfo {author} {\bibfnamefont
  {S.~J.}\ \bibnamefont {Bader}}, \bibinfo {author} {\bibfnamefont {J.~C.~M.}\
  \bibnamefont {Hwang}}, \bibinfo {author} {\bibfnamefont {H.~G.}\ \bibnamefont
  {Xing}},\ and\ \bibinfo {author} {\bibfnamefont {D.}~\bibnamefont {Jena}},\
  }\bibfield  {title} {\enquote {\bibinfo {title} {First {{RF Power Operation}}
  of {{AlN}}/{{GaN}}/{{AlN HEMTs With}} {$>$}3 {{A}}/mm and 3 {{W}}/mm at 10
  {{GHz}}},}\ }\href {https://doi.org/10.1109/JEDS.2020.3042050} {\bibfield
  {journal} {\bibinfo  {journal} {IEEE J. Electron Devices Soc.}\ }\textbf
  {\bibinfo {volume} {9}},\ \bibinfo {pages} {121--124} (\bibinfo {year}
  {2021})}\BibitemShut {NoStop}%
\bibitem [{\citenamefont {Kim}\ \emph {et~al.}(2023)\citenamefont {Kim},
  \citenamefont {Zhang}, \citenamefont {Encomendero}, \citenamefont {Singhal},
  \citenamefont {Nomoto}, \citenamefont {Hickman}, \citenamefont {Wang},
  \citenamefont {Fay}, \citenamefont {Toita}, \citenamefont {Jena},\ and\
  \citenamefont {Xing}}]{kimNpolarGaNAlGaN2023}%
  \BibitemOpen
  \bibfield  {author} {\bibinfo {author} {\bibfnamefont {E.}~\bibnamefont
  {Kim}}, \bibinfo {author} {\bibfnamefont {Z.}~\bibnamefont {Zhang}}, \bibinfo
  {author} {\bibfnamefont {J.}~\bibnamefont {Encomendero}}, \bibinfo {author}
  {\bibfnamefont {J.}~\bibnamefont {Singhal}}, \bibinfo {author} {\bibfnamefont
  {K.}~\bibnamefont {Nomoto}}, \bibinfo {author} {\bibfnamefont
  {A.}~\bibnamefont {Hickman}}, \bibinfo {author} {\bibfnamefont
  {C.}~\bibnamefont {Wang}}, \bibinfo {author} {\bibfnamefont {P.}~\bibnamefont
  {Fay}}, \bibinfo {author} {\bibfnamefont {M.}~\bibnamefont {Toita}}, \bibinfo
  {author} {\bibfnamefont {D.}~\bibnamefont {Jena}},\ and\ \bibinfo {author}
  {\bibfnamefont {H.~G.}\ \bibnamefont {Xing}},\ }\bibfield  {title} {\enquote
  {\bibinfo {title} {N-polar {{GaN}}/{{AlGaN}}/{{AlN}} high electron mobility
  transistors on single-crystal bulk {{AlN}} substrates},}\ }\href
  {https://doi.org/10.1063/5.0138939} {\bibfield  {journal} {\bibinfo
  {journal} {Appl. Phys. Lett.}\ }\textbf {\bibinfo {volume} {122}},\ \bibinfo
  {pages} {092104} (\bibinfo {year} {2023})}\BibitemShut {NoStop}%
\bibitem [{\citenamefont {Amano}\ \emph {et~al.}(2020)\citenamefont {Amano},
  \citenamefont {Collazo}, \citenamefont {Santi}, \citenamefont {Einfeldt},
  \citenamefont {Funato}, \citenamefont {Glaab}, \citenamefont {Hagedorn},
  \citenamefont {Hirano}, \citenamefont {Hirayama}, \citenamefont {Ishii},
  \citenamefont {Kashima}, \citenamefont {Kawakami}, \citenamefont {Kirste},
  \citenamefont {Kneissl}, \citenamefont {Martin}, \citenamefont {Mehnke},
  \citenamefont {Meneghini}, \citenamefont {Ougazzaden}, \citenamefont
  {Parbrook}, \citenamefont {Rajan}, \citenamefont {Reddy}, \citenamefont
  {R{\"o}mer}, \citenamefont {Ruschel}, \citenamefont {Sarkar}, \citenamefont
  {Scholz}, \citenamefont {Schowalter}, \citenamefont {Shields}, \citenamefont
  {Sitar}, \citenamefont {Sulmoni}, \citenamefont {Wang}, \citenamefont
  {Wernicke}, \citenamefont {Weyers}, \citenamefont {Witzigmann}, \citenamefont
  {Wu}, \citenamefont {Wunderer},\ and\ \citenamefont
  {Zhang}}]{amano2020UVEmitter2020}%
  \BibitemOpen
  \bibfield  {author} {\bibinfo {author} {\bibfnamefont {H.}~\bibnamefont
  {Amano}}, \bibinfo {author} {\bibfnamefont {R.}~\bibnamefont {Collazo}},
  \bibinfo {author} {\bibfnamefont {C.~D.}\ \bibnamefont {Santi}}, \bibinfo
  {author} {\bibfnamefont {S.}~\bibnamefont {Einfeldt}}, \bibinfo {author}
  {\bibfnamefont {M.}~\bibnamefont {Funato}}, \bibinfo {author} {\bibfnamefont
  {J.}~\bibnamefont {Glaab}}, \bibinfo {author} {\bibfnamefont
  {S.}~\bibnamefont {Hagedorn}}, \bibinfo {author} {\bibfnamefont
  {A.}~\bibnamefont {Hirano}}, \bibinfo {author} {\bibfnamefont
  {H.}~\bibnamefont {Hirayama}}, \bibinfo {author} {\bibfnamefont
  {R.}~\bibnamefont {Ishii}}, \bibinfo {author} {\bibfnamefont
  {Y.}~\bibnamefont {Kashima}}, \bibinfo {author} {\bibfnamefont
  {Y.}~\bibnamefont {Kawakami}}, \bibinfo {author} {\bibfnamefont
  {R.}~\bibnamefont {Kirste}}, \bibinfo {author} {\bibfnamefont
  {M.}~\bibnamefont {Kneissl}}, \bibinfo {author} {\bibfnamefont
  {R.}~\bibnamefont {Martin}}, \bibinfo {author} {\bibfnamefont
  {F.}~\bibnamefont {Mehnke}}, \bibinfo {author} {\bibfnamefont
  {M.}~\bibnamefont {Meneghini}}, \bibinfo {author} {\bibfnamefont
  {A.}~\bibnamefont {Ougazzaden}}, \bibinfo {author} {\bibfnamefont {P.~J.}\
  \bibnamefont {Parbrook}}, \bibinfo {author} {\bibfnamefont {S.}~\bibnamefont
  {Rajan}}, \bibinfo {author} {\bibfnamefont {P.}~\bibnamefont {Reddy}},
  \bibinfo {author} {\bibfnamefont {F.}~\bibnamefont {R{\"o}mer}}, \bibinfo
  {author} {\bibfnamefont {J.}~\bibnamefont {Ruschel}}, \bibinfo {author}
  {\bibfnamefont {B.}~\bibnamefont {Sarkar}}, \bibinfo {author} {\bibfnamefont
  {F.}~\bibnamefont {Scholz}}, \bibinfo {author} {\bibfnamefont {L.~J.}\
  \bibnamefont {Schowalter}}, \bibinfo {author} {\bibfnamefont
  {P.}~\bibnamefont {Shields}}, \bibinfo {author} {\bibfnamefont
  {Z.}~\bibnamefont {Sitar}}, \bibinfo {author} {\bibfnamefont
  {L.}~\bibnamefont {Sulmoni}}, \bibinfo {author} {\bibfnamefont
  {T.}~\bibnamefont {Wang}}, \bibinfo {author} {\bibfnamefont {T.}~\bibnamefont
  {Wernicke}}, \bibinfo {author} {\bibfnamefont {M.}~\bibnamefont {Weyers}},
  \bibinfo {author} {\bibfnamefont {B.}~\bibnamefont {Witzigmann}}, \bibinfo
  {author} {\bibfnamefont {Y.-R.}\ \bibnamefont {Wu}}, \bibinfo {author}
  {\bibfnamefont {T.}~\bibnamefont {Wunderer}},\ and\ \bibinfo {author}
  {\bibfnamefont {Y.}~\bibnamefont {Zhang}},\ }\bibfield  {title} {\enquote
  {\bibinfo {title} {The 2020 {{UV}} emitter roadmap},}\ }\href
  {https://doi.org/10.1088/1361-6463/aba64c} {\bibfield  {journal} {\bibinfo
  {journal} {J. Phys. D: Appl. Phys.}\ }\textbf {\bibinfo {volume} {53}},\
  \bibinfo {pages} {503001} (\bibinfo {year} {2020})}\BibitemShut {NoStop}%
\bibitem [{\citenamefont {Zhang}\ \emph {et~al.}(2019)\citenamefont {Zhang},
  \citenamefont {Kushimoto}, \citenamefont {Sakai}, \citenamefont {Sugiyama},
  \citenamefont {Schowalter}, \citenamefont {Sasaoka},\ and\ \citenamefont
  {Amano}}]{zhang271NmDeepultraviolet2019}%
  \BibitemOpen
  \bibfield  {author} {\bibinfo {author} {\bibfnamefont {Z.}~\bibnamefont
  {Zhang}}, \bibinfo {author} {\bibfnamefont {M.}~\bibnamefont {Kushimoto}},
  \bibinfo {author} {\bibfnamefont {T.}~\bibnamefont {Sakai}}, \bibinfo
  {author} {\bibfnamefont {N.}~\bibnamefont {Sugiyama}}, \bibinfo {author}
  {\bibfnamefont {L.~J.}\ \bibnamefont {Schowalter}}, \bibinfo {author}
  {\bibfnamefont {C.}~\bibnamefont {Sasaoka}},\ and\ \bibinfo {author}
  {\bibfnamefont {H.}~\bibnamefont {Amano}},\ }\bibfield  {title} {\enquote
  {\bibinfo {title} {A 271.8 nm deep-ultraviolet laser diode for room
  temperature operation},}\ }\href {https://doi.org/10.7567/1882-0786/ab50e0}
  {\bibfield  {journal} {\bibinfo  {journal} {Appl. Phys. Express}\ }\textbf
  {\bibinfo {volume} {12}},\ \bibinfo {pages} {124003} (\bibinfo {year}
  {2019})}\BibitemShut {NoStop}%
\bibitem [{\citenamefont {Zhang}\ \emph
  {et~al.}(2022{\natexlab{a}})\citenamefont {Zhang}, \citenamefont {Kushimoto},
  \citenamefont {Yoshikawa}, \citenamefont {Aoto}, \citenamefont {Sasaoka},
  \citenamefont {Schowalter},\ and\ \citenamefont
  {Amano}}]{zhangKeyTemperaturedependentCharacteristics2022}%
  \BibitemOpen
  \bibfield  {author} {\bibinfo {author} {\bibfnamefont {Z.}~\bibnamefont
  {Zhang}}, \bibinfo {author} {\bibfnamefont {M.}~\bibnamefont {Kushimoto}},
  \bibinfo {author} {\bibfnamefont {A.}~\bibnamefont {Yoshikawa}}, \bibinfo
  {author} {\bibfnamefont {K.}~\bibnamefont {Aoto}}, \bibinfo {author}
  {\bibfnamefont {C.}~\bibnamefont {Sasaoka}}, \bibinfo {author} {\bibfnamefont
  {L.~J.}\ \bibnamefont {Schowalter}},\ and\ \bibinfo {author} {\bibfnamefont
  {H.}~\bibnamefont {Amano}},\ }\bibfield  {title} {\enquote {\bibinfo {title}
  {Key temperature-dependent characteristics of {{AlGaN-based UV-C}} laser
  diode and demonstration of room-temperature continuous-wave lasing},}\ }\href
  {https://doi.org/10.1063/5.0124480} {\bibfield  {journal} {\bibinfo
  {journal} {Appl. Phys. Lett.}\ }\textbf {\bibinfo {volume} {121}},\ \bibinfo
  {pages} {222103} (\bibinfo {year} {2022}{\natexlab{a}})}\BibitemShut
  {NoStop}%
\bibitem [{\citenamefont {{van Deurzen}}\ \emph {et~al.}(2022)\citenamefont
  {{van Deurzen}}, \citenamefont {Page}, \citenamefont {Protasenko},
  \citenamefont {Nomoto}, \citenamefont {Xing},\ and\ \citenamefont
  {Jena}}]{vandeurzenOpticallyPumpedDeepUV2022}%
  \BibitemOpen
  \bibfield  {author} {\bibinfo {author} {\bibfnamefont {L.}~\bibnamefont {{van
  Deurzen}}}, \bibinfo {author} {\bibfnamefont {R.}~\bibnamefont {Page}},
  \bibinfo {author} {\bibfnamefont {V.}~\bibnamefont {Protasenko}}, \bibinfo
  {author} {\bibfnamefont {K.}~\bibnamefont {Nomoto}}, \bibinfo {author}
  {\bibfnamefont {H.~G.}\ \bibnamefont {Xing}},\ and\ \bibinfo {author}
  {\bibfnamefont {D.}~\bibnamefont {Jena}},\ }\bibfield  {title} {\enquote
  {\bibinfo {title} {Optically pumped deep-{{UV}} multimode lasing in {{AlGaN}}
  double heterostructure grown by molecular beam homoepitaxy},}\ }\href
  {https://doi.org/10.1063/5.0085365} {\bibfield  {journal} {\bibinfo
  {journal} {AIP Adv.}\ }\textbf {\bibinfo {volume} {12}},\ \bibinfo {pages}
  {035023} (\bibinfo {year} {2022})}\BibitemShut {NoStop}%
\bibitem [{\citenamefont
  {Ginzburg}(1976)}]{ginzburg_key_problems_physics_astrophysics}%
  \BibitemOpen
  \bibfield  {author} {\bibinfo {author} {\bibfnamefont {V.}~\bibnamefont
  {Ginzburg}},\ }\href@noop {} {\emph {\bibinfo {title} {Key problems of
  physics and astrophysics}}}\ (\bibinfo  {publisher} {Mir Publishers},\
  \bibinfo {address} {Moscow},\ \bibinfo {year} {1976})\BibitemShut {NoStop}%
\bibitem [{\citenamefont {Zhou}\ \emph {et~al.}(2020)\citenamefont {Zhou},
  \citenamefont {Zhang}, \citenamefont {Li}, \citenamefont {Golovynskyi},
  \citenamefont {Tang}, \citenamefont {Wu}, \citenamefont {Wang},\ and\
  \citenamefont {Li}}]{zhouBandgapPhotoluminescenceAlN2020}%
  \BibitemOpen
  \bibfield  {author} {\bibinfo {author} {\bibfnamefont {Q.}~\bibnamefont
  {Zhou}}, \bibinfo {author} {\bibfnamefont {Z.}~\bibnamefont {Zhang}},
  \bibinfo {author} {\bibfnamefont {H.}~\bibnamefont {Li}}, \bibinfo {author}
  {\bibfnamefont {S.}~\bibnamefont {Golovynskyi}}, \bibinfo {author}
  {\bibfnamefont {X.}~\bibnamefont {Tang}}, \bibinfo {author} {\bibfnamefont
  {H.}~\bibnamefont {Wu}}, \bibinfo {author} {\bibfnamefont {J.}~\bibnamefont
  {Wang}},\ and\ \bibinfo {author} {\bibfnamefont {B.}~\bibnamefont {Li}},\
  }\bibfield  {title} {\enquote {\bibinfo {title} {Below bandgap
  photoluminescence of an {{AlN}} crystal: {{Co-existence}} of two different
  charging states of a defect center},}\ }\href
  {https://doi.org/10.1063/5.0012685} {\bibfield  {journal} {\bibinfo
  {journal} {APL Mater.}\ }\textbf {\bibinfo {volume} {8}},\ \bibinfo {pages}
  {081107} (\bibinfo {year} {2020})}\BibitemShut {NoStop}%
\bibitem [{\citenamefont {Feneberg}\ \emph {et~al.}(2010)\citenamefont
  {Feneberg}, \citenamefont {Leute}, \citenamefont {Neuschl}, \citenamefont
  {Thonke},\ and\ \citenamefont
  {Bickermann}}]{fenebergHighexcitationHighresolutionPhotoluminescence2010}%
  \BibitemOpen
  \bibfield  {author} {\bibinfo {author} {\bibfnamefont {M.}~\bibnamefont
  {Feneberg}}, \bibinfo {author} {\bibfnamefont {R.~A.~R.}\ \bibnamefont
  {Leute}}, \bibinfo {author} {\bibfnamefont {B.}~\bibnamefont {Neuschl}},
  \bibinfo {author} {\bibfnamefont {K.}~\bibnamefont {Thonke}},\ and\ \bibinfo
  {author} {\bibfnamefont {M.}~\bibnamefont {Bickermann}},\ }\bibfield  {title}
  {\enquote {\bibinfo {title} {High-excitation and high-resolution
  photoluminescence spectra of bulk {{AlN}}},}\ }\href
  {https://doi.org/10.1103/PhysRevB.82.075208} {\bibfield  {journal} {\bibinfo
  {journal} {Phys. Rev. B}\ }\textbf {\bibinfo {volume} {82}},\ \bibinfo
  {pages} {075208} (\bibinfo {year} {2010})}\BibitemShut {NoStop}%
\bibitem [{\citenamefont {Chichibu}\ \emph {et~al.}(2013)\citenamefont
  {Chichibu}, \citenamefont {Hazu}, \citenamefont {Ishikawa}, \citenamefont
  {Tashiro}, \citenamefont {Ohtomo}, \citenamefont {Furusawa}, \citenamefont
  {Uedono}, \citenamefont {Mita}, \citenamefont {Xie}, \citenamefont
  {Collazo},\ and\ \citenamefont
  {Sitar}}]{chichibuExcitonicEmissionDynamics2013}%
  \BibitemOpen
  \bibfield  {author} {\bibinfo {author} {\bibfnamefont {S.~F.}\ \bibnamefont
  {Chichibu}}, \bibinfo {author} {\bibfnamefont {K.}~\bibnamefont {Hazu}},
  \bibinfo {author} {\bibfnamefont {Y.}~\bibnamefont {Ishikawa}}, \bibinfo
  {author} {\bibfnamefont {M.}~\bibnamefont {Tashiro}}, \bibinfo {author}
  {\bibfnamefont {T.}~\bibnamefont {Ohtomo}}, \bibinfo {author} {\bibfnamefont
  {K.}~\bibnamefont {Furusawa}}, \bibinfo {author} {\bibfnamefont
  {A.}~\bibnamefont {Uedono}}, \bibinfo {author} {\bibfnamefont
  {S.}~\bibnamefont {Mita}}, \bibinfo {author} {\bibfnamefont {J.}~\bibnamefont
  {Xie}}, \bibinfo {author} {\bibfnamefont {R.}~\bibnamefont {Collazo}},\ and\
  \bibinfo {author} {\bibfnamefont {Z.}~\bibnamefont {Sitar}},\ }\bibfield
  {title} {\enquote {\bibinfo {title} {Excitonic emission dynamics in
  homoepitaxial {{AlN}} films studied using polarized and spatio-time-resolved
  cathodoluminescence measurements},}\ }\href
  {https://doi.org/10.1063/1.4823826} {\bibfield  {journal} {\bibinfo
  {journal} {Appl. Phys. Lett.}\ }\textbf {\bibinfo {volume} {103}},\ \bibinfo
  {pages} {142103} (\bibinfo {year} {2013})}\BibitemShut {NoStop}%
\bibitem [{\citenamefont {Thonke}\ \emph {et~al.}(2017)\citenamefont {Thonke},
  \citenamefont {Lamprecht}, \citenamefont {Collazo},\ and\ \citenamefont
  {Sitar}}]{thonkeOpticalSignaturesSilicon2017}%
  \BibitemOpen
  \bibfield  {author} {\bibinfo {author} {\bibfnamefont {K.}~\bibnamefont
  {Thonke}}, \bibinfo {author} {\bibfnamefont {M.}~\bibnamefont {Lamprecht}},
  \bibinfo {author} {\bibfnamefont {R.}~\bibnamefont {Collazo}},\ and\ \bibinfo
  {author} {\bibfnamefont {Z.}~\bibnamefont {Sitar}},\ }\bibfield  {title}
  {\enquote {\bibinfo {title} {Optical signatures of silicon and oxygen related
  {{DX}} centers in {{AlN}}},}\ }\href {https://doi.org/10.1002/pssa.201600749}
  {\bibfield  {journal} {\bibinfo  {journal} {Phys. Status Solidi A}\ }\textbf
  {\bibinfo {volume} {214}},\ \bibinfo {pages} {1600749} (\bibinfo {year}
  {2017})}\BibitemShut {NoStop}%
\bibitem [{\citenamefont {Ishii}\ \emph {et~al.}(2022)\citenamefont {Ishii},
  \citenamefont {Nagashima}, \citenamefont {Yamamoto}, \citenamefont {Hitomi},
  \citenamefont {Funato},\ and\ \citenamefont
  {Kawakami}}]{ishiiStimulatedEmissionMechanism2022}%
  \BibitemOpen
  \bibfield  {author} {\bibinfo {author} {\bibfnamefont {R.}~\bibnamefont
  {Ishii}}, \bibinfo {author} {\bibfnamefont {T.}~\bibnamefont {Nagashima}},
  \bibinfo {author} {\bibfnamefont {R.}~\bibnamefont {Yamamoto}}, \bibinfo
  {author} {\bibfnamefont {T.}~\bibnamefont {Hitomi}}, \bibinfo {author}
  {\bibfnamefont {M.}~\bibnamefont {Funato}},\ and\ \bibinfo {author}
  {\bibfnamefont {Y.}~\bibnamefont {Kawakami}},\ }\bibfield  {title} {\enquote
  {\bibinfo {title} {Stimulated emission mechanism of aluminum nitride},}\
  }\href {https://doi.org/10.1103/PhysRevB.105.205206} {\bibfield  {journal}
  {\bibinfo  {journal} {Phys. Rev. B}\ }\textbf {\bibinfo {volume} {105}},\
  \bibinfo {pages} {205206} (\bibinfo {year} {2022})}\BibitemShut {NoStop}%
\bibitem [{\citenamefont {Funato}\ \emph {et~al.}(2012)\citenamefont {Funato},
  \citenamefont {Matsuda}, \citenamefont {Banal}, \citenamefont {Ishii},\ and\
  \citenamefont {Kawakami}}]{funatoHomoepitaxyPhotoluminescenceProperties2012}%
  \BibitemOpen
  \bibfield  {author} {\bibinfo {author} {\bibfnamefont {M.}~\bibnamefont
  {Funato}}, \bibinfo {author} {\bibfnamefont {K.}~\bibnamefont {Matsuda}},
  \bibinfo {author} {\bibfnamefont {R.~G.}\ \bibnamefont {Banal}}, \bibinfo
  {author} {\bibfnamefont {R.}~\bibnamefont {Ishii}},\ and\ \bibinfo {author}
  {\bibfnamefont {Y.}~\bibnamefont {Kawakami}},\ }\bibfield  {title} {\enquote
  {\bibinfo {title} {Homoepitaxy and {{Photoluminescence Properties}} of (0001)
  {{AlN}}},}\ }\href {https://doi.org/10.1143/APEX.5.082001} {\bibfield
  {journal} {\bibinfo  {journal} {Appl. Phys. Express}\ }\textbf {\bibinfo
  {volume} {5}},\ \bibinfo {pages} {082001} (\bibinfo {year}
  {2012})}\BibitemShut {NoStop}%
\bibitem [{\citenamefont {Leute}\ \emph {et~al.}(2009)\citenamefont {Leute},
  \citenamefont {Feneberg}, \citenamefont {Sauer}, \citenamefont {Thonke},
  \citenamefont {Thapa}, \citenamefont {Scholz}, \citenamefont {Taniyasu},\
  and\ \citenamefont {Kasu}}]{leutePhotoluminescenceHighlyExcited2009}%
  \BibitemOpen
  \bibfield  {author} {\bibinfo {author} {\bibfnamefont {R.~a.~R.}\
  \bibnamefont {Leute}}, \bibinfo {author} {\bibfnamefont {M.}~\bibnamefont
  {Feneberg}}, \bibinfo {author} {\bibfnamefont {R.}~\bibnamefont {Sauer}},
  \bibinfo {author} {\bibfnamefont {K.}~\bibnamefont {Thonke}}, \bibinfo
  {author} {\bibfnamefont {S.~B.}\ \bibnamefont {Thapa}}, \bibinfo {author}
  {\bibfnamefont {F.}~\bibnamefont {Scholz}}, \bibinfo {author} {\bibfnamefont
  {Y.}~\bibnamefont {Taniyasu}},\ and\ \bibinfo {author} {\bibfnamefont
  {M.}~\bibnamefont {Kasu}},\ }\bibfield  {title} {\enquote {\bibinfo {title}
  {Photoluminescence of highly excited {{AlN}}: {{Biexcitons}} and
  exciton-exciton scattering},}\ }\href {https://doi.org/10.1063/1.3186044}
  {\bibfield  {journal} {\bibinfo  {journal} {Appl. Phys. Lett.}\ }\textbf
  {\bibinfo {volume} {95}},\ \bibinfo {pages} {031903} (\bibinfo {year}
  {2009})}\BibitemShut {NoStop}%
\bibitem [{\citenamefont {Feneberg}\ \emph {et~al.}(2011)\citenamefont
  {Feneberg}, \citenamefont {Neuschl}, \citenamefont {Thonke}, \citenamefont
  {Collazo}, \citenamefont {Rice}, \citenamefont {Sitar}, \citenamefont
  {Dalmau}, \citenamefont {Xie}, \citenamefont {Mita},\ and\ \citenamefont
  {Goldhahn}}]{fenebergSharpBoundFree2011}%
  \BibitemOpen
  \bibfield  {author} {\bibinfo {author} {\bibfnamefont {M.}~\bibnamefont
  {Feneberg}}, \bibinfo {author} {\bibfnamefont {B.}~\bibnamefont {Neuschl}},
  \bibinfo {author} {\bibfnamefont {K.}~\bibnamefont {Thonke}}, \bibinfo
  {author} {\bibfnamefont {R.}~\bibnamefont {Collazo}}, \bibinfo {author}
  {\bibfnamefont {A.}~\bibnamefont {Rice}}, \bibinfo {author} {\bibfnamefont
  {Z.}~\bibnamefont {Sitar}}, \bibinfo {author} {\bibfnamefont
  {R.}~\bibnamefont {Dalmau}}, \bibinfo {author} {\bibfnamefont
  {J.}~\bibnamefont {Xie}}, \bibinfo {author} {\bibfnamefont {S.}~\bibnamefont
  {Mita}},\ and\ \bibinfo {author} {\bibfnamefont {R.}~\bibnamefont
  {Goldhahn}},\ }\bibfield  {title} {\enquote {\bibinfo {title} {Sharp bound
  and free exciton lines from homoepitaxial {{AlN}}},}\ }\href
  {https://doi.org/10.1002/pssa.201000947} {\bibfield  {journal} {\bibinfo
  {journal} {Phys. Rev. B}\ }\textbf {\bibinfo {volume} {208}},\ \bibinfo
  {pages} {1520--1522} (\bibinfo {year} {2011})}\BibitemShut {NoStop}%
\bibitem [{\citenamefont {Chichibu}\ \emph {et~al.}(2010)\citenamefont
  {Chichibu}, \citenamefont {Onuma}, \citenamefont {Hazu},\ and\ \citenamefont
  {Uedono}}]{chichibuMajorImpactsPoint2010}%
  \BibitemOpen
  \bibfield  {author} {\bibinfo {author} {\bibfnamefont {S.~F.}\ \bibnamefont
  {Chichibu}}, \bibinfo {author} {\bibfnamefont {T.}~\bibnamefont {Onuma}},
  \bibinfo {author} {\bibfnamefont {K.}~\bibnamefont {Hazu}},\ and\ \bibinfo
  {author} {\bibfnamefont {A.}~\bibnamefont {Uedono}},\ }\bibfield  {title}
  {\enquote {\bibinfo {title} {Major impacts of point defects and impurities on
  the carrier recombination dynamics in {{AlN}}},}\ }\href
  {https://doi.org/10.1063/1.3517484} {\bibfield  {journal} {\bibinfo
  {journal} {Appl. Phys. Lett.}\ }\textbf {\bibinfo {volume} {97}},\ \bibinfo
  {pages} {201904} (\bibinfo {year} {2010})}\BibitemShut {NoStop}%
\bibitem [{\citenamefont {Bryan}\ \emph {et~al.}(2014)\citenamefont {Bryan},
  \citenamefont {Bryan}, \citenamefont {Bobea}, \citenamefont {Hussey},
  \citenamefont {Kirste}, \citenamefont {Sitar},\ and\ \citenamefont
  {Collazo}}]{bryanExcitonTransitionsOxygen2014}%
  \BibitemOpen
  \bibfield  {author} {\bibinfo {author} {\bibfnamefont {Z.}~\bibnamefont
  {Bryan}}, \bibinfo {author} {\bibfnamefont {I.}~\bibnamefont {Bryan}},
  \bibinfo {author} {\bibfnamefont {M.}~\bibnamefont {Bobea}}, \bibinfo
  {author} {\bibfnamefont {L.}~\bibnamefont {Hussey}}, \bibinfo {author}
  {\bibfnamefont {R.}~\bibnamefont {Kirste}}, \bibinfo {author} {\bibfnamefont
  {Z.}~\bibnamefont {Sitar}},\ and\ \bibinfo {author} {\bibfnamefont
  {R.}~\bibnamefont {Collazo}},\ }\bibfield  {title} {\enquote {\bibinfo
  {title} {Exciton transitions and oxygen as a donor in m-plane {{AlN}}
  homoepitaxial films},}\ }\href {https://doi.org/10.1063/1.4870284} {\bibfield
   {journal} {\bibinfo  {journal} {J. Appl. Phys.}\ }\textbf {\bibinfo {volume}
  {115}},\ \bibinfo {pages} {133503} (\bibinfo {year} {2014})}\BibitemShut
  {NoStop}%
\bibitem [{\citenamefont {Paskov}\ \emph {et~al.}(2001)\citenamefont {Paskov},
  \citenamefont {Paskova}, \citenamefont {Holtz},\ and\ \citenamefont
  {Monemar}}]{paskovSpinexchangeSplittingExcitons2001}%
  \BibitemOpen
  \bibfield  {author} {\bibinfo {author} {\bibfnamefont {P.~P.}\ \bibnamefont
  {Paskov}}, \bibinfo {author} {\bibfnamefont {T.}~\bibnamefont {Paskova}},
  \bibinfo {author} {\bibfnamefont {P.~O.}\ \bibnamefont {Holtz}},\ and\
  \bibinfo {author} {\bibfnamefont {B.}~\bibnamefont {Monemar}},\ }\bibfield
  {title} {\enquote {\bibinfo {title} {Spin-exchange splitting of excitons in
  {{GaN}}},}\ }\href {https://doi.org/10.1103/PhysRevB.64.115201} {\bibfield
  {journal} {\bibinfo  {journal} {Phys. Rev. B}\ }\textbf {\bibinfo {volume}
  {64}},\ \bibinfo {pages} {115201} (\bibinfo {year} {2001})}\BibitemShut
  {NoStop}%
\bibitem [{\citenamefont {Neuschl}\ \emph {et~al.}(2012)\citenamefont
  {Neuschl}, \citenamefont {Thonke}, \citenamefont {Feneberg}, \citenamefont
  {Mita}, \citenamefont {Xie}, \citenamefont {Dalmau}, \citenamefont
  {Collazo},\ and\ \citenamefont
  {Sitar}}]{neuschlOpticalIdentificationSilicon2012}%
  \BibitemOpen
  \bibfield  {author} {\bibinfo {author} {\bibfnamefont {B.}~\bibnamefont
  {Neuschl}}, \bibinfo {author} {\bibfnamefont {K.}~\bibnamefont {Thonke}},
  \bibinfo {author} {\bibfnamefont {M.}~\bibnamefont {Feneberg}}, \bibinfo
  {author} {\bibfnamefont {S.}~\bibnamefont {Mita}}, \bibinfo {author}
  {\bibfnamefont {J.}~\bibnamefont {Xie}}, \bibinfo {author} {\bibfnamefont
  {R.}~\bibnamefont {Dalmau}}, \bibinfo {author} {\bibfnamefont
  {R.}~\bibnamefont {Collazo}},\ and\ \bibinfo {author} {\bibfnamefont
  {Z.}~\bibnamefont {Sitar}},\ }\bibfield  {title} {\enquote {\bibinfo {title}
  {Optical identification of silicon as a shallow donor in {{MOVPE}} grown
  homoepitaxial {{AlN}}},}\ }\href {https://doi.org/10.1002/pssb.201100381}
  {\bibfield  {journal} {\bibinfo  {journal} {Phys. Status Solidi B}\ }\textbf
  {\bibinfo {volume} {249}},\ \bibinfo {pages} {511--515} (\bibinfo {year}
  {2012})}\BibitemShut {NoStop}%
\bibitem [{\citenamefont {Tatarczak}\ \emph {et~al.}(2021)\citenamefont
  {Tatarczak}, \citenamefont {Turski}, \citenamefont {Korona}, \citenamefont
  {Grzanka}, \citenamefont {Skierbiszewski},\ and\ \citenamefont
  {Wysmo{\l}ek}}]{tatarczakOpticalPropertiesNpolar2021}%
  \BibitemOpen
  \bibfield  {author} {\bibinfo {author} {\bibfnamefont {P.}~\bibnamefont
  {Tatarczak}}, \bibinfo {author} {\bibfnamefont {H.}~\bibnamefont {Turski}},
  \bibinfo {author} {\bibfnamefont {K.~P.}\ \bibnamefont {Korona}}, \bibinfo
  {author} {\bibfnamefont {E.}~\bibnamefont {Grzanka}}, \bibinfo {author}
  {\bibfnamefont {C.}~\bibnamefont {Skierbiszewski}},\ and\ \bibinfo {author}
  {\bibfnamefont {A.}~\bibnamefont {Wysmo{\l}ek}},\ }\bibfield  {title}
  {\enquote {\bibinfo {title} {Optical properties of {{N-polar GaN}}: {{The}}
  possible role of nitrogen vacancy-related defects},}\ }\href
  {https://doi.org/10.1016/j.apsusc.2021.150734} {\bibfield  {journal}
  {\bibinfo  {journal} {Appl. Surf. Sci.}\ }\textbf {\bibinfo {volume} {566}},\
  \bibinfo {pages} {150734} (\bibinfo {year} {2021})}\BibitemShut {NoStop}%
\bibitem [{\citenamefont {Li}\ \emph {et~al.}(2011)\citenamefont {Li},
  \citenamefont {Lestradet}, \citenamefont {Xiao},\ and\ \citenamefont
  {Li}}]{li_Phys.StatusSolidiA_2011}%
  \BibitemOpen
  \bibfield  {author} {\bibinfo {author} {\bibfnamefont {Z.~Q.}\ \bibnamefont
  {Li}}, \bibinfo {author} {\bibfnamefont {M.}~\bibnamefont {Lestradet}},
  \bibinfo {author} {\bibfnamefont {Y.~G.}\ \bibnamefont {Xiao}},\ and\
  \bibinfo {author} {\bibfnamefont {S.}~\bibnamefont {Li}},\ }\bibfield
  {title} {\enquote {\bibinfo {title} {Effects of polarization charge on the
  photovoltaic properties of {{InGaN}} solar cells: {{Effects}} of polarization
  charge on photovoltaic properties of {{InGaN}} solar cells},}\ }\href
  {https://doi.org/10.1002/pssa.201026489} {\bibfield  {journal} {\bibinfo
  {journal} {Phys. Status Solidi (a)}\ }\textbf {\bibinfo {volume} {208}},\
  \bibinfo {pages} {928--931} (\bibinfo {year} {2011})}\BibitemShut {NoStop}%
\bibitem [{\citenamefont {Akyol}\ \emph {et~al.}(2011)\citenamefont {Akyol},
  \citenamefont {Nath}, \citenamefont {G{\"u}r}, \citenamefont {Park},\ and\
  \citenamefont {Rajan}}]{akyol_Jpn.J.Appl.Phys._2011}%
  \BibitemOpen
  \bibfield  {author} {\bibinfo {author} {\bibfnamefont {F.}~\bibnamefont
  {Akyol}}, \bibinfo {author} {\bibfnamefont {D.~N.}\ \bibnamefont {Nath}},
  \bibinfo {author} {\bibfnamefont {E.}~\bibnamefont {G{\"u}r}}, \bibinfo
  {author} {\bibfnamefont {P.~S.}\ \bibnamefont {Park}},\ and\ \bibinfo
  {author} {\bibfnamefont {S.}~\bibnamefont {Rajan}},\ }\bibfield  {title}
  {\enquote {\bibinfo {title} {N-{{Polar III}}\textendash{{Nitride Green}} (540
  nm) {{Light Emitting Diode}}},}\ }\href
  {https://doi.org/10.1143/JJAP.50.052101} {\bibfield  {journal} {\bibinfo
  {journal} {Jpn. J. Appl. Phys.}\ }\textbf {\bibinfo {volume} {50}},\ \bibinfo
  {pages} {052101} (\bibinfo {year} {2011})}\BibitemShut {NoStop}%
\bibitem [{\citenamefont {Dong}\ \emph {et~al.}(2012)\citenamefont {Dong},
  \citenamefont {Chen}, \citenamefont {Liu}, \citenamefont {Lu}, \citenamefont
  {Chen}, \citenamefont {Zhang},\ and\ \citenamefont
  {Zheng}}]{dong_Appl.Phys.Lett._2012}%
  \BibitemOpen
  \bibfield  {author} {\bibinfo {author} {\bibfnamefont {K.}~\bibnamefont
  {Dong}}, \bibinfo {author} {\bibfnamefont {D.}~\bibnamefont {Chen}}, \bibinfo
  {author} {\bibfnamefont {B.}~\bibnamefont {Liu}}, \bibinfo {author}
  {\bibfnamefont {H.}~\bibnamefont {Lu}}, \bibinfo {author} {\bibfnamefont
  {P.}~\bibnamefont {Chen}}, \bibinfo {author} {\bibfnamefont {R.}~\bibnamefont
  {Zhang}},\ and\ \bibinfo {author} {\bibfnamefont {Y.}~\bibnamefont {Zheng}},\
  }\bibfield  {title} {\enquote {\bibinfo {title} {Characteristics of
  polarization-doped {{N-face III-nitride}} light-emitting diodes},}\ }\href
  {https://doi.org/10.1063/1.3687181} {\bibfield  {journal} {\bibinfo
  {journal} {Appl. Phys. Lett.}\ }\textbf {\bibinfo {volume} {100}},\ \bibinfo
  {pages} {073507} (\bibinfo {year} {2012})}\BibitemShut {NoStop}%
\bibitem [{\citenamefont {Han}\ \emph {et~al.}(2012)\citenamefont {Han},
  \citenamefont {Lee}, \citenamefont {Lim}, \citenamefont {Lee}, \citenamefont
  {Kim}, \citenamefont {Kim}, \citenamefont {Kim},\ and\ \citenamefont
  {Park}}]{han_Jpn.J.Appl.Phys._2012}%
  \BibitemOpen
  \bibfield  {author} {\bibinfo {author} {\bibfnamefont {S.-H.}\ \bibnamefont
  {Han}}, \bibinfo {author} {\bibfnamefont {D.-Y.}\ \bibnamefont {Lee}},
  \bibinfo {author} {\bibfnamefont {J.-Y.}\ \bibnamefont {Lim}}, \bibinfo
  {author} {\bibfnamefont {J.~W.}\ \bibnamefont {Lee}}, \bibinfo {author}
  {\bibfnamefont {D.-J.}\ \bibnamefont {Kim}}, \bibinfo {author} {\bibfnamefont
  {Y.~S.}\ \bibnamefont {Kim}}, \bibinfo {author} {\bibfnamefont {S.-T.}\
  \bibnamefont {Kim}},\ and\ \bibinfo {author} {\bibfnamefont {S.-J.}\
  \bibnamefont {Park}},\ }\bibfield  {title} {\enquote {\bibinfo {title}
  {Effect of {{Internal Electric Field}} in {{Well Layer}} of {{InGaN}}/{{GaN
  Multiple Quantum Well Light-Emitting Diodes}} on {{Efficiency Droop}}},}\
  }\href {https://doi.org/10.1143/JJAP.51.100201} {\bibfield  {journal}
  {\bibinfo  {journal} {Jpn. J. Appl. Phys.}\ }\textbf {\bibinfo {volume}
  {51}},\ \bibinfo {pages} {100201} (\bibinfo {year} {2012})}\BibitemShut
  {NoStop}%
\bibitem [{\citenamefont {Wong}\ \emph {et~al.}(2013)\citenamefont {Wong},
  \citenamefont {Keller}, \citenamefont {Dasgupta}, \citenamefont
  {Denninghoff}, \citenamefont {Kolluri}, \citenamefont {Brown}, \citenamefont
  {Lu}, \citenamefont {Fichtenbaum}, \citenamefont {Ahmadi}, \citenamefont
  {Singisetti}, \citenamefont {Chini}, \citenamefont {Rajan}, \citenamefont
  {DenBaars}, \citenamefont {Speck},\ and\ \citenamefont
  {Mishra}}]{wong_Semicond.Sci.Technol._2013}%
  \BibitemOpen
  \bibfield  {author} {\bibinfo {author} {\bibfnamefont {M.~H.}\ \bibnamefont
  {Wong}}, \bibinfo {author} {\bibfnamefont {S.}~\bibnamefont {Keller}},
  \bibinfo {author} {\bibfnamefont {N.}~\bibnamefont {Dasgupta}, \bibfnamefont
  {Sansaptak}}, \bibinfo {author} {\bibfnamefont {D.~J.}\ \bibnamefont
  {Denninghoff}}, \bibinfo {author} {\bibfnamefont {S.}~\bibnamefont
  {Kolluri}}, \bibinfo {author} {\bibfnamefont {D.~F.}\ \bibnamefont {Brown}},
  \bibinfo {author} {\bibfnamefont {J.}~\bibnamefont {Lu}}, \bibinfo {author}
  {\bibfnamefont {N.~A.}\ \bibnamefont {Fichtenbaum}}, \bibinfo {author}
  {\bibfnamefont {E.}~\bibnamefont {Ahmadi}}, \bibinfo {author} {\bibfnamefont
  {U.}~\bibnamefont {Singisetti}}, \bibinfo {author} {\bibfnamefont
  {A.}~\bibnamefont {Chini}}, \bibinfo {author} {\bibfnamefont
  {S.}~\bibnamefont {Rajan}}, \bibinfo {author} {\bibfnamefont {S.~P.}\
  \bibnamefont {DenBaars}}, \bibinfo {author} {\bibfnamefont {J.~S.}\
  \bibnamefont {Speck}},\ and\ \bibinfo {author} {\bibfnamefont {U.~K.}\
  \bibnamefont {Mishra}},\ }\bibfield  {title} {\enquote {\bibinfo {title}
  {N-polar {{GaN}} epitaxy and high electron mobility transistors},}\ }\href
  {https://doi.org/10.1088/0268-1242/28/7/074009} {\bibfield  {journal}
  {\bibinfo  {journal} {Semicond. Sci. Technol.}\ }\textbf {\bibinfo {volume}
  {28}},\ \bibinfo {pages} {074009} (\bibinfo {year} {2013})}\BibitemShut
  {NoStop}%
\bibitem [{\citenamefont {Feng}\ \emph {et~al.}(2015)\citenamefont {Feng},
  \citenamefont {Liao}, \citenamefont {Leung}, \citenamefont {Han},
  \citenamefont {Yang},\ and\ \citenamefont {Wang}}]{feng_J.Appl.Phys._2015}%
  \BibitemOpen
  \bibfield  {author} {\bibinfo {author} {\bibfnamefont {S.-W.}\ \bibnamefont
  {Feng}}, \bibinfo {author} {\bibfnamefont {P.-H.}\ \bibnamefont {Liao}},
  \bibinfo {author} {\bibfnamefont {B.}~\bibnamefont {Leung}}, \bibinfo
  {author} {\bibfnamefont {J.}~\bibnamefont {Han}}, \bibinfo {author}
  {\bibfnamefont {F.-W.}\ \bibnamefont {Yang}},\ and\ \bibinfo {author}
  {\bibfnamefont {H.-C.}\ \bibnamefont {Wang}},\ }\bibfield  {title} {\enquote
  {\bibinfo {title} {Efficient carrier relaxation and fast carrier
  recombination of {{{\emph{N}}}}-polar {{InGaN}}/{{GaN}} light emitting
  diodes},}\ }\href {https://doi.org/10.1063/1.4927421} {\bibfield  {journal}
  {\bibinfo  {journal} {J. Appl. Phys.}\ }\textbf {\bibinfo {volume} {118}},\
  \bibinfo {pages} {043104} (\bibinfo {year} {2015})}\BibitemShut {NoStop}%
\bibitem [{\citenamefont {van Deurzen}\ \emph {et~al.}(2021)\citenamefont {van
  Deurzen}, \citenamefont {Bharadwaj}, \citenamefont {Lee}, \citenamefont
  {Protasenko}, \citenamefont {Turski}, \citenamefont {Xing},\ and\
  \citenamefont {Jena}}]{vandeurzenlenEnhancedEfficiencyBottom2021}%
  \BibitemOpen
  \bibfield  {author} {\bibinfo {author} {\bibfnamefont {L.}~\bibnamefont {van
  Deurzen}}, \bibinfo {author} {\bibfnamefont {S.}~\bibnamefont {Bharadwaj}},
  \bibinfo {author} {\bibfnamefont {K.}~\bibnamefont {Lee}}, \bibinfo {author}
  {\bibfnamefont {V.}~\bibnamefont {Protasenko}}, \bibinfo {author}
  {\bibfnamefont {H.}~\bibnamefont {Turski}}, \bibinfo {author} {\bibfnamefont
  {H.~G.}\ \bibnamefont {Xing}},\ and\ \bibinfo {author} {\bibfnamefont
  {D.}~\bibnamefont {Jena}},\ }\bibfield  {title} {\enquote {\bibinfo {title}
  {Enhanced efficiency in bottom tunnel junction {{InGaN}} blue {{LEDs}}},}\
  }in\ \href {https://doi.org/10.1117/12.2582439} {\emph {\bibinfo {booktitle}
  {Light-{{Emitting Devices}}, {{Materials}}, and {{Applications XXV}}}}},\
  Vol.\ \bibinfo {volume} {11706}\ (\bibinfo  {publisher} {{SPIE}},\ \bibinfo
  {year} {2021})\ pp.\ \bibinfo {pages} {30--35}\BibitemShut {NoStop}%
\bibitem [{\citenamefont {Cho}\ \emph {et~al.}(2019)\citenamefont {Cho},
  \citenamefont {Bharadwaj}, \citenamefont {Hu}, \citenamefont {Nomoto},
  \citenamefont {Jahn}, \citenamefont {Xing},\ and\ \citenamefont
  {Jena}}]{choBlueGaLightemitting2019}%
  \BibitemOpen
  \bibfield  {author} {\bibinfo {author} {\bibfnamefont {Y.}~\bibnamefont
  {Cho}}, \bibinfo {author} {\bibfnamefont {S.}~\bibnamefont {Bharadwaj}},
  \bibinfo {author} {\bibfnamefont {Z.}~\bibnamefont {Hu}}, \bibinfo {author}
  {\bibfnamefont {K.}~\bibnamefont {Nomoto}}, \bibinfo {author} {\bibfnamefont
  {U.}~\bibnamefont {Jahn}}, \bibinfo {author} {\bibfnamefont {H.~G.}\
  \bibnamefont {Xing}},\ and\ \bibinfo {author} {\bibfnamefont
  {D.}~\bibnamefont {Jena}},\ }\bibfield  {title} {\enquote {\bibinfo {title}
  {Blue ({{In}},{{Ga}}){{N}} light-emitting diodes with buried n+\textendash p+
  tunnel junctions by plasma-assisted molecular beam epitaxy},}\ }\href
  {https://doi.org/10.7567/1347-4065/ab1e78} {\bibfield  {journal} {\bibinfo
  {journal} {Jpn. J. Appl. Phys.}\ }\textbf {\bibinfo {volume} {58}},\ \bibinfo
  {pages} {060914} (\bibinfo {year} {2019})}\BibitemShut {NoStop}%
\bibitem [{\citenamefont {Bharadwaj}\ \emph
  {et~al.}(2020{\natexlab{a}})\citenamefont {Bharadwaj}, \citenamefont
  {Miller}, \citenamefont {Lee}, \citenamefont {Lederman}, \citenamefont
  {Siekacz}, \citenamefont {Xing}, \citenamefont {Jena}, \citenamefont
  {Skierbiszewski},\ and\ \citenamefont
  {Turski}}]{bharadwajEnhancedInjectionEfficiency2020}%
  \BibitemOpen
  \bibfield  {author} {\bibinfo {author} {\bibfnamefont {S.}~\bibnamefont
  {Bharadwaj}}, \bibinfo {author} {\bibfnamefont {J.}~\bibnamefont {Miller}},
  \bibinfo {author} {\bibfnamefont {K.}~\bibnamefont {Lee}}, \bibinfo {author}
  {\bibfnamefont {J.}~\bibnamefont {Lederman}}, \bibinfo {author}
  {\bibfnamefont {M.}~\bibnamefont {Siekacz}}, \bibinfo {author} {\bibfnamefont
  {H.~G.}\ \bibnamefont {Xing}}, \bibinfo {author} {\bibfnamefont
  {D.}~\bibnamefont {Jena}}, \bibinfo {author} {\bibfnamefont {C.}~\bibnamefont
  {Skierbiszewski}},\ and\ \bibinfo {author} {\bibfnamefont {H.}~\bibnamefont
  {Turski}},\ }\bibfield  {title} {\enquote {\bibinfo {title} {Enhanced
  injection efficiency and light output in bottom tunnel-junction
  light-emitting diodes},}\ }\href {https://doi.org/10.1364/OE.384021}
  {\bibfield  {journal} {\bibinfo  {journal} {Opt. Express}\ }\textbf {\bibinfo
  {volume} {28}},\ \bibinfo {pages} {4489--4500} (\bibinfo {year}
  {2020}{\natexlab{a}})}\BibitemShut {NoStop}%
\bibitem [{\citenamefont {Lee}\ \emph {et~al.}(2020{\natexlab{a}})\citenamefont
  {Lee}, \citenamefont {Bharadwaj}, \citenamefont {Shao}, \citenamefont {van
  Deurzen}, \citenamefont {Protasenko}, \citenamefont {Muller}, \citenamefont
  {Xing},\ and\ \citenamefont {Jena}}]{leeLightemittingDiodesAlN2020}%
  \BibitemOpen
  \bibfield  {author} {\bibinfo {author} {\bibfnamefont {K.}~\bibnamefont
  {Lee}}, \bibinfo {author} {\bibfnamefont {S.}~\bibnamefont {Bharadwaj}},
  \bibinfo {author} {\bibfnamefont {Y.-T.}\ \bibnamefont {Shao}}, \bibinfo
  {author} {\bibfnamefont {L.}~\bibnamefont {van Deurzen}}, \bibinfo {author}
  {\bibfnamefont {V.}~\bibnamefont {Protasenko}}, \bibinfo {author}
  {\bibfnamefont {D.~A.}\ \bibnamefont {Muller}}, \bibinfo {author}
  {\bibfnamefont {H.~G.}\ \bibnamefont {Xing}},\ and\ \bibinfo {author}
  {\bibfnamefont {D.}~\bibnamefont {Jena}},\ }\bibfield  {title} {\enquote
  {\bibinfo {title} {Light-emitting diodes with {{AlN}} polarization-induced
  buried tunnel junctions: {{A}} second look},}\ }\href
  {https://doi.org/10.1063/5.0015097} {\bibfield  {journal} {\bibinfo
  {journal} {Appl. Phys. Lett.}\ }\textbf {\bibinfo {volume} {117}},\ \bibinfo
  {pages} {061104} (\bibinfo {year} {2020}{\natexlab{a}})}\BibitemShut
  {NoStop}%
\bibitem [{\citenamefont {Bharadwaj}\ \emph
  {et~al.}(2020{\natexlab{b}})\citenamefont {Bharadwaj}, \citenamefont {Lee},
  \citenamefont {Nomoto}, \citenamefont {Hickman}, \citenamefont {{van
  Deurzen}}, \citenamefont {Protasenko}, \citenamefont {Xing},\ and\
  \citenamefont {Jena}}]{bharadwajBottomTunnelJunction2020}%
  \BibitemOpen
  \bibfield  {author} {\bibinfo {author} {\bibfnamefont {S.}~\bibnamefont
  {Bharadwaj}}, \bibinfo {author} {\bibfnamefont {K.}~\bibnamefont {Lee}},
  \bibinfo {author} {\bibfnamefont {K.}~\bibnamefont {Nomoto}}, \bibinfo
  {author} {\bibfnamefont {A.}~\bibnamefont {Hickman}}, \bibinfo {author}
  {\bibfnamefont {L.}~\bibnamefont {{van Deurzen}}}, \bibinfo {author}
  {\bibfnamefont {V.}~\bibnamefont {Protasenko}}, \bibinfo {author}
  {\bibfnamefont {H.~G.}\ \bibnamefont {Xing}},\ and\ \bibinfo {author}
  {\bibfnamefont {D.}~\bibnamefont {Jena}},\ }\bibfield  {title} {\enquote
  {\bibinfo {title} {Bottom tunnel junction blue light-emitting field-effect
  transistors},}\ }\href {https://doi.org/10.1063/5.0009430} {\bibfield
  {journal} {\bibinfo  {journal} {Appl. Phys. Lett.}\ }\textbf {\bibinfo
  {volume} {117}},\ \bibinfo {pages} {031107} (\bibinfo {year}
  {2020}{\natexlab{b}})}\BibitemShut {NoStop}%
\bibitem [{\citenamefont {Cho}\ \emph {et~al.}(2020)\citenamefont {Cho},
  \citenamefont {Chang}, \citenamefont {Lee}, \citenamefont {Gong},
  \citenamefont {Nomoto}, \citenamefont {Toita}, \citenamefont {Schowalter},
  \citenamefont {Muller}, \citenamefont {Jena},\ and\ \citenamefont
  {Xing}}]{choMolecularBeamHomoepitaxy2020}%
  \BibitemOpen
  \bibfield  {author} {\bibinfo {author} {\bibfnamefont {Y.}~\bibnamefont
  {Cho}}, \bibinfo {author} {\bibfnamefont {C.~S.}\ \bibnamefont {Chang}},
  \bibinfo {author} {\bibfnamefont {K.}~\bibnamefont {Lee}}, \bibinfo {author}
  {\bibfnamefont {M.}~\bibnamefont {Gong}}, \bibinfo {author} {\bibfnamefont
  {K.}~\bibnamefont {Nomoto}}, \bibinfo {author} {\bibfnamefont
  {M.}~\bibnamefont {Toita}}, \bibinfo {author} {\bibfnamefont {L.~J.}\
  \bibnamefont {Schowalter}}, \bibinfo {author} {\bibfnamefont {D.~A.}\
  \bibnamefont {Muller}}, \bibinfo {author} {\bibfnamefont {D.}~\bibnamefont
  {Jena}},\ and\ \bibinfo {author} {\bibfnamefont {H.~G.}\ \bibnamefont
  {Xing}},\ }\bibfield  {title} {\enquote {\bibinfo {title} {Molecular beam
  homoepitaxy on bulk {{AlN}} enabled by aluminum-assisted surface cleaning},}\
  }\href {https://doi.org/10.1063/1.5143968} {\bibfield  {journal} {\bibinfo
  {journal} {Appl. Phys. Lett.}\ }\textbf {\bibinfo {volume} {116}},\ \bibinfo
  {pages} {172106} (\bibinfo {year} {2020})}\BibitemShut {NoStop}%
\bibitem [{\citenamefont {Lee}\ \emph {et~al.}(2020{\natexlab{b}})\citenamefont
  {Lee}, \citenamefont {Cho}, \citenamefont {Schowalter}, \citenamefont
  {Toita}, \citenamefont {Xing},\ and\ \citenamefont
  {Jena}}]{leeSurfaceControlMBE2020}%
  \BibitemOpen
  \bibfield  {author} {\bibinfo {author} {\bibfnamefont {K.}~\bibnamefont
  {Lee}}, \bibinfo {author} {\bibfnamefont {Y.}~\bibnamefont {Cho}}, \bibinfo
  {author} {\bibfnamefont {L.~J.}\ \bibnamefont {Schowalter}}, \bibinfo
  {author} {\bibfnamefont {M.}~\bibnamefont {Toita}}, \bibinfo {author}
  {\bibfnamefont {H.~G.}\ \bibnamefont {Xing}},\ and\ \bibinfo {author}
  {\bibfnamefont {D.}~\bibnamefont {Jena}},\ }\bibfield  {title} {\enquote
  {\bibinfo {title} {Surface control and {{MBE}} growth diagram for homoepitaxy
  on single-crystal {{AlN}} substrates},}\ }\href
  {https://doi.org/10.1063/5.0010813} {\bibfield  {journal} {\bibinfo
  {journal} {Appl. Phys. Lett.}\ }\textbf {\bibinfo {volume} {116}},\ \bibinfo
  {pages} {262102} (\bibinfo {year} {2020}{\natexlab{b}})}\BibitemShut
  {NoStop}%
\bibitem [{\citenamefont {Singhal}\ \emph {et~al.}(2022)\citenamefont
  {Singhal}, \citenamefont {Encomendero}, \citenamefont {Cho}, \citenamefont
  {{van Deurzen}}, \citenamefont {Zhang}, \citenamefont {Nomoto}, \citenamefont
  {Toita}, \citenamefont {Xing},\ and\ \citenamefont
  {Jena}}]{singhalMolecularBeamHomoepitaxy2022}%
  \BibitemOpen
  \bibfield  {author} {\bibinfo {author} {\bibfnamefont {J.}~\bibnamefont
  {Singhal}}, \bibinfo {author} {\bibfnamefont {J.}~\bibnamefont
  {Encomendero}}, \bibinfo {author} {\bibfnamefont {Y.}~\bibnamefont {Cho}},
  \bibinfo {author} {\bibfnamefont {L.}~\bibnamefont {{van Deurzen}}}, \bibinfo
  {author} {\bibfnamefont {Z.}~\bibnamefont {Zhang}}, \bibinfo {author}
  {\bibfnamefont {K.}~\bibnamefont {Nomoto}}, \bibinfo {author} {\bibfnamefont
  {M.}~\bibnamefont {Toita}}, \bibinfo {author} {\bibfnamefont {H.~G.}\
  \bibnamefont {Xing}},\ and\ \bibinfo {author} {\bibfnamefont
  {D.}~\bibnamefont {Jena}},\ }\bibfield  {title} {\enquote {\bibinfo {title}
  {Molecular beam homoepitaxy of {{N-polar AlN}} on bulk {{AlN}} substrates},}\
  }\href {https://doi.org/10.1063/5.0100225} {\bibfield  {journal} {\bibinfo
  {journal} {AIP Adv.}\ }\textbf {\bibinfo {volume} {12}},\ \bibinfo {pages}
  {095314} (\bibinfo {year} {2022})}\BibitemShut {NoStop}%
\bibitem [{\citenamefont {Zhang}\ \emph
  {et~al.}(2022{\natexlab{b}})\citenamefont {Zhang}, \citenamefont {Hayashi},
  \citenamefont {Tohei}, \citenamefont {Sakai}, \citenamefont {Protasenko},
  \citenamefont {Singhal}, \citenamefont {Miyake}, \citenamefont {Xing},
  \citenamefont {Jena},\ and\ \citenamefont
  {Cho}}]{zhangMolecularBeamHomoepitaxy2022}%
  \BibitemOpen
  \bibfield  {author} {\bibinfo {author} {\bibfnamefont {Z.}~\bibnamefont
  {Zhang}}, \bibinfo {author} {\bibfnamefont {Y.}~\bibnamefont {Hayashi}},
  \bibinfo {author} {\bibfnamefont {T.}~\bibnamefont {Tohei}}, \bibinfo
  {author} {\bibfnamefont {A.}~\bibnamefont {Sakai}}, \bibinfo {author}
  {\bibfnamefont {V.}~\bibnamefont {Protasenko}}, \bibinfo {author}
  {\bibfnamefont {J.}~\bibnamefont {Singhal}}, \bibinfo {author} {\bibfnamefont
  {H.}~\bibnamefont {Miyake}}, \bibinfo {author} {\bibfnamefont {H.~G.}\
  \bibnamefont {Xing}}, \bibinfo {author} {\bibfnamefont {D.}~\bibnamefont
  {Jena}},\ and\ \bibinfo {author} {\bibfnamefont {Y.}~\bibnamefont {Cho}},\
  }\bibfield  {title} {\enquote {\bibinfo {title} {Molecular beam homoepitaxy
  of {{N-polar AlN}}: {{Enabling}} role of aluminum-assisted surface
  cleaning},}\ }\href {https://doi.org/10.1126/sciadv.abo6408} {\bibfield
  {journal} {\bibinfo  {journal} {Sci. Adv.}\ }\textbf {\bibinfo {volume}
  {8}},\ \bibinfo {pages} {eabo6408} (\bibinfo {year}
  {2022}{\natexlab{b}})}\BibitemShut {NoStop}%
\bibitem [{\citenamefont {L\"{a}hnemann}\ \emph {et~al.}()\citenamefont
  {L\"{a}hnemann}, \citenamefont {Orri}, \citenamefont {Prestat}, \citenamefont
  {{\AA}nes}, \citenamefont {Johnstone},\ and\ \citenamefont
  {Tappy}}]{LumiSpy2023}%
  \BibitemOpen
  \bibfield  {author} {\bibinfo {author} {\bibfnamefont {J.}~\bibnamefont
  {L\"{a}hnemann}}, \bibinfo {author} {\bibfnamefont {J.~F.}\ \bibnamefont
  {Orri}}, \bibinfo {author} {\bibfnamefont {E.}~\bibnamefont {Prestat}},
  \bibinfo {author} {\bibfnamefont {H.~W.}\ \bibnamefont {{\AA}nes}}, \bibinfo
  {author} {\bibfnamefont {D.}~\bibnamefont {Johnstone}},\ and\ \bibinfo
  {author} {\bibfnamefont {N.}~\bibnamefont {Tappy}},\ }\bibfield  {title}
  {\enquote {\bibinfo {title} {Lumispy v0.2.1},}\ }\href
  {https://doi.org/10.5281/zenodo.4640445} {\
  10.5281/zenodo.4640445}\BibitemShut {NoStop}%
\bibitem [{\citenamefont {Ye}\ \emph {et~al.}(2008)\citenamefont {Ye},
  \citenamefont {Chen}, \citenamefont {Zhu},\ and\ \citenamefont
  {Wei}}]{yeAsymmetryAdsorptionOxygen2008}%
  \BibitemOpen
  \bibfield  {author} {\bibinfo {author} {\bibfnamefont {H.}~\bibnamefont
  {Ye}}, \bibinfo {author} {\bibfnamefont {G.}~\bibnamefont {Chen}}, \bibinfo
  {author} {\bibfnamefont {Y.}~\bibnamefont {Zhu}},\ and\ \bibinfo {author}
  {\bibfnamefont {S.-H.}\ \bibnamefont {Wei}},\ }\bibfield  {title} {\enquote
  {\bibinfo {title} {Asymmetry of adsorption of oxygen at wurtzite {{AlN}}
  (0001) and $(000\overline{1})$ surfaces: {{First-principles}}
  calculations},}\ }\href {https://doi.org/10.1103/PhysRevB.77.033302}
  {\bibfield  {journal} {\bibinfo  {journal} {Phys. Rev. B}\ }\textbf {\bibinfo
  {volume} {77}},\ \bibinfo {pages} {033302} (\bibinfo {year}
  {2008})}\BibitemShut {NoStop}%
\bibitem [{\citenamefont {Miao}\ \emph {et~al.}(2010)\citenamefont {Miao},
  \citenamefont {Moses}, \citenamefont {Weber}, \citenamefont {Janotti},\ and\
  \citenamefont {de~Walle}}]{miaoEffectsSurfaceReconstructions2010}%
  \BibitemOpen
  \bibfield  {author} {\bibinfo {author} {\bibfnamefont {M.~S.}\ \bibnamefont
  {Miao}}, \bibinfo {author} {\bibfnamefont {P.~G.}\ \bibnamefont {Moses}},
  \bibinfo {author} {\bibfnamefont {J.~R.}\ \bibnamefont {Weber}}, \bibinfo
  {author} {\bibfnamefont {A.}~\bibnamefont {Janotti}},\ and\ \bibinfo {author}
  {\bibfnamefont {C.~G.~V.}\ \bibnamefont {de~Walle}},\ }\bibfield  {title}
  {\enquote {\bibinfo {title} {Effects of surface reconstructions on oxygen
  adsorption at {{AlN}} polar surfaces},}\ }\href
  {https://doi.org/10.1209/0295-5075/89/56004} {\bibfield  {journal} {\bibinfo
  {journal} {Europhys. Lett.}\ }\textbf {\bibinfo {volume} {89}},\ \bibinfo
  {pages} {56004} (\bibinfo {year} {2010})}\BibitemShut {NoStop}%
\bibitem [{\citenamefont {Li}\ \emph {et~al.}(2003)\citenamefont {Li},
  \citenamefont {Nam}, \citenamefont {Nakarmi}, \citenamefont {Lin},
  \citenamefont {Jiang}, \citenamefont {Carrier},\ and\ \citenamefont
  {Wei}}]{liBandStructureFundamental2003}%
  \BibitemOpen
  \bibfield  {author} {\bibinfo {author} {\bibfnamefont {J.}~\bibnamefont
  {Li}}, \bibinfo {author} {\bibfnamefont {K.~B.}\ \bibnamefont {Nam}},
  \bibinfo {author} {\bibfnamefont {M.~L.}\ \bibnamefont {Nakarmi}}, \bibinfo
  {author} {\bibfnamefont {J.~Y.}\ \bibnamefont {Lin}}, \bibinfo {author}
  {\bibfnamefont {H.~X.}\ \bibnamefont {Jiang}}, \bibinfo {author}
  {\bibfnamefont {P.}~\bibnamefont {Carrier}},\ and\ \bibinfo {author}
  {\bibfnamefont {S.-H.}\ \bibnamefont {Wei}},\ }\bibfield  {title} {\enquote
  {\bibinfo {title} {Band structure and fundamental optical transitions in
  wurtzite {{AlN}}},}\ }\href {https://doi.org/10.1063/1.1633965} {\bibfield
  {journal} {\bibinfo  {journal} {Appl. Phys. Lett.}\ }\textbf {\bibinfo
  {volume} {83}},\ \bibinfo {pages} {5163--5165} (\bibinfo {year}
  {2003})}\BibitemShut {NoStop}%
\bibitem [{\citenamefont {Cho}(1976)}]{choUnifiedTheorySymmetrybreaking1976}%
  \BibitemOpen
  \bibfield  {author} {\bibinfo {author} {\bibfnamefont {K.}~\bibnamefont
  {Cho}},\ }\bibfield  {title} {\enquote {\bibinfo {title} {Unified theory of
  symmetry-breaking effects on excitons in cubic and wurtzite structures},}\
  }\href {https://doi.org/10.1103/PhysRevB.14.4463} {\bibfield  {journal}
  {\bibinfo  {journal} {Phys. Rev. B}\ }\textbf {\bibinfo {volume} {14}},\
  \bibinfo {pages} {4463--4482} (\bibinfo {year} {1976})}\BibitemShut {NoStop}%
\bibitem [{\citenamefont {Feneberg}\ \emph {et~al.}(2013)\citenamefont
  {Feneberg}, \citenamefont {F{\'a}tima~Romero}, \citenamefont {Neuschl},
  \citenamefont {Thonke}, \citenamefont {R{\"o}ppischer}, \citenamefont
  {Cobet}, \citenamefont {Esser}, \citenamefont {Bickermann},\ and\
  \citenamefont {Goldhahn}}]{feneberg_Appl.Phys.Lett._2013}%
  \BibitemOpen
  \bibfield  {author} {\bibinfo {author} {\bibfnamefont {M.}~\bibnamefont
  {Feneberg}}, \bibinfo {author} {\bibfnamefont {M.}~\bibnamefont
  {F{\'a}tima~Romero}}, \bibinfo {author} {\bibfnamefont {B.}~\bibnamefont
  {Neuschl}}, \bibinfo {author} {\bibfnamefont {K.}~\bibnamefont {Thonke}},
  \bibinfo {author} {\bibfnamefont {M.}~\bibnamefont {R{\"o}ppischer}},
  \bibinfo {author} {\bibfnamefont {C.}~\bibnamefont {Cobet}}, \bibinfo
  {author} {\bibfnamefont {N.}~\bibnamefont {Esser}}, \bibinfo {author}
  {\bibfnamefont {M.}~\bibnamefont {Bickermann}},\ and\ \bibinfo {author}
  {\bibfnamefont {R.}~\bibnamefont {Goldhahn}},\ }\bibfield  {title} {\enquote
  {\bibinfo {title} {Negative spin-exchange splitting in the exciton fine
  structure of {{AlN}}},}\ }\href {https://doi.org/10.1063/1.4790645}
  {\bibfield  {journal} {\bibinfo  {journal} {Appl. Phys. Lett.}\ }\textbf
  {\bibinfo {volume} {102}},\ \bibinfo {pages} {052112} (\bibinfo {year}
  {2013})}\BibitemShut {NoStop}%
\bibitem [{\citenamefont {Ishii}, \citenamefont {Funato},\ and\ \citenamefont
  {Kawakami}(2020)}]{ishiiLongrangeElectronholeExchange2020}%
  \BibitemOpen
  \bibfield  {author} {\bibinfo {author} {\bibfnamefont {R.}~\bibnamefont
  {Ishii}}, \bibinfo {author} {\bibfnamefont {M.}~\bibnamefont {Funato}},\ and\
  \bibinfo {author} {\bibfnamefont {Y.}~\bibnamefont {Kawakami}},\ }\bibfield
  {title} {\enquote {\bibinfo {title} {Long-range electron-hole exchange
  interaction in aluminum nitride},}\ }\href
  {https://doi.org/10.1103/PhysRevB.102.155202} {\bibfield  {journal} {\bibinfo
   {journal} {Phys. Rev. B}\ }\textbf {\bibinfo {volume} {102}},\ \bibinfo
  {pages} {155202} (\bibinfo {year} {2020})}\BibitemShut {NoStop}%
\bibitem [{\citenamefont {Chichibu}\ \emph {et~al.}(2019)\citenamefont
  {Chichibu}, \citenamefont {Kojima}, \citenamefont {Hazu}, \citenamefont
  {Ishikawa}, \citenamefont {Furusawa}, \citenamefont {Mita}, \citenamefont
  {Collazo}, \citenamefont {Sitar},\ and\ \citenamefont
  {Uedono}}]{chichibuInplaneOpticalPolarization2019}%
  \BibitemOpen
  \bibfield  {author} {\bibinfo {author} {\bibfnamefont {S.~F.}\ \bibnamefont
  {Chichibu}}, \bibinfo {author} {\bibfnamefont {K.}~\bibnamefont {Kojima}},
  \bibinfo {author} {\bibfnamefont {K.}~\bibnamefont {Hazu}}, \bibinfo {author}
  {\bibfnamefont {Y.}~\bibnamefont {Ishikawa}}, \bibinfo {author}
  {\bibfnamefont {K.}~\bibnamefont {Furusawa}}, \bibinfo {author}
  {\bibfnamefont {S.}~\bibnamefont {Mita}}, \bibinfo {author} {\bibfnamefont
  {R.}~\bibnamefont {Collazo}}, \bibinfo {author} {\bibfnamefont
  {Z.}~\bibnamefont {Sitar}},\ and\ \bibinfo {author} {\bibfnamefont
  {A.}~\bibnamefont {Uedono}},\ }\bibfield  {title} {\enquote {\bibinfo {title}
  {In-plane optical polarization and dynamic properties of the near-band-edge
  emission of an m-plane freestanding {{AlN}} substrate and a homoepitaxial
  film},}\ }\href {https://doi.org/10.1063/1.5116900} {\bibfield  {journal}
  {\bibinfo  {journal} {Applied Physics Letters}\ }\textbf {\bibinfo {volume}
  {115}},\ \bibinfo {pages} {151903} (\bibinfo {year} {2019})}\BibitemShut
  {NoStop}%
\bibitem [{\citenamefont {Ishii}, \citenamefont {Funato},\ and\ \citenamefont
  {Kawakami}(2013)}]{ishiiHugeElectronholeExchange2013}%
  \BibitemOpen
  \bibfield  {author} {\bibinfo {author} {\bibfnamefont {R.}~\bibnamefont
  {Ishii}}, \bibinfo {author} {\bibfnamefont {M.}~\bibnamefont {Funato}},\ and\
  \bibinfo {author} {\bibfnamefont {Y.}~\bibnamefont {Kawakami}},\ }\bibfield
  {title} {\enquote {\bibinfo {title} {Huge electron-hole exchange interaction
  in aluminum nitride},}\ }\href {https://doi.org/10.1103/PhysRevB.87.161204}
  {\bibfield  {journal} {\bibinfo  {journal} {Phys. Rev. B}\ }\textbf {\bibinfo
  {volume} {87}},\ \bibinfo {pages} {161204} (\bibinfo {year}
  {2013})}\BibitemShut {NoStop}%
\bibitem [{\citenamefont {Koda}\ and\ \citenamefont
  {Langer}(1968)}]{koda_Phys.Rev.Lett._1968}%
  \BibitemOpen
  \bibfield  {author} {\bibinfo {author} {\bibfnamefont {T.}~\bibnamefont
  {Koda}}\ and\ \bibinfo {author} {\bibfnamefont {D.~W.}\ \bibnamefont
  {Langer}},\ }\bibfield  {title} {\enquote {\bibinfo {title} {Splitting of
  {{Exciton Lines}} in {{Wurtzite-Type II-VI Crystals}} by {{Uniaxial
  Stress}}},}\ }\href {https://doi.org/10.1103/PhysRevLett.20.50} {\bibfield
  {journal} {\bibinfo  {journal} {Phys. Rev. Lett.}\ }\textbf {\bibinfo
  {volume} {20}},\ \bibinfo {pages} {50--53} (\bibinfo {year}
  {1968})}\BibitemShut {NoStop}%
\bibitem [{\citenamefont {Akimoto}\ and\ \citenamefont
  {Hasegawa}(1968)}]{akimoto_Phys.Rev.Lett._1968}%
  \BibitemOpen
  \bibfield  {author} {\bibinfo {author} {\bibfnamefont {O.}~\bibnamefont
  {Akimoto}}\ and\ \bibinfo {author} {\bibfnamefont {H.}~\bibnamefont
  {Hasegawa}},\ }\bibfield  {title} {\enquote {\bibinfo {title}
  {Strain-{{Induced Splitting}} and {{Polarization}} of {{Excitons Due}} to
  {{Exchange Interaction}}},}\ }\href
  {https://doi.org/10.1103/PhysRevLett.20.916} {\bibfield  {journal} {\bibinfo
  {journal} {Phys. Rev. Lett.}\ }\textbf {\bibinfo {volume} {20}},\ \bibinfo
  {pages} {916--918} (\bibinfo {year} {1968})}\BibitemShut {NoStop}%
\bibitem [{\citenamefont {Nilsson}, \citenamefont {Janz{\'e}n},\ and\
  \citenamefont
  {{Kakanakova-Georgieva}}(2016)}]{nilsson_J.Phys.Appl.Phys._2016}%
  \BibitemOpen
  \bibfield  {author} {\bibinfo {author} {\bibfnamefont {D.}~\bibnamefont
  {Nilsson}}, \bibinfo {author} {\bibfnamefont {E.}~\bibnamefont
  {Janz{\'e}n}},\ and\ \bibinfo {author} {\bibfnamefont {A.}~\bibnamefont
  {{Kakanakova-Georgieva}}},\ }\bibfield  {title} {\enquote {\bibinfo {title}
  {Lattice parameters of {{AlN}} bulk, homoepitaxial and heteroepitaxial
  material},}\ }\href {https://doi.org/10.1088/0022-3727/49/17/175108}
  {\bibfield  {journal} {\bibinfo  {journal} {J. Phys. D: Appl. Phys.}\
  }\textbf {\bibinfo {volume} {49}},\ \bibinfo {pages} {175108} (\bibinfo
  {year} {2016})}\BibitemShut {NoStop}%
\bibitem [{\citenamefont {Neuschl}\ \emph {et~al.}(2013)\citenamefont
  {Neuschl}, \citenamefont {Thonke}, \citenamefont {Feneberg}, \citenamefont
  {Goldhahn}, \citenamefont {Wunderer}, \citenamefont {Yang}, \citenamefont
  {Johnson}, \citenamefont {Xie}, \citenamefont {Mita}, \citenamefont {Rice},
  \citenamefont {Collazo},\ and\ \citenamefont
  {Sitar}}]{neuschlDirectDeterminationSilicon2013}%
  \BibitemOpen
  \bibfield  {author} {\bibinfo {author} {\bibfnamefont {B.}~\bibnamefont
  {Neuschl}}, \bibinfo {author} {\bibfnamefont {K.}~\bibnamefont {Thonke}},
  \bibinfo {author} {\bibfnamefont {M.}~\bibnamefont {Feneberg}}, \bibinfo
  {author} {\bibfnamefont {R.}~\bibnamefont {Goldhahn}}, \bibinfo {author}
  {\bibfnamefont {T.}~\bibnamefont {Wunderer}}, \bibinfo {author}
  {\bibfnamefont {Z.}~\bibnamefont {Yang}}, \bibinfo {author} {\bibfnamefont
  {N.~M.}\ \bibnamefont {Johnson}}, \bibinfo {author} {\bibfnamefont
  {J.}~\bibnamefont {Xie}}, \bibinfo {author} {\bibfnamefont {S.}~\bibnamefont
  {Mita}}, \bibinfo {author} {\bibfnamefont {A.}~\bibnamefont {Rice}}, \bibinfo
  {author} {\bibfnamefont {R.}~\bibnamefont {Collazo}},\ and\ \bibinfo {author}
  {\bibfnamefont {Z.}~\bibnamefont {Sitar}},\ }\bibfield  {title} {\enquote
  {\bibinfo {title} {Direct determination of the silicon donor ionization
  energy in homoepitaxial {{AlN}} from photoluminescence two-electron
  transitions},}\ }\href {https://doi.org/10.1063/1.4821183} {\bibfield
  {journal} {\bibinfo  {journal} {Appl. Phys. Lett.}\ }\textbf {\bibinfo
  {volume} {103}},\ \bibinfo {pages} {122105} (\bibinfo {year}
  {2013})}\BibitemShut {NoStop}%
\bibitem [{\citenamefont {Zywietz}, \citenamefont {Neugebauer},\ and\
  \citenamefont {Scheffler}(1999)}]{zywietzAdsorptionOxygenGaN1999}%
  \BibitemOpen
  \bibfield  {author} {\bibinfo {author} {\bibfnamefont {T.~K.}\ \bibnamefont
  {Zywietz}}, \bibinfo {author} {\bibfnamefont {J.}~\bibnamefont
  {Neugebauer}},\ and\ \bibinfo {author} {\bibfnamefont {M.}~\bibnamefont
  {Scheffler}},\ }\bibfield  {title} {\enquote {\bibinfo {title} {The
  adsorption of oxygen at {{GaN}} surfaces},}\ }\href
  {https://doi.org/10.1063/1.123658} {\bibfield  {journal} {\bibinfo  {journal}
  {Appl. Phys. Lett.}\ }\textbf {\bibinfo {volume} {74}},\ \bibinfo {pages}
  {1695--1697} (\bibinfo {year} {1999})}\BibitemShut {NoStop}%
\bibitem [{\citenamefont {{Fern{\'a}ndez-Garrido}}\ \emph
  {et~al.}(2016)\citenamefont {{Fern{\'a}ndez-Garrido}}, \citenamefont
  {L{\"a}hnemann}, \citenamefont {Hauswald}, \citenamefont {Korytov},
  \citenamefont {Albrecht}, \citenamefont {Ch{\`e}ze}, \citenamefont
  {Skierbiszewski},\ and\ \citenamefont
  {Brandt}}]{fernandez-garrido_Phys.Rev.Appl._2016}%
  \BibitemOpen
  \bibfield  {author} {\bibinfo {author} {\bibfnamefont {S.}~\bibnamefont
  {{Fern{\'a}ndez-Garrido}}}, \bibinfo {author} {\bibfnamefont
  {J.}~\bibnamefont {L{\"a}hnemann}}, \bibinfo {author} {\bibfnamefont
  {C.}~\bibnamefont {Hauswald}}, \bibinfo {author} {\bibfnamefont
  {M.}~\bibnamefont {Korytov}}, \bibinfo {author} {\bibfnamefont
  {M.}~\bibnamefont {Albrecht}}, \bibinfo {author} {\bibfnamefont
  {C.}~\bibnamefont {Ch{\`e}ze}}, \bibinfo {author} {\bibfnamefont
  {C.}~\bibnamefont {Skierbiszewski}},\ and\ \bibinfo {author} {\bibfnamefont
  {O.}~\bibnamefont {Brandt}},\ }\bibfield  {title} {\enquote {\bibinfo {title}
  {Comparison of the luminous efficiencies of {{Ga-}} and {{N-polar
  In$_x$Ga$_{1-x}$N/In$_y$Ga$_{1-y}$N}} quantum wells grown by plasma-assisted
  molecular beam epitaxy},}\ }\href
  {https://doi.org/10.1103/physrevapplied.6.034017} {\bibfield  {journal}
  {\bibinfo  {journal} {Phys. Rev. Applied}\ }\textbf {\bibinfo {volume} {6}},\
  \bibinfo {pages} {034017} (\bibinfo {year} {2016})}\BibitemShut {NoStop}%
\bibitem [{\citenamefont {Ch{\`e}ze}\ \emph {et~al.}(2018)\citenamefont
  {Ch{\`e}ze}, \citenamefont {Feix}, \citenamefont {L{\"a}hnemann},
  \citenamefont {Flissikowski}, \citenamefont {Kry{\'s}ko}, \citenamefont
  {Wolny}, \citenamefont {Turski}, \citenamefont {Skierbiszewski},\ and\
  \citenamefont {Brandt}}]{cheze_Appl.Phys.Lett._2018}%
  \BibitemOpen
  \bibfield  {author} {\bibinfo {author} {\bibfnamefont {C.}~\bibnamefont
  {Ch{\`e}ze}}, \bibinfo {author} {\bibfnamefont {F.}~\bibnamefont {Feix}},
  \bibinfo {author} {\bibfnamefont {J.}~\bibnamefont {L{\"a}hnemann}}, \bibinfo
  {author} {\bibfnamefont {T.}~\bibnamefont {Flissikowski}}, \bibinfo {author}
  {\bibfnamefont {M.}~\bibnamefont {Kry{\'s}ko}}, \bibinfo {author}
  {\bibfnamefont {P.}~\bibnamefont {Wolny}}, \bibinfo {author} {\bibfnamefont
  {H.}~\bibnamefont {Turski}}, \bibinfo {author} {\bibfnamefont
  {C.}~\bibnamefont {Skierbiszewski}},\ and\ \bibinfo {author} {\bibfnamefont
  {O.}~\bibnamefont {Brandt}},\ }\bibfield  {title} {\enquote {\bibinfo {title}
  {Luminescent {{N-polar}} ({{In}},{{Ga}}){{N}}/{{GaN}} quantum wells achieved
  by plasma-assisted molecular beam epitaxy at temperatures exceeding 700
  \textdegree{{C}}},}\ }\href {https://doi.org/10.1063/1.5009184} {\bibfield
  {journal} {\bibinfo  {journal} {Appl. Phys. Lett.}\ }\textbf {\bibinfo
  {volume} {112}},\ \bibinfo {pages} {022102} (\bibinfo {year}
  {2018})}\BibitemShut {NoStop}%
\bibitem [{\citenamefont {Auzelle}\ \emph {et~al.}(2022)\citenamefont
  {Auzelle}, \citenamefont {Sinito}, \citenamefont {L{\"a}hnemann},
  \citenamefont {Gao}, \citenamefont {Flissikowski}, \citenamefont {Trampert},
  \citenamefont {{Fern{\'a}ndez-Garrido}},\ and\ \citenamefont
  {Brandt}}]{auzelle_Phys.Rev.Appl._2022}%
  \BibitemOpen
  \bibfield  {author} {\bibinfo {author} {\bibfnamefont {T.}~\bibnamefont
  {Auzelle}}, \bibinfo {author} {\bibfnamefont {C.}~\bibnamefont {Sinito}},
  \bibinfo {author} {\bibfnamefont {J.}~\bibnamefont {L{\"a}hnemann}}, \bibinfo
  {author} {\bibfnamefont {G.}~\bibnamefont {Gao}}, \bibinfo {author}
  {\bibfnamefont {T.}~\bibnamefont {Flissikowski}}, \bibinfo {author}
  {\bibfnamefont {A.}~\bibnamefont {Trampert}}, \bibinfo {author}
  {\bibfnamefont {S.}~\bibnamefont {{Fern{\'a}ndez-Garrido}}},\ and\ \bibinfo
  {author} {\bibfnamefont {O.}~\bibnamefont {Brandt}},\ }\bibfield  {title}
  {\enquote {\bibinfo {title} {Interface {{Recombination}} in {{Ga-}} and
  {{N-Polar GaN}}/({{Al}},{{Ga}}){{N Quantum Wells Grown}} by {{Molecular Beam
  Epitaxy}}},}\ }\href {https://doi.org/10.1103/PhysRevApplied.17.044030}
  {\bibfield  {journal} {\bibinfo  {journal} {Phys. Rev. Applied}\ }\textbf
  {\bibinfo {volume} {17}},\ \bibinfo {pages} {044030} (\bibinfo {year}
  {2022})}\BibitemShut {NoStop}%
\bibitem [{\citenamefont {Murotani}\ \emph {et~al.}(2009)\citenamefont
  {Murotani}, \citenamefont {Kuronaka}, \citenamefont {Yamada}, \citenamefont
  {Taguchi}, \citenamefont {Okada},\ and\ \citenamefont
  {Amano}}]{murotaniTemperatureDependenceExcitonic2009}%
  \BibitemOpen
  \bibfield  {author} {\bibinfo {author} {\bibfnamefont {H.}~\bibnamefont
  {Murotani}}, \bibinfo {author} {\bibfnamefont {T.}~\bibnamefont {Kuronaka}},
  \bibinfo {author} {\bibfnamefont {Y.}~\bibnamefont {Yamada}}, \bibinfo
  {author} {\bibfnamefont {T.}~\bibnamefont {Taguchi}}, \bibinfo {author}
  {\bibfnamefont {N.}~\bibnamefont {Okada}},\ and\ \bibinfo {author}
  {\bibfnamefont {H.}~\bibnamefont {Amano}},\ }\bibfield  {title} {\enquote
  {\bibinfo {title} {Temperature dependence of excitonic transitions in a-plane
  {{AlN}} epitaxial layers},}\ }\href {https://doi.org/10.1063/1.3116183}
  {\bibfield  {journal} {\bibinfo  {journal} {J. Appl. Phys.}\ }\textbf
  {\bibinfo {volume} {105}},\ \bibinfo {pages} {083533} (\bibinfo {year}
  {2009})}\BibitemShut {NoStop}%
\bibitem [{\citenamefont {Baldereschi}\ and\ \citenamefont
  {Diaz}(1970)}]{baldereschi_IlNuovoCimentoBSer.10_1970}%
  \BibitemOpen
  \bibfield  {author} {\bibinfo {author} {\bibfnamefont {A.}~\bibnamefont
  {Baldereschi}}\ and\ \bibinfo {author} {\bibfnamefont {M.~G.}\ \bibnamefont
  {Diaz}},\ }\bibfield  {title} {\enquote {\bibinfo {title} {Anisotropy of
  excitons in semiconductors},}\ }\href {https://doi.org/10.1007/BF02710415}
  {\bibfield  {journal} {\bibinfo  {journal} {Nuov. Cim. B}\ }\textbf {\bibinfo
  {volume} {68}},\ \bibinfo {pages} {217--229} (\bibinfo {year}
  {1970})}\BibitemShut {NoStop}%
\bibitem [{\citenamefont {Muljarov}\ \emph {et~al.}(2000)\citenamefont
  {Muljarov}, \citenamefont {Yablonskii}, \citenamefont {Tikhodeev},
  \citenamefont {Bulatov},\ and\ \citenamefont
  {Birman}}]{muljarov_J.Math.Phys._2000}%
  \BibitemOpen
  \bibfield  {author} {\bibinfo {author} {\bibfnamefont {E.~A.}\ \bibnamefont
  {Muljarov}}, \bibinfo {author} {\bibfnamefont {A.~L.}\ \bibnamefont
  {Yablonskii}}, \bibinfo {author} {\bibfnamefont {S.~G.}\ \bibnamefont
  {Tikhodeev}}, \bibinfo {author} {\bibfnamefont {A.~E.}\ \bibnamefont
  {Bulatov}},\ and\ \bibinfo {author} {\bibfnamefont {J.~L.}\ \bibnamefont
  {Birman}},\ }\bibfield  {title} {\enquote {\bibinfo {title} {Hyperspherical
  theory of anisotropic exciton},}\ }\href {https://doi.org/10.1063/1.1286772}
  {\bibfield  {journal} {\bibinfo  {journal} {J. Math. Phys.}\ }\textbf
  {\bibinfo {volume} {41}},\ \bibinfo {pages} {6026--6041} (\bibinfo {year}
  {2000})}\BibitemShut {NoStop}%
\bibitem [{\citenamefont {Gil}\ \emph {et~al.}(2012)\citenamefont {Gil},
  \citenamefont {Felbacq}, \citenamefont {Guizal},\ and\ \citenamefont
  {Bouchitt{\'e}}}]{gil_Phys.StatusSolidiB_2012}%
  \BibitemOpen
  \bibfield  {author} {\bibinfo {author} {\bibfnamefont {B.}~\bibnamefont
  {Gil}}, \bibinfo {author} {\bibfnamefont {D.}~\bibnamefont {Felbacq}},
  \bibinfo {author} {\bibfnamefont {B.}~\bibnamefont {Guizal}},\ and\ \bibinfo
  {author} {\bibfnamefont {G.}~\bibnamefont {Bouchitt{\'e}}},\ }\bibfield
  {title} {\enquote {\bibinfo {title} {Excitonic states and their wave
  functions in anisotropic materials: {{A}} computation using the
  finite-element method and its application to {{AlN}}},}\ }\href
  {https://doi.org/10.1002/pssb.201100142} {\bibfield  {journal} {\bibinfo
  {journal} {Phys. Status Solidi B}\ }\textbf {\bibinfo {volume} {249}},\
  \bibinfo {pages} {455--458} (\bibinfo {year} {2012})}\BibitemShut {NoStop}%
\bibitem [{\citenamefont {Ishii}, \citenamefont {Funato},\ and\ \citenamefont
  {Kawakami}(2014)}]{ishiiEffectsStrongElectron2014}%
  \BibitemOpen
  \bibfield  {author} {\bibinfo {author} {\bibfnamefont {R.}~\bibnamefont
  {Ishii}}, \bibinfo {author} {\bibfnamefont {M.}~\bibnamefont {Funato}},\ and\
  \bibinfo {author} {\bibfnamefont {Y.}~\bibnamefont {Kawakami}},\ }\bibfield
  {title} {\enquote {\bibinfo {title} {Effects of strong electron\textendash
  hole exchange and exciton\textendash phonon interactions on the exciton
  binding energy of aluminum nitride},}\ }\href
  {https://doi.org/10.7567/JJAP.53.091001} {\bibfield  {journal} {\bibinfo
  {journal} {Jpn. J. Appl. Phys.}\ }\textbf {\bibinfo {volume} {53}},\ \bibinfo
  {pages} {091001} (\bibinfo {year} {2014})}\BibitemShut {NoStop}%
\bibitem [{\citenamefont {Davydov}\ \emph {et~al.}(1998)\citenamefont
  {Davydov}, \citenamefont {Kitaev}, \citenamefont {Goncharuk}, \citenamefont
  {Smirnov}, \citenamefont {Graul}, \citenamefont {Semchinova}, \citenamefont
  {Uffmann}, \citenamefont {Smirnov}, \citenamefont {Mirgorodsky},\ and\
  \citenamefont {Evarestov}}]{davydovPhononDispersionRaman1998}%
  \BibitemOpen
  \bibfield  {author} {\bibinfo {author} {\bibfnamefont {V.~{\relax Yu}.}\
  \bibnamefont {Davydov}}, \bibinfo {author} {\bibfnamefont {{\relax Yu}.~E.}\
  \bibnamefont {Kitaev}}, \bibinfo {author} {\bibfnamefont {I.~N.}\
  \bibnamefont {Goncharuk}}, \bibinfo {author} {\bibfnamefont {A.~N.}\
  \bibnamefont {Smirnov}}, \bibinfo {author} {\bibfnamefont {J.}~\bibnamefont
  {Graul}}, \bibinfo {author} {\bibfnamefont {O.}~\bibnamefont {Semchinova}},
  \bibinfo {author} {\bibfnamefont {D.}~\bibnamefont {Uffmann}}, \bibinfo
  {author} {\bibfnamefont {M.~B.}\ \bibnamefont {Smirnov}}, \bibinfo {author}
  {\bibfnamefont {A.~P.}\ \bibnamefont {Mirgorodsky}},\ and\ \bibinfo {author}
  {\bibfnamefont {R.~A.}\ \bibnamefont {Evarestov}},\ }\bibfield  {title}
  {\enquote {\bibinfo {title} {Phonon dispersion and {{Raman}} scattering in
  hexagonal {{GaN}} and {{AlN}}},}\ }\href
  {https://doi.org/10.1103/PhysRevB.58.12899} {\bibfield  {journal} {\bibinfo
  {journal} {Phys. Rev. B}\ }\textbf {\bibinfo {volume} {58}},\ \bibinfo
  {pages} {12899--12907} (\bibinfo {year} {1998})}\BibitemShut {NoStop}%
\bibitem [{\citenamefont {Bieker}\ \emph
  {et~al.}(2015{\natexlab{a}})\citenamefont {Bieker}, \citenamefont {Henn},
  \citenamefont {Kiessling}, \citenamefont {Ossau},\ and\ \citenamefont
  {Molenkamp}}]{bieker_Phys.Rev.Lett._2015}%
  \BibitemOpen
  \bibfield  {author} {\bibinfo {author} {\bibfnamefont {S.}~\bibnamefont
  {Bieker}}, \bibinfo {author} {\bibfnamefont {T.}~\bibnamefont {Henn}},
  \bibinfo {author} {\bibfnamefont {T.}~\bibnamefont {Kiessling}}, \bibinfo
  {author} {\bibfnamefont {W.}~\bibnamefont {Ossau}},\ and\ \bibinfo {author}
  {\bibfnamefont {L.~W.}\ \bibnamefont {Molenkamp}},\ }\bibfield  {title}
  {\enquote {\bibinfo {title} {Spatially resolved thermodynamics of the
  partially ionized exciton gas in {{GaAs}}},}\ }\href
  {https://doi.org/10.1103/physrevlett.114.227402} {\bibfield  {journal}
  {\bibinfo  {journal} {Phys. Rev. Lett.}\ }\textbf {\bibinfo {volume} {114}},\
  \bibinfo {pages} {227402} (\bibinfo {year} {2015}{\natexlab{a}})}\BibitemShut
  {NoStop}%
\bibitem [{\citenamefont {Jahn}\ \emph {et~al.}(2022)\citenamefont {Jahn},
  \citenamefont {Kaganer}, \citenamefont {Sabelfeld}, \citenamefont {Kireeva},
  \citenamefont {L{\"a}hnemann}, \citenamefont {Pf{\"u}ller}, \citenamefont
  {Flissikowski}, \citenamefont {Ch{\`e}ze}, \citenamefont {Biermann},
  \citenamefont {Calarco},\ and\ \citenamefont
  {Brandt}}]{jahn_Phys.Rev.Appl._2022}%
  \BibitemOpen
  \bibfield  {author} {\bibinfo {author} {\bibfnamefont {U.}~\bibnamefont
  {Jahn}}, \bibinfo {author} {\bibfnamefont {V.~M.}\ \bibnamefont {Kaganer}},
  \bibinfo {author} {\bibfnamefont {K.~K.}\ \bibnamefont {Sabelfeld}}, \bibinfo
  {author} {\bibfnamefont {A.~E.}\ \bibnamefont {Kireeva}}, \bibinfo {author}
  {\bibfnamefont {J.}~\bibnamefont {L{\"a}hnemann}}, \bibinfo {author}
  {\bibfnamefont {C.}~\bibnamefont {Pf{\"u}ller}}, \bibinfo {author}
  {\bibfnamefont {T.}~\bibnamefont {Flissikowski}}, \bibinfo {author}
  {\bibfnamefont {C.}~\bibnamefont {Ch{\`e}ze}}, \bibinfo {author}
  {\bibfnamefont {K.}~\bibnamefont {Biermann}}, \bibinfo {author}
  {\bibfnamefont {R.}~\bibnamefont {Calarco}},\ and\ \bibinfo {author}
  {\bibfnamefont {O.}~\bibnamefont {Brandt}},\ }\bibfield  {title} {\enquote
  {\bibinfo {title} {Carrier {{Diffusion}} in {{GaN}} : {{A Cathodoluminescence
  Study}}. {{I}}. {{Temperature-Dependent Generation Volume}}},}\ }\href
  {https://doi.org/10.1103/PhysRevApplied.17.024017} {\bibfield  {journal}
  {\bibinfo  {journal} {Phys. Rev. Applied}\ }\textbf {\bibinfo {volume}
  {17}},\ \bibinfo {pages} {024017} (\bibinfo {year} {2022})}\BibitemShut
  {NoStop}%
\bibitem [{\citenamefont {Lugli}\ \emph {et~al.}(1987)\citenamefont {Lugli},
  \citenamefont {Jacoboni}, \citenamefont {Reggiani},\ and\ \citenamefont
  {Kocevar}}]{lugli_Appl.Phys.Lett._1987}%
  \BibitemOpen
  \bibfield  {author} {\bibinfo {author} {\bibfnamefont {P.}~\bibnamefont
  {Lugli}}, \bibinfo {author} {\bibfnamefont {C.}~\bibnamefont {Jacoboni}},
  \bibinfo {author} {\bibfnamefont {L.}~\bibnamefont {Reggiani}},\ and\
  \bibinfo {author} {\bibfnamefont {P.}~\bibnamefont {Kocevar}},\ }\bibfield
  {title} {\enquote {\bibinfo {title} {Monte {{Carlo}} algorithm for hot
  phonons in polar semiconductors},}\ }\href {https://doi.org/10.1063/1.97925}
  {\bibfield  {journal} {\bibinfo  {journal} {Appl. Phys. Lett.}\ }\textbf
  {\bibinfo {volume} {50}},\ \bibinfo {pages} {1251--1253} (\bibinfo {year}
  {1987})}\BibitemShut {NoStop}%
\bibitem [{\citenamefont {Bieker}\ \emph
  {et~al.}(2015{\natexlab{b}})\citenamefont {Bieker}, \citenamefont
  {Kiessling}, \citenamefont {Ossau},\ and\ \citenamefont
  {Molenkamp}}]{bieker_Phys.Rev.B_2015}%
  \BibitemOpen
  \bibfield  {author} {\bibinfo {author} {\bibfnamefont {S.}~\bibnamefont
  {Bieker}}, \bibinfo {author} {\bibfnamefont {T.}~\bibnamefont {Kiessling}},
  \bibinfo {author} {\bibfnamefont {W.}~\bibnamefont {Ossau}},\ and\ \bibinfo
  {author} {\bibfnamefont {L.~W.}\ \bibnamefont {Molenkamp}},\ }\bibfield
  {title} {\enquote {\bibinfo {title} {Correct determination of low-temperature
  free-exciton diffusion profiles in {{GaAs}}},}\ }\href
  {https://doi.org/10.1103/physrevb.92.121201} {\bibfield  {journal} {\bibinfo
  {journal} {Phys. Rev. B}\ }\textbf {\bibinfo {volume} {92}},\ \bibinfo
  {pages} {121201(R)} (\bibinfo {year} {2015}{\natexlab{b}})}\BibitemShut
  {NoStop}%
\end{thebibliography}%


\begin{thebibliography}{20}%
\makeatletter
\providecommand \@ifxundefined [1]{%
 \@ifx{#1\undefined}
}%
\providecommand \@ifnum [1]{%
 \ifnum #1\expandafter \@firstoftwo
 \else \expandafter \@secondoftwo
 \fi
}%
\providecommand \@ifx [1]{%
 \ifx #1\expandafter \@firstoftwo
 \else \expandafter \@secondoftwo
 \fi
}%
\providecommand \natexlab [1]{#1}%
\providecommand \enquote  [1]{``#1''}%
\providecommand \bibnamefont  [1]{#1}%
\providecommand \bibfnamefont [1]{#1}%
\providecommand \citenamefont [1]{#1}%
\providecommand \href@noop [0]{\@secondoftwo}%
\providecommand \href [0]{\begingroup \@sanitize@url \@href}%
\providecommand \@href[1]{\@@startlink{#1}\@@href}%
\providecommand \@@href[1]{\endgroup#1\@@endlink}%
\providecommand \@sanitize@url [0]{\catcode `\\12\catcode `\$12\catcode
  `\&12\catcode `\#12\catcode `\^12\catcode `\_12\catcode `\%12\relax}%
\providecommand \@@startlink[1]{}%
\providecommand \@@endlink[0]{}%
\providecommand \url  [0]{\begingroup\@sanitize@url \@url }%
\providecommand \@url [1]{\endgroup\@href {#1}{\urlprefix }}%
\providecommand \urlprefix  [0]{URL }%
\providecommand \Eprint [0]{\href }%
\providecommand \doibase [0]{https://doi.org/}%
\providecommand \selectlanguage [0]{\@gobble}%
\providecommand \bibinfo  [0]{\@secondoftwo}%
\providecommand \bibfield  [0]{\@secondoftwo}%
\providecommand \translation [1]{[#1]}%
\providecommand \BibitemOpen [0]{}%
\providecommand \bibitemStop [0]{}%
\providecommand \bibitemNoStop [0]{.\EOS\space}%
\providecommand \EOS [0]{\spacefactor3000\relax}%
\providecommand \BibitemShut  [1]{\csname bibitem#1\endcsname}%
\let\auto@bib@innerbib\@empty
\bibitem [{\citenamefont {Jahn}\ \emph {et~al.}(2022)\citenamefont {Jahn},
  \citenamefont {Kaganer}, \citenamefont {Sabelfeld}, \citenamefont {Kireeva},
  \citenamefont {L{\"a}hnemann}, \citenamefont {Pf{\"u}ller}, \citenamefont
  {Flissikowski}, \citenamefont {Ch{\`e}ze}, \citenamefont {Biermann},
  \citenamefont {Calarco},\ and\ \citenamefont
  {Brandt}}]{jahnCarrierDiffusionGaN2022}%
  \BibitemOpen
  \bibfield  {author} {\bibinfo {author} {\bibfnamefont {U.}~\bibnamefont
  {Jahn}}, \bibinfo {author} {\bibfnamefont {V.~M.}\ \bibnamefont {Kaganer}},
  \bibinfo {author} {\bibfnamefont {K.~K.}\ \bibnamefont {Sabelfeld}}, \bibinfo
  {author} {\bibfnamefont {A.~E.}\ \bibnamefont {Kireeva}}, \bibinfo {author}
  {\bibfnamefont {J.}~\bibnamefont {L{\"a}hnemann}}, \bibinfo {author}
  {\bibfnamefont {C.}~\bibnamefont {Pf{\"u}ller}}, \bibinfo {author}
  {\bibfnamefont {T.}~\bibnamefont {Flissikowski}}, \bibinfo {author}
  {\bibfnamefont {C.}~\bibnamefont {Ch{\`e}ze}}, \bibinfo {author}
  {\bibfnamefont {K.}~\bibnamefont {Biermann}}, \bibinfo {author}
  {\bibfnamefont {R.}~\bibnamefont {Calarco}},\ and\ \bibinfo {author}
  {\bibfnamefont {O.}~\bibnamefont {Brandt}},\ }\bibfield  {title} {\enquote
  {\bibinfo {title} {Carrier {{Diffusion}} in {{GaN}}: {{A Cathodoluminescence
  Study}}. {{I}}. {{Temperature-Dependent Generation Volume}}},}\ }\href
  {https://doi.org/10.1103/PhysRevApplied.17.024017} {\bibfield  {journal}
  {\bibinfo  {journal} {Physical Review Applied}\ }\textbf {\bibinfo {volume}
  {17}},\ \bibinfo {pages} {024017} (\bibinfo {year} {2022})}\BibitemShut
  {NoStop}%
\bibitem [{\citenamefont {Drouin}\ \emph {et~al.}(2007)\citenamefont {Drouin},
  \citenamefont {Couture}, \citenamefont {Joly}, \citenamefont {Tastet},
  \citenamefont {Aimez},\ and\ \citenamefont {Gauvin}}]{drouinCASINOV2422007}%
  \BibitemOpen
  \bibfield  {author} {\bibinfo {author} {\bibfnamefont {D.}~\bibnamefont
  {Drouin}}, \bibinfo {author} {\bibfnamefont {A.~R.}\ \bibnamefont {Couture}},
  \bibinfo {author} {\bibfnamefont {D.}~\bibnamefont {Joly}}, \bibinfo {author}
  {\bibfnamefont {X.}~\bibnamefont {Tastet}}, \bibinfo {author} {\bibfnamefont
  {V.}~\bibnamefont {Aimez}},\ and\ \bibinfo {author} {\bibfnamefont
  {R.}~\bibnamefont {Gauvin}},\ }\bibfield  {title} {\enquote {\bibinfo {title}
  {{{CASINO V2}}.42\textemdash{{A Fast}} and {{Easy-to-use Modeling Tool}} for
  {{Scanning Electron Microscopy}} and {{Microanalysis Users}}},}\ }\href
  {https://doi.org/10.1002/sca.20000} {\bibfield  {journal} {\bibinfo
  {journal} {Scanning}\ }\textbf {\bibinfo {volume} {29}},\ \bibinfo {pages}
  {92--101} (\bibinfo {year} {2007})}\BibitemShut {NoStop}%
\bibitem [{\citenamefont {Wu}\ and\ \citenamefont
  {Wittry}(1978)}]{wuInvestigationMinorityCarrier1978}%
  \BibitemOpen
  \bibfield  {author} {\bibinfo {author} {\bibfnamefont {C.~J.}\ \bibnamefont
  {Wu}}\ and\ \bibinfo {author} {\bibfnamefont {D.~B.}\ \bibnamefont
  {Wittry}},\ }\bibfield  {title} {\enquote {\bibinfo {title} {Investigation of
  minority-carrier diffusion lengths by electron bombardment of {{Schottky}}
  barriers},}\ }\href {https://doi.org/10.1063/1.325163} {\bibfield  {journal}
  {\bibinfo  {journal} {Journal of Applied Physics}\ }\textbf {\bibinfo
  {volume} {49}},\ \bibinfo {pages} {2827--2836} (\bibinfo {year}
  {1978})}\BibitemShut {NoStop}%
\bibitem [{\citenamefont {Jahn}\ \emph {et~al.}(2003)\citenamefont {Jahn},
  \citenamefont {Dhar}, \citenamefont {Brandt}, \citenamefont {Grahn},
  \citenamefont {Ploog},\ and\ \citenamefont {Watson}}]{Jahn_jap_2003}%
  \BibitemOpen
  \bibfield  {author} {\bibinfo {author} {\bibfnamefont {U.}~\bibnamefont
  {Jahn}}, \bibinfo {author} {\bibfnamefont {S.}~\bibnamefont {Dhar}}, \bibinfo
  {author} {\bibfnamefont {O.}~\bibnamefont {Brandt}}, \bibinfo {author}
  {\bibfnamefont {H.~T.}\ \bibnamefont {Grahn}}, \bibinfo {author}
  {\bibfnamefont {K.~H.}\ \bibnamefont {Ploog}},\ and\ \bibinfo {author}
  {\bibfnamefont {I.~M.}\ \bibnamefont {Watson}},\ }\bibfield  {title}
  {\enquote {\bibinfo {title} {Exciton localization and quantum efficiency --
  {{A}} comparative cathodoluminescence study of ({{In}},{{Ga}}){{N}}/{{GaN}}
  and {{GaN}}/({{Al}},{{Ga}}){{N}} quantum wells},}\ }\href
  {https://doi.org/10.1063/1.1529993} {\bibfield  {journal} {\bibinfo
  {journal} {Journal of Applied Physics}\ }\textbf {\bibinfo {volume} {93}},\
  \bibinfo {pages} {1048--1053} (\bibinfo {year} {2003})}\BibitemShut {NoStop}%
\bibitem [{\citenamefont {Leute}\ \emph {et~al.}(2009)\citenamefont {Leute},
  \citenamefont {Feneberg}, \citenamefont {Sauer}, \citenamefont {Thonke},
  \citenamefont {Thapa}, \citenamefont {Scholz}, \citenamefont {Taniyasu},\
  and\ \citenamefont {Kasu}}]{leutePhotoluminescenceHighlyExcited2009}%
  \BibitemOpen
  \bibfield  {author} {\bibinfo {author} {\bibfnamefont {R.~a.~R.}\
  \bibnamefont {Leute}}, \bibinfo {author} {\bibfnamefont {M.}~\bibnamefont
  {Feneberg}}, \bibinfo {author} {\bibfnamefont {R.}~\bibnamefont {Sauer}},
  \bibinfo {author} {\bibfnamefont {K.}~\bibnamefont {Thonke}}, \bibinfo
  {author} {\bibfnamefont {S.~B.}\ \bibnamefont {Thapa}}, \bibinfo {author}
  {\bibfnamefont {F.}~\bibnamefont {Scholz}}, \bibinfo {author} {\bibfnamefont
  {Y.}~\bibnamefont {Taniyasu}},\ and\ \bibinfo {author} {\bibfnamefont
  {M.}~\bibnamefont {Kasu}},\ }\bibfield  {title} {\enquote {\bibinfo {title}
  {Photoluminescence of highly excited {{AlN}}: {{Biexcitons}} and
  exciton-exciton scattering},}\ }\href {https://doi.org/10.1063/1.3186044}
  {\bibfield  {journal} {\bibinfo  {journal} {Applied Physics Letters}\
  }\textbf {\bibinfo {volume} {95}},\ \bibinfo {pages} {031903} (\bibinfo
  {year} {2009})}\BibitemShut {NoStop}%
\bibitem [{\citenamefont {Ishii}\ \emph {et~al.}(2022)\citenamefont {Ishii},
  \citenamefont {Nagashima}, \citenamefont {Yamamoto}, \citenamefont {Hitomi},
  \citenamefont {Funato},\ and\ \citenamefont
  {Kawakami}}]{ishiiStimulatedEmissionMechanism2022}%
  \BibitemOpen
  \bibfield  {author} {\bibinfo {author} {\bibfnamefont {R.}~\bibnamefont
  {Ishii}}, \bibinfo {author} {\bibfnamefont {T.}~\bibnamefont {Nagashima}},
  \bibinfo {author} {\bibfnamefont {R.}~\bibnamefont {Yamamoto}}, \bibinfo
  {author} {\bibfnamefont {T.}~\bibnamefont {Hitomi}}, \bibinfo {author}
  {\bibfnamefont {M.}~\bibnamefont {Funato}},\ and\ \bibinfo {author}
  {\bibfnamefont {Y.}~\bibnamefont {Kawakami}},\ }\bibfield  {title} {\enquote
  {\bibinfo {title} {Stimulated emission mechanism of aluminum nitride},}\
  }\href {https://doi.org/10.1103/PhysRevB.105.205206} {\bibfield  {journal}
  {\bibinfo  {journal} {Physical Review B}\ }\textbf {\bibinfo {volume}
  {105}},\ \bibinfo {pages} {205206} (\bibinfo {year} {2022})}\BibitemShut
  {NoStop}%
\bibitem [{\citenamefont {Neuschl}\ \emph {et~al.}(2013)\citenamefont
  {Neuschl}, \citenamefont {Thonke}, \citenamefont {Feneberg}, \citenamefont
  {Goldhahn}, \citenamefont {Wunderer}, \citenamefont {Yang}, \citenamefont
  {Johnson}, \citenamefont {Xie}, \citenamefont {Mita}, \citenamefont {Rice},
  \citenamefont {Collazo},\ and\ \citenamefont
  {Sitar}}]{neuschlDirectDeterminationSilicon2013}%
  \BibitemOpen
  \bibfield  {author} {\bibinfo {author} {\bibfnamefont {B.}~\bibnamefont
  {Neuschl}}, \bibinfo {author} {\bibfnamefont {K.}~\bibnamefont {Thonke}},
  \bibinfo {author} {\bibfnamefont {M.}~\bibnamefont {Feneberg}}, \bibinfo
  {author} {\bibfnamefont {R.}~\bibnamefont {Goldhahn}}, \bibinfo {author}
  {\bibfnamefont {T.}~\bibnamefont {Wunderer}}, \bibinfo {author}
  {\bibfnamefont {Z.}~\bibnamefont {Yang}}, \bibinfo {author} {\bibfnamefont
  {N.~M.}\ \bibnamefont {Johnson}}, \bibinfo {author} {\bibfnamefont
  {J.}~\bibnamefont {Xie}}, \bibinfo {author} {\bibfnamefont {S.}~\bibnamefont
  {Mita}}, \bibinfo {author} {\bibfnamefont {A.}~\bibnamefont {Rice}}, \bibinfo
  {author} {\bibfnamefont {R.}~\bibnamefont {Collazo}},\ and\ \bibinfo {author}
  {\bibfnamefont {Z.}~\bibnamefont {Sitar}},\ }\bibfield  {title} {\enquote
  {\bibinfo {title} {Direct determination of the silicon donor ionization
  energy in homoepitaxial {{AlN}} from photoluminescence two-electron
  transitions},}\ }\href {https://doi.org/10.1063/1.4821183} {\bibfield
  {journal} {\bibinfo  {journal} {Applied Physics Letters}\ }\textbf {\bibinfo
  {volume} {103}},\ \bibinfo {pages} {122105} (\bibinfo {year}
  {2013})}\BibitemShut {NoStop}%
\bibitem [{\citenamefont {Nepal}\ \emph {et~al.}(2006)\citenamefont {Nepal},
  \citenamefont {Nakarmi}, \citenamefont {Jang}, \citenamefont {Lin},\ and\
  \citenamefont {Jiang}}]{nepalGrowthPhotoluminescenceStudies2006}%
  \BibitemOpen
  \bibfield  {author} {\bibinfo {author} {\bibfnamefont {N.}~\bibnamefont
  {Nepal}}, \bibinfo {author} {\bibfnamefont {M.~L.}\ \bibnamefont {Nakarmi}},
  \bibinfo {author} {\bibfnamefont {H.~U.}\ \bibnamefont {Jang}}, \bibinfo
  {author} {\bibfnamefont {J.~Y.}\ \bibnamefont {Lin}},\ and\ \bibinfo {author}
  {\bibfnamefont {H.~X.}\ \bibnamefont {Jiang}},\ }\bibfield  {title} {\enquote
  {\bibinfo {title} {Growth and photoluminescence studies of {{Zn-doped AlN}}
  epilayers},}\ }\href {https://doi.org/10.1063/1.2387869} {\bibfield
  {journal} {\bibinfo  {journal} {Applied Physics Letters}\ }\textbf {\bibinfo
  {volume} {89}},\ \bibinfo {pages} {192111} (\bibinfo {year}
  {2006})}\BibitemShut {NoStop}%
\bibitem [{\citenamefont {Nepal}\ \emph {et~al.}(2004)\citenamefont {Nepal},
  \citenamefont {Nakarmi}, \citenamefont {Nam}, \citenamefont {Lin},\ and\
  \citenamefont {Jiang}}]{nepalAcceptorboundExcitonTransition2004}%
  \BibitemOpen
  \bibfield  {author} {\bibinfo {author} {\bibfnamefont {N.}~\bibnamefont
  {Nepal}}, \bibinfo {author} {\bibfnamefont {M.~L.}\ \bibnamefont {Nakarmi}},
  \bibinfo {author} {\bibfnamefont {K.~B.}\ \bibnamefont {Nam}}, \bibinfo
  {author} {\bibfnamefont {J.~Y.}\ \bibnamefont {Lin}},\ and\ \bibinfo {author}
  {\bibfnamefont {H.~X.}\ \bibnamefont {Jiang}},\ }\bibfield  {title} {\enquote
  {\bibinfo {title} {Acceptor-bound exciton transition in {{Mg-doped AlN}}
  epilayer},}\ }\href {https://doi.org/10.1063/1.1796521} {\bibfield  {journal}
  {\bibinfo  {journal} {Applied Physics Letters}\ }\textbf {\bibinfo {volume}
  {85}},\ \bibinfo {pages} {2271--2273} (\bibinfo {year} {2004})}\BibitemShut
  {NoStop}%
\bibitem [{\citenamefont {Nakarmi}\ \emph {et~al.}(2006)\citenamefont
  {Nakarmi}, \citenamefont {Nepal}, \citenamefont {Ugolini}, \citenamefont
  {Altahtamouni}, \citenamefont {Lin},\ and\ \citenamefont
  {Jiang}}]{nakarmiCorrelationOpticalElectrical2006}%
  \BibitemOpen
  \bibfield  {author} {\bibinfo {author} {\bibfnamefont {M.~L.}\ \bibnamefont
  {Nakarmi}}, \bibinfo {author} {\bibfnamefont {N.}~\bibnamefont {Nepal}},
  \bibinfo {author} {\bibfnamefont {C.}~\bibnamefont {Ugolini}}, \bibinfo
  {author} {\bibfnamefont {T.~M.}\ \bibnamefont {Altahtamouni}}, \bibinfo
  {author} {\bibfnamefont {J.~Y.}\ \bibnamefont {Lin}},\ and\ \bibinfo {author}
  {\bibfnamefont {H.~X.}\ \bibnamefont {Jiang}},\ }\bibfield  {title} {\enquote
  {\bibinfo {title} {Correlation between optical and electrical properties of
  {{Mg-doped AlN}} epilayers},}\ }\href {https://doi.org/10.1063/1.2362582}
  {\bibfield  {journal} {\bibinfo  {journal} {Applied Physics Letters}\
  }\textbf {\bibinfo {volume} {89}},\ \bibinfo {pages} {152120} (\bibinfo
  {year} {2006})}\BibitemShut {NoStop}%
\bibitem [{\citenamefont {Sedhain}\ \emph {et~al.}(2008)\citenamefont
  {Sedhain}, \citenamefont {Al~Tahtamouni}, \citenamefont {Li}, \citenamefont
  {Lin},\ and\ \citenamefont {Jiang}}]{sedhainBerylliumAcceptorBinding2008}%
  \BibitemOpen
  \bibfield  {author} {\bibinfo {author} {\bibfnamefont {A.}~\bibnamefont
  {Sedhain}}, \bibinfo {author} {\bibfnamefont {T.~M.}\ \bibnamefont
  {Al~Tahtamouni}}, \bibinfo {author} {\bibfnamefont {J.}~\bibnamefont {Li}},
  \bibinfo {author} {\bibfnamefont {J.~Y.}\ \bibnamefont {Lin}},\ and\ \bibinfo
  {author} {\bibfnamefont {H.~X.}\ \bibnamefont {Jiang}},\ }\bibfield  {title}
  {\enquote {\bibinfo {title} {Beryllium acceptor binding energy in {{AlN}}},}\
  }\href {https://doi.org/10.1063/1.2996977} {\bibfield  {journal} {\bibinfo
  {journal} {Applied Physics Letters}\ }\textbf {\bibinfo {volume} {93}},\
  \bibinfo {pages} {141104} (\bibinfo {year} {2008})}\BibitemShut {NoStop}%
\bibitem [{\citenamefont {Bryan}\ \emph {et~al.}(2014)\citenamefont {Bryan},
  \citenamefont {Bryan}, \citenamefont {Bobea}, \citenamefont {Hussey},
  \citenamefont {Kirste}, \citenamefont {Sitar},\ and\ \citenamefont
  {Collazo}}]{bryanExcitonTransitionsOxygen2014}%
  \BibitemOpen
  \bibfield  {author} {\bibinfo {author} {\bibfnamefont {Z.}~\bibnamefont
  {Bryan}}, \bibinfo {author} {\bibfnamefont {I.}~\bibnamefont {Bryan}},
  \bibinfo {author} {\bibfnamefont {M.}~\bibnamefont {Bobea}}, \bibinfo
  {author} {\bibfnamefont {L.}~\bibnamefont {Hussey}}, \bibinfo {author}
  {\bibfnamefont {R.}~\bibnamefont {Kirste}}, \bibinfo {author} {\bibfnamefont
  {Z.}~\bibnamefont {Sitar}},\ and\ \bibinfo {author} {\bibfnamefont
  {R.}~\bibnamefont {Collazo}},\ }\bibfield  {title} {\enquote {\bibinfo
  {title} {Exciton transitions and oxygen as a donor in m-plane {{AlN}}
  homoepitaxial films},}\ }\href {https://doi.org/10.1063/1.4870284} {\bibfield
   {journal} {\bibinfo  {journal} {Journal of Applied Physics}\ }\textbf
  {\bibinfo {volume} {115}},\ \bibinfo {pages} {133503} (\bibinfo {year}
  {2014})}\BibitemShut {NoStop}%
\bibitem [{\citenamefont {Funato}\ \emph {et~al.}(2012)\citenamefont {Funato},
  \citenamefont {Matsuda}, \citenamefont {Banal}, \citenamefont {Ishii},\ and\
  \citenamefont {Kawakami}}]{funatoHomoepitaxyPhotoluminescenceProperties2012}%
  \BibitemOpen
  \bibfield  {author} {\bibinfo {author} {\bibfnamefont {M.}~\bibnamefont
  {Funato}}, \bibinfo {author} {\bibfnamefont {K.}~\bibnamefont {Matsuda}},
  \bibinfo {author} {\bibfnamefont {R.~G.}\ \bibnamefont {Banal}}, \bibinfo
  {author} {\bibfnamefont {R.}~\bibnamefont {Ishii}},\ and\ \bibinfo {author}
  {\bibfnamefont {Y.}~\bibnamefont {Kawakami}},\ }\bibfield  {title} {\enquote
  {\bibinfo {title} {Homoepitaxy and {{Photoluminescence Properties}} of (0001)
  {{AlN}}},}\ }\href {https://doi.org/10.1143/APEX.5.082001} {\bibfield
  {journal} {\bibinfo  {journal} {Applied Physics Express}\ }\textbf {\bibinfo
  {volume} {5}},\ \bibinfo {pages} {082001} (\bibinfo {year}
  {2012})}\BibitemShut {NoStop}%
\bibitem [{\citenamefont {Chichibu}\ \emph {et~al.}(2019)\citenamefont
  {Chichibu}, \citenamefont {Kojima}, \citenamefont {Hazu}, \citenamefont
  {Ishikawa}, \citenamefont {Furusawa}, \citenamefont {Mita}, \citenamefont
  {Collazo}, \citenamefont {Sitar},\ and\ \citenamefont
  {Uedono}}]{chichibu2019}%
  \BibitemOpen
  \bibfield  {author} {\bibinfo {author} {\bibfnamefont {S.~F.}\ \bibnamefont
  {Chichibu}}, \bibinfo {author} {\bibfnamefont {K.}~\bibnamefont {Kojima}},
  \bibinfo {author} {\bibfnamefont {K.}~\bibnamefont {Hazu}}, \bibinfo {author}
  {\bibfnamefont {Y.}~\bibnamefont {Ishikawa}}, \bibinfo {author}
  {\bibfnamefont {K.}~\bibnamefont {Furusawa}}, \bibinfo {author}
  {\bibfnamefont {S.}~\bibnamefont {Mita}}, \bibinfo {author} {\bibfnamefont
  {R.}~\bibnamefont {Collazo}}, \bibinfo {author} {\bibfnamefont
  {Z.}~\bibnamefont {Sitar}},\ and\ \bibinfo {author} {\bibfnamefont
  {A.}~\bibnamefont {Uedono}},\ }\bibfield  {title} {\enquote {\bibinfo {title}
  {In-plane optical polarization and dynamic properties of the near-band-edge
  emission of an m-plane freestanding {{AlN}} substrate and a homoepitaxial
  film},}\ }\href {https://doi.org/10.1063/1.5116900} {\bibfield  {journal}
  {\bibinfo  {journal} {Appl. Phys. Lett.}\ }\textbf {\bibinfo {volume}
  {115}},\ \bibinfo {pages} {151903} (\bibinfo {year} {2019})}\BibitemShut
  {NoStop}%
\bibitem [{\citenamefont {Murotani}\ \emph {et~al.}(2009)\citenamefont
  {Murotani}, \citenamefont {Kuronaka}, \citenamefont {Yamada}, \citenamefont
  {Taguchi}, \citenamefont {Okada},\ and\ \citenamefont
  {Amano}}]{murotaniTemperatureDependenceExcitonic2009}%
  \BibitemOpen
  \bibfield  {author} {\bibinfo {author} {\bibfnamefont {H.}~\bibnamefont
  {Murotani}}, \bibinfo {author} {\bibfnamefont {T.}~\bibnamefont {Kuronaka}},
  \bibinfo {author} {\bibfnamefont {Y.}~\bibnamefont {Yamada}}, \bibinfo
  {author} {\bibfnamefont {T.}~\bibnamefont {Taguchi}}, \bibinfo {author}
  {\bibfnamefont {N.}~\bibnamefont {Okada}},\ and\ \bibinfo {author}
  {\bibfnamefont {H.}~\bibnamefont {Amano}},\ }\bibfield  {title} {\enquote
  {\bibinfo {title} {Temperature dependence of excitonic transitions in a-plane
  {{AlN}} epitaxial layers},}\ }\href {https://doi.org/10.1063/1.3116183}
  {\bibfield  {journal} {\bibinfo  {journal} {Journal of Applied Physics}\
  }\textbf {\bibinfo {volume} {105}},\ \bibinfo {pages} {083533} (\bibinfo
  {year} {2009})}\BibitemShut {NoStop}%
\bibitem [{\citenamefont {Feneberg}\ \emph {et~al.}(2010)\citenamefont
  {Feneberg}, \citenamefont {Leute}, \citenamefont {Neuschl}, \citenamefont
  {Thonke},\ and\ \citenamefont
  {Bickermann}}]{fenebergHighexcitationHighresolutionPhotoluminescence2010}%
  \BibitemOpen
  \bibfield  {author} {\bibinfo {author} {\bibfnamefont {M.}~\bibnamefont
  {Feneberg}}, \bibinfo {author} {\bibfnamefont {R.~A.~R.}\ \bibnamefont
  {Leute}}, \bibinfo {author} {\bibfnamefont {B.}~\bibnamefont {Neuschl}},
  \bibinfo {author} {\bibfnamefont {K.}~\bibnamefont {Thonke}},\ and\ \bibinfo
  {author} {\bibfnamefont {M.}~\bibnamefont {Bickermann}},\ }\bibfield  {title}
  {\enquote {\bibinfo {title} {High-excitation and high-resolution
  photoluminescence spectra of bulk {{AlN}}},}\ }\href
  {https://doi.org/10.1103/PhysRevB.82.075208} {\bibfield  {journal} {\bibinfo
  {journal} {Physical Review B}\ }\textbf {\bibinfo {volume} {82}},\ \bibinfo
  {pages} {075208} (\bibinfo {year} {2010})}\BibitemShut {NoStop}%
\bibitem [{\citenamefont {Neuschl}\ \emph {et~al.}(2012)\citenamefont
  {Neuschl}, \citenamefont {Thonke}, \citenamefont {Feneberg}, \citenamefont
  {Mita}, \citenamefont {Xie}, \citenamefont {Dalmau}, \citenamefont
  {Collazo},\ and\ \citenamefont
  {Sitar}}]{neuschlOpticalIdentificationSilicon2012}%
  \BibitemOpen
  \bibfield  {author} {\bibinfo {author} {\bibfnamefont {B.}~\bibnamefont
  {Neuschl}}, \bibinfo {author} {\bibfnamefont {K.}~\bibnamefont {Thonke}},
  \bibinfo {author} {\bibfnamefont {M.}~\bibnamefont {Feneberg}}, \bibinfo
  {author} {\bibfnamefont {S.}~\bibnamefont {Mita}}, \bibinfo {author}
  {\bibfnamefont {J.}~\bibnamefont {Xie}}, \bibinfo {author} {\bibfnamefont
  {R.}~\bibnamefont {Dalmau}}, \bibinfo {author} {\bibfnamefont
  {R.}~\bibnamefont {Collazo}},\ and\ \bibinfo {author} {\bibfnamefont
  {Z.}~\bibnamefont {Sitar}},\ }\bibfield  {title} {\enquote {\bibinfo {title}
  {Optical identification of silicon as a shallow donor in {{MOVPE}} grown
  homoepitaxial {{AlN}}},}\ }\href {https://doi.org/10.1002/pssb.201100381}
  {\bibfield  {journal} {\bibinfo  {journal} {physica status solidi (b)}\
  }\textbf {\bibinfo {volume} {249}},\ \bibinfo {pages} {511--515} (\bibinfo
  {year} {2012})}\BibitemShut {NoStop}%
\bibitem [{\citenamefont {Feneberg}\ \emph {et~al.}(2011)\citenamefont
  {Feneberg}, \citenamefont {Neuschl}, \citenamefont {Thonke}, \citenamefont
  {Collazo}, \citenamefont {Rice}, \citenamefont {Sitar}, \citenamefont
  {Dalmau}, \citenamefont {Xie}, \citenamefont {Mita},\ and\ \citenamefont
  {Goldhahn}}]{fenebergSharpBoundFree2011}%
  \BibitemOpen
  \bibfield  {author} {\bibinfo {author} {\bibfnamefont {M.}~\bibnamefont
  {Feneberg}}, \bibinfo {author} {\bibfnamefont {B.}~\bibnamefont {Neuschl}},
  \bibinfo {author} {\bibfnamefont {K.}~\bibnamefont {Thonke}}, \bibinfo
  {author} {\bibfnamefont {R.}~\bibnamefont {Collazo}}, \bibinfo {author}
  {\bibfnamefont {A.}~\bibnamefont {Rice}}, \bibinfo {author} {\bibfnamefont
  {Z.}~\bibnamefont {Sitar}}, \bibinfo {author} {\bibfnamefont
  {R.}~\bibnamefont {Dalmau}}, \bibinfo {author} {\bibfnamefont
  {J.}~\bibnamefont {Xie}}, \bibinfo {author} {\bibfnamefont {S.}~\bibnamefont
  {Mita}},\ and\ \bibinfo {author} {\bibfnamefont {R.}~\bibnamefont
  {Goldhahn}},\ }\bibfield  {title} {\enquote {\bibinfo {title} {Sharp bound
  and free exciton lines from homoepitaxial {{AlN}}},}\ }\href
  {https://doi.org/10.1002/pssa.201000947} {\bibfield  {journal} {\bibinfo
  {journal} {physica status solidi (a)}\ }\textbf {\bibinfo {volume} {208}},\
  \bibinfo {pages} {1520--1522} (\bibinfo {year} {2011})}\BibitemShut {NoStop}%
\bibitem [{\citenamefont {Ishii}, \citenamefont {Funato},\ and\ \citenamefont
  {Kawakami}(2020)}]{ishiiLongrangeElectronholeExchange2020}%
  \BibitemOpen
  \bibfield  {author} {\bibinfo {author} {\bibfnamefont {R.}~\bibnamefont
  {Ishii}}, \bibinfo {author} {\bibfnamefont {M.}~\bibnamefont {Funato}},\ and\
  \bibinfo {author} {\bibfnamefont {Y.}~\bibnamefont {Kawakami}},\ }\bibfield
  {title} {\enquote {\bibinfo {title} {Long-range electron-hole exchange
  interaction in aluminum nitride},}\ }\href
  {https://doi.org/10.1103/PhysRevB.102.155202} {\bibfield  {journal} {\bibinfo
   {journal} {Phys. Rev. B}\ }\textbf {\bibinfo {volume} {102}},\ \bibinfo
  {pages} {155202} (\bibinfo {year} {2020})}\BibitemShut {NoStop}%
\bibitem [{\citenamefont {Chichibu}\ \emph {et~al.}(2013)\citenamefont
  {Chichibu}, \citenamefont {Hazu}, \citenamefont {Ishikawa}, \citenamefont
  {Tashiro}, \citenamefont {Ohtomo}, \citenamefont {Furusawa}, \citenamefont
  {Uedono}, \citenamefont {Mita}, \citenamefont {Xie}, \citenamefont
  {Collazo},\ and\ \citenamefont
  {Sitar}}]{chichibuExcitonicEmissionDynamics2013}%
  \BibitemOpen
  \bibfield  {author} {\bibinfo {author} {\bibfnamefont {S.~F.}\ \bibnamefont
  {Chichibu}}, \bibinfo {author} {\bibfnamefont {K.}~\bibnamefont {Hazu}},
  \bibinfo {author} {\bibfnamefont {Y.}~\bibnamefont {Ishikawa}}, \bibinfo
  {author} {\bibfnamefont {M.}~\bibnamefont {Tashiro}}, \bibinfo {author}
  {\bibfnamefont {T.}~\bibnamefont {Ohtomo}}, \bibinfo {author} {\bibfnamefont
  {K.}~\bibnamefont {Furusawa}}, \bibinfo {author} {\bibfnamefont
  {A.}~\bibnamefont {Uedono}}, \bibinfo {author} {\bibfnamefont
  {S.}~\bibnamefont {Mita}}, \bibinfo {author} {\bibfnamefont {J.}~\bibnamefont
  {Xie}}, \bibinfo {author} {\bibfnamefont {R.}~\bibnamefont {Collazo}},\ and\
  \bibinfo {author} {\bibfnamefont {Z.}~\bibnamefont {Sitar}},\ }\bibfield
  {title} {\enquote {\bibinfo {title} {Excitonic emission dynamics in
  homoepitaxial {{AlN}} films studied using polarized and spatio-time-resolved
  cathodoluminescence measurements},}\ }\href
  {https://doi.org/10.1063/1.4823826} {\bibfield  {journal} {\bibinfo
  {journal} {Applied Physics Letters}\ }\textbf {\bibinfo {volume} {103}},\
  \bibinfo {pages} {142103} (\bibinfo {year} {2013})}\BibitemShut {NoStop}%
\end{thebibliography}%

\end{document}


\graphicspath{{./figures/}}

\title{\large{Supplementary Material: Excitonic and deep-level emission in N- and Al-polar homoepitaxial AlN grown by molecular beam epitaxy}}

\author{L.~van~Deurzen}
\email[Electronic mail: ]{lhv9@cornell.edu}
\affiliation{School of Applied and Engineering Physics, Cornell University, Ithaca, New York 14853, USA}
\author{J.~Singhal}
\author{J.~Encomendero}
\affiliation{Department of Electrical and Computer Engineering, Cornell University, Ithaca, New York 14853, USA}
\author{N.~Pieczulewski}
\affiliation{Department of Materials Science and Engineering, Cornell University, Ithaca, New York 14853, USA}
\author{C.S.~Chang}
\affiliation{School of Applied and Engineering Physics, Cornell University, Ithaca, New York 14853, USA}
\affiliation{Research Laboratory of Electronics, Massachusetts Institute of Technology, MA 02139, USA}
\author{Y.~Cho}
\affiliation{Department of Electrical and Computer Engineering, Cornell University, Ithaca, New York 14853, USA}
\author{D.A.~Muller}
\affiliation{School of Applied and Engineering Physics, Cornell University, Ithaca, New York 14853, USA}
\affiliation{Kavli Institute at Cornell for Nanoscale Science, Cornell University, Ithaca, New York 14853, USA}
\author{H.G.~Xing}
\affiliation{Department of Electrical and Computer Engineering, Cornell University, Ithaca, New York 14853, USA}
\affiliation{Department of Materials Science and Engineering, Cornell University, Ithaca, New York 14853, USA}
\affiliation{Kavli Institute at Cornell for Nanoscale Science, Cornell University, Ithaca, New York 14853, USA}
\author{D.~Jena}
\affiliation{School of Applied and Engineering Physics, Cornell University, Ithaca, New York 14853, USA}
\affiliation{Department of Electrical and Computer Engineering, Cornell University, Ithaca, New York 14853, USA}
\affiliation{Department of Materials Science and Engineering, Cornell University, Ithaca, New York 14853, USA}
\affiliation{Kavli Institute at Cornell for Nanoscale Science, Cornell University, Ithaca, New York 14853, USA}
\author{O.~Brandt}
\author{J.~L\"{a}hnemann}
\affiliation{Paul-Drude-Institut f\"ur Festk\"orperelektronik, Leibniz-Institut im Forschungsverbund Berlin e.V., 10117 Berlin, Germany}

\begin{abstract}
The supplementary material includes:\\
Figures S1--S8, Table ST1, and References 1--20.
\end{abstract}

\maketitle

\noindent\rule{17.5cm}{0.4pt}

\section{Reciprocal space maps and $\omega$ scans for samples I and II}

\begin{figure}[H]\includegraphics[width=\textwidth]{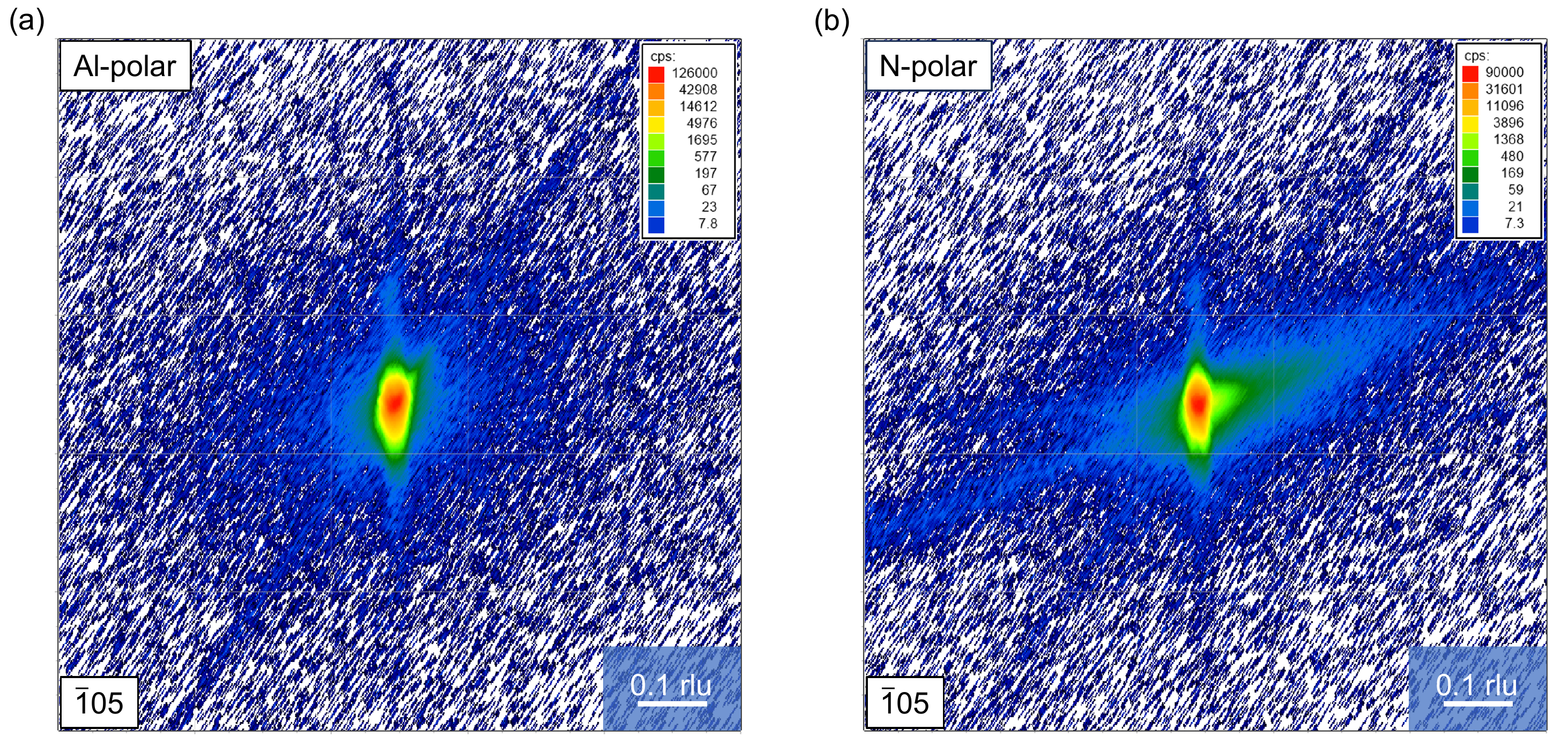}
\caption{Reciprocal space maps around the asymmetric $\bar{1}05$ reflection for (a) sample I  and (b) sample II. For both samples, a single intense and narrow $\bar{1}05$ reflection is observed, confirming the absence of strain in the homoepitaxial layer. The sharp reflection allows the observation of the streaks visible in both maps. The vertical and horizontal streaks correspond to the crystal truncation rods from the \{0001\} top and the \{10$\bar{1}$0\} side surfaces, the sharp diagonal one to the monochromator streak.} 
\label{Figure_S1}
\end{figure}

\newpage

\begin{figure}[H]\includegraphics[width=\textwidth]{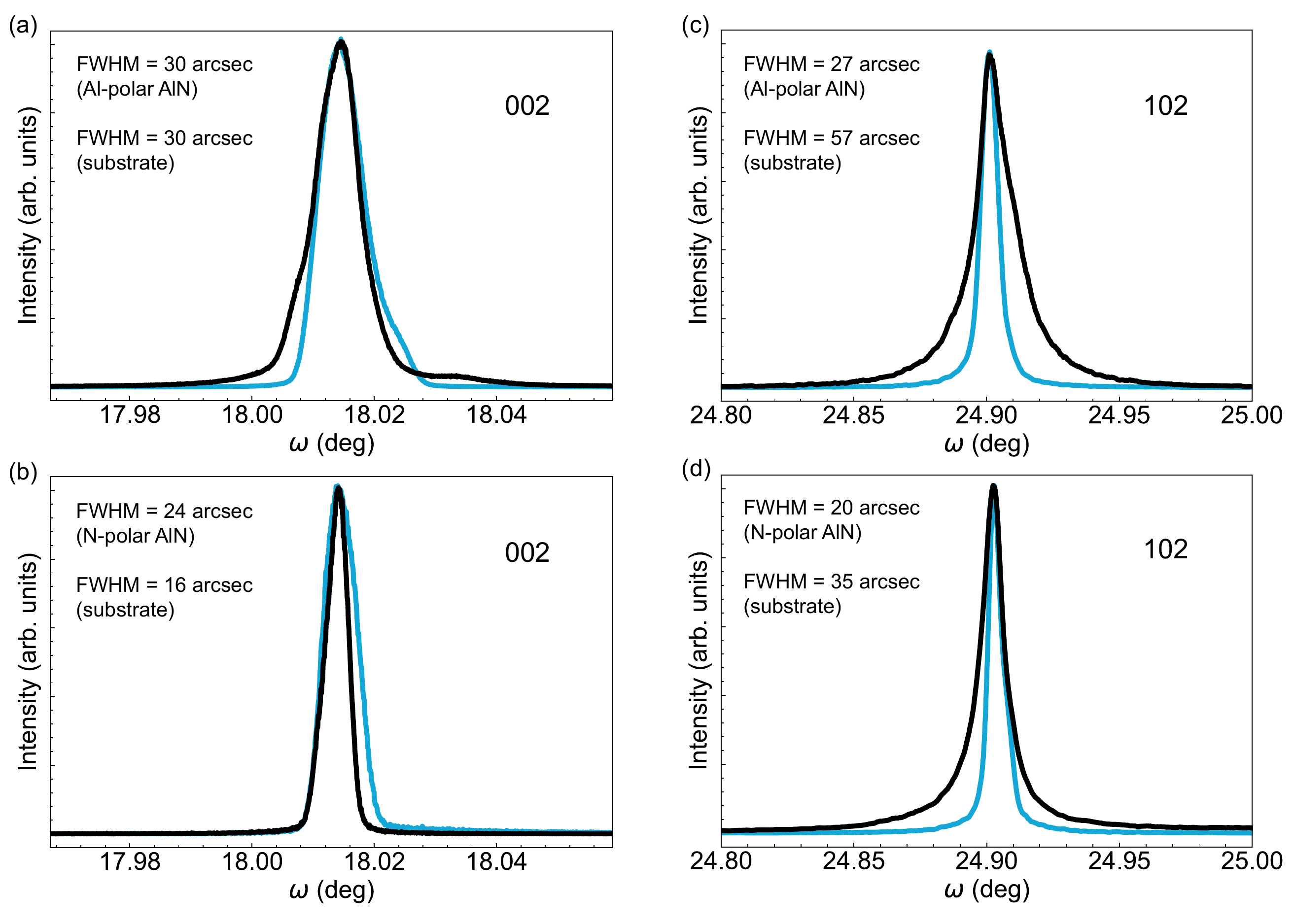}
\caption{[(a) and (b)] Symmetric 002 and [(c) and (d)] skew-geometry asymmetric 102 x-ray $\omega$ scans of samples I (Al-polar AlN, top row) and II (N-polar AlN, bottom row) along with their corresponding bulk AlN substrates measured from the back. The blue curves correspond to the MBE layers and the black curves correspond to the bulk AlN substrates.} 
\label{Figure_S2}
\end{figure}


\section{Characterization of N-polar \texorpdfstring{A\MakeLowercase{l}N}{AlN} sample with \texorpdfstring{G\MakeLowercase{a}N}{GaN} cap (sample III)}

\begin{figure}[H]\includegraphics[width=10cm]{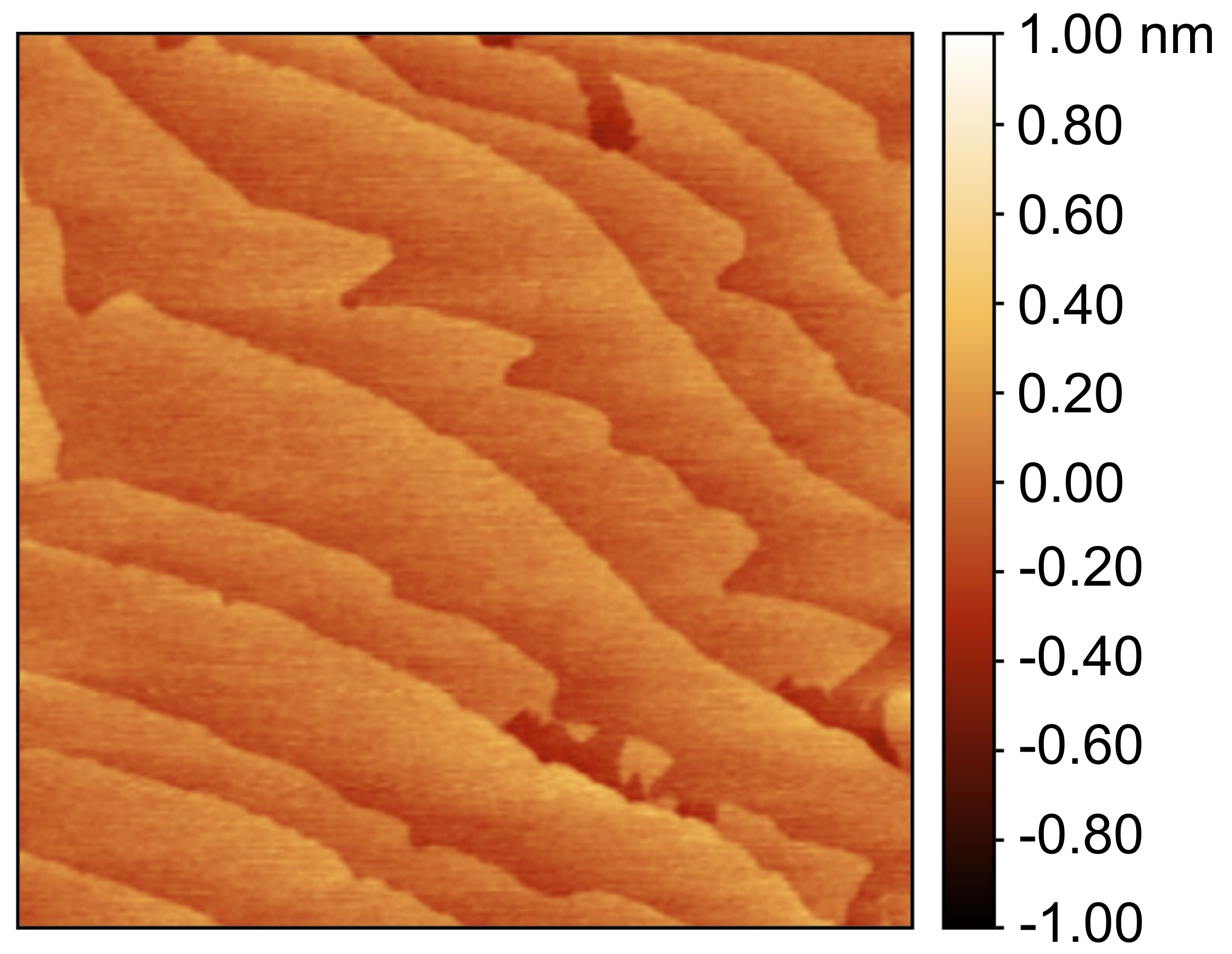}
\caption{$2 \times 2$~\textmu m$^{2}$ atomic force topograph of sample III. The root-mean-square roughness of the as-grown N-polar GaN $c$-plane surface is 130~pm.} 
\label{Figure_S3}
\end{figure}

\begin{figure}[H]\includegraphics[width=12cm]{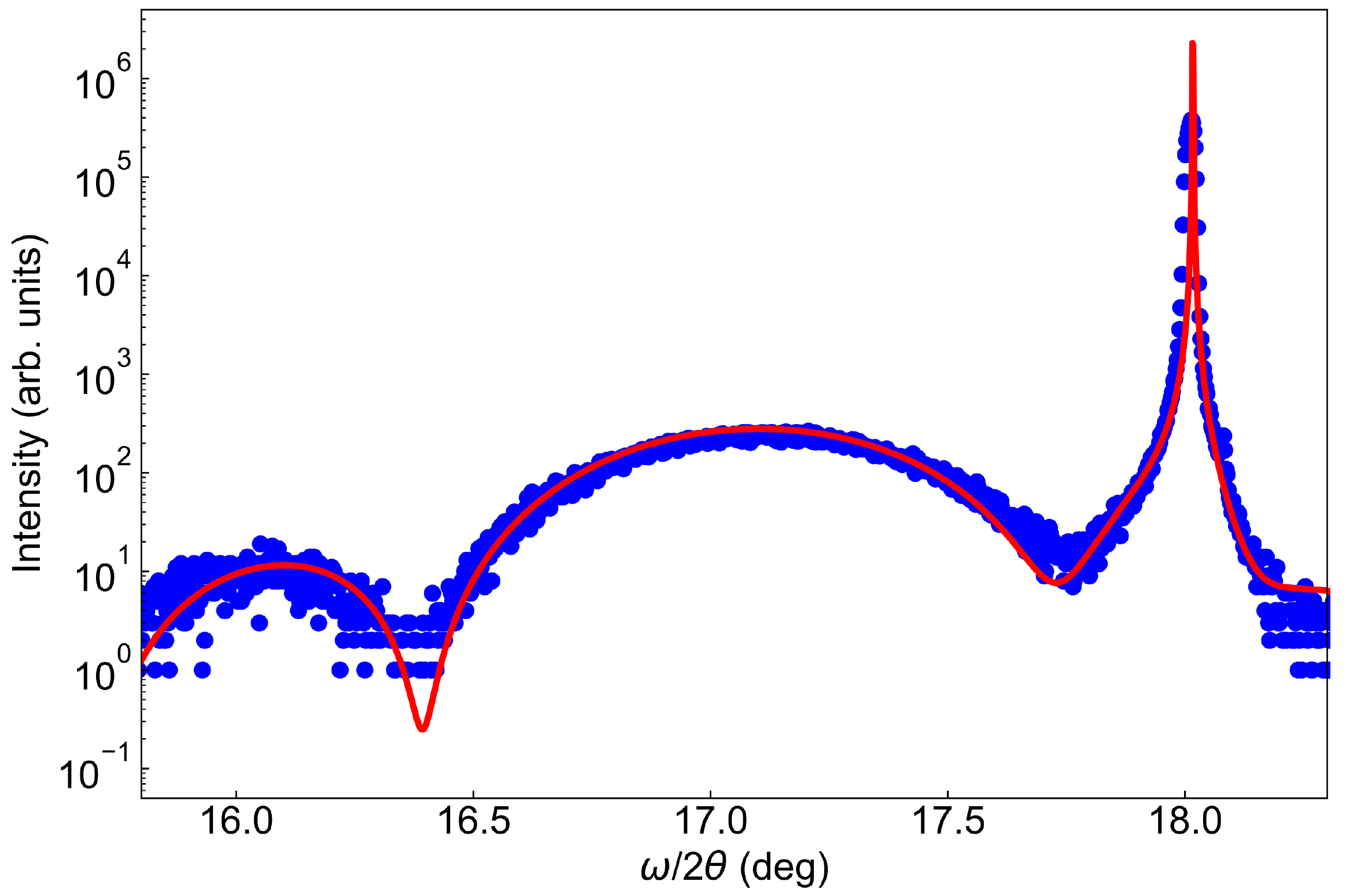}
\caption{$\omega/2\theta$ scan across the 002 reflection of sample III. The symbols and the solid line show the experimental data and the corresponding simulation (assuming a coherently strained 6.7-nm-thick GaN cap layer on AlN), respectively.}\label{Figure_S4}
\end{figure}
\begin{figure}[H]\includegraphics[width=11cm]{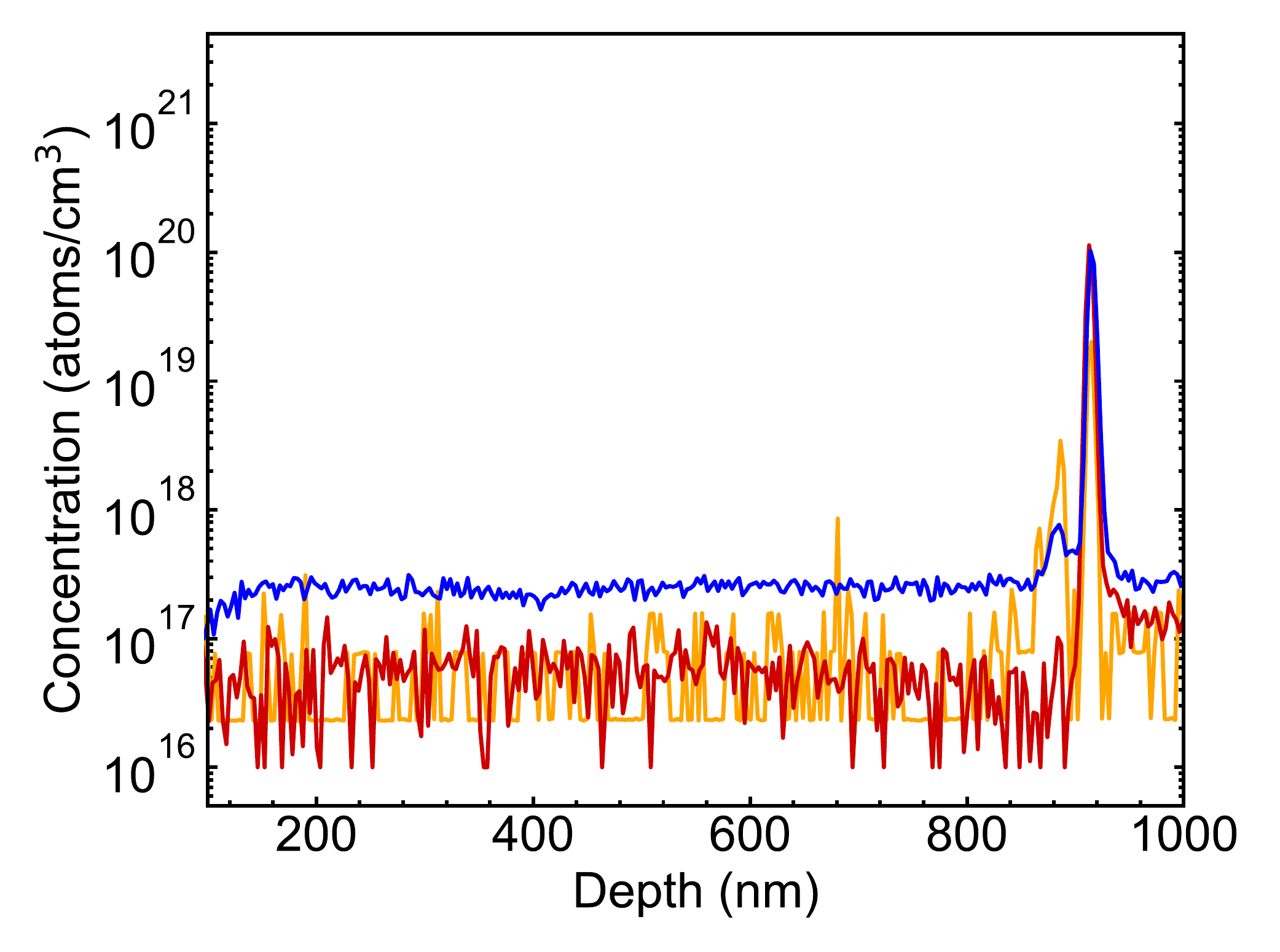}
\caption{Concentration of C (red), Si (orange), and O (blue) in sample III as measured by secondary ion-mass spectroscopy (SIMS). Si and C are at the detection limit, whereas O is measured near the detection limit, at an average level of approximately 2~$\times$~10$^{17}$~cm$^{-3}$. The H signal was also at its detection limit of 3~$\times$~10$^{17}$~cm$^{-3}$.}
\label{Figure_S5}
\end{figure}

\section{Ordering of free excitons in \texorpdfstring{A\MakeLowercase{l}N}{AlN}}

\begin{figure}[H]\includegraphics[width=12cm]{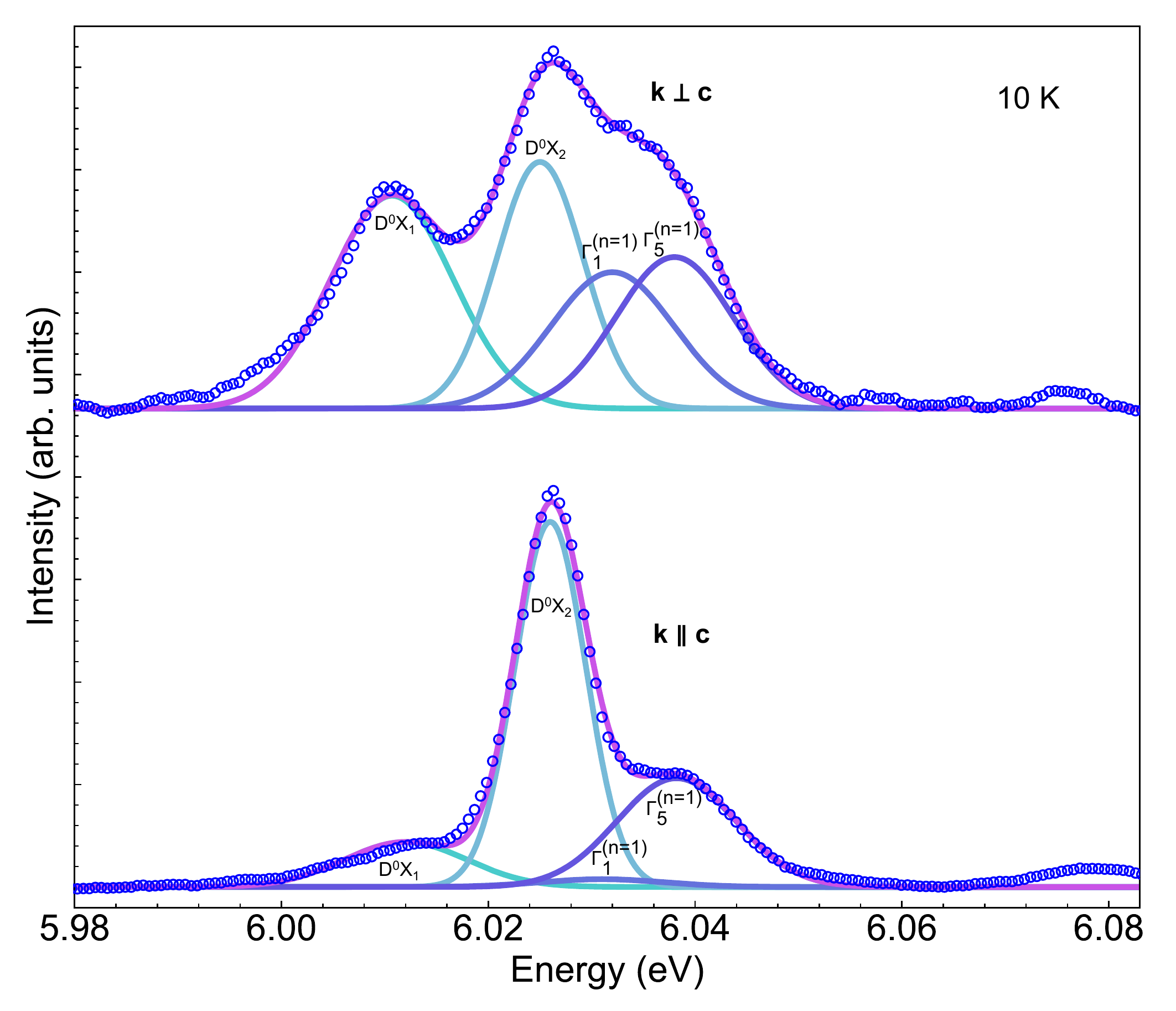}
\caption{Low-temperature near-bandedge cathodoluminescence spectra of Sample I, for emission from both the sample cross-section ($\mathbf{k} \perp \mathbf{c}$, top spectrum) and layer surface ($\mathbf{k} \| \mathbf{c}$, bottom spectrum), for excitation with an 11 keV electron beam. The line in both spectra is a fit of the data with four Gaussians (also shown).}
\label{Figure_S6}
\end{figure}

To infer the ordering of the free excitons in AlN, we compare cathodoluminescence spectra excited and taken from both the sample cross-section ($\mathbf{k} \perp \mathbf{c}$) and layer surface ($\mathbf{k} \parallel \mathbf{c}$) shown in Fig.~\ref{Figure_S6}. For AlN, exciton states of $\Gamma^c_{7} \otimes \Gamma^v_{7+}$ symmetry (often called A-excitons) are split further by the spin-exchange interaction into excitons with irreducible representation:

\begin{equation*}
 \Gamma^c_{7} \otimes \Gamma^v_{7+}  \rightarrow \Gamma_{1} \oplus \Gamma_{2} \oplus \Gamma_{5}(2) ,
\end{equation*}
where $\Gamma_{1}$ is an exciton singlet satisfying only $\mathbf{k} \perp \mathbf{c}$, $\Gamma_{2}$ is a dipole-forbidden paraexciton, and $\Gamma_{5}$ is a doubly degenerate exciton triplet satisfying both $\mathbf{k} \perp \mathbf{c}$ and $\mathbf{k} \parallel \mathbf{c}$. For both orientations, that is, side emission ($\mathbf{k} \perp \mathbf{c}$) and surface emission ($\mathbf{k} \parallel \mathbf{c}$), we resolve two donor-bound exciton transitions $\mathrm{D^{0}X_{1}}$ and $\mathrm{D^{0}X_{2}}$. The contributions of $\mathrm{D^{0}X_{1}}$, most notably for the $\mathbf{k} \perp \mathbf{c}$ spectrum, is due to the partial excitation of the substrate due to the appreciable carrier-generation volume from the 11~keV primary electrons. Additionally, two other transitions are resolved at $6.031 \pm 0.001$~eV and $6.038 \pm 0.001$~eV, which we assign to free excitons. Since the transition at 6.031~eV is detected only for the $\mathbf{k} \perp \mathbf{c}$ spectrum, we assign this peak to the $\Gamma_{1}^{n=1}$ transition. The peak at 6.038~eV is observed for both orientations and thus corresponds to the $\Gamma_{5}^{n=1}$ transition. Since the energy of $\Gamma_{1}^{n=1}$ falls below that of $\Gamma_{5}^{n=1}$, this implies a negative spin-exchange splitting in our AlN layers. Finally, the transitions near or just below 6.08~eV (not fitted) in both spectra are assigned to the $\Gamma_{1}^{n=2}$ and $\Gamma_{5}^{n=2}$ transitions. The highest-energy transition in Fig.~3(b) in the main manuscript, which corresponds to free electron-hole recombination, implies an exciton binding energy of 57~meV for $\Gamma_{5}$ and 64~meV for $\Gamma_{1}$.

\section{Generation volume in cathodoluminescence spectroscopy}

\begin{figure}[H]\includegraphics[width=12cm]{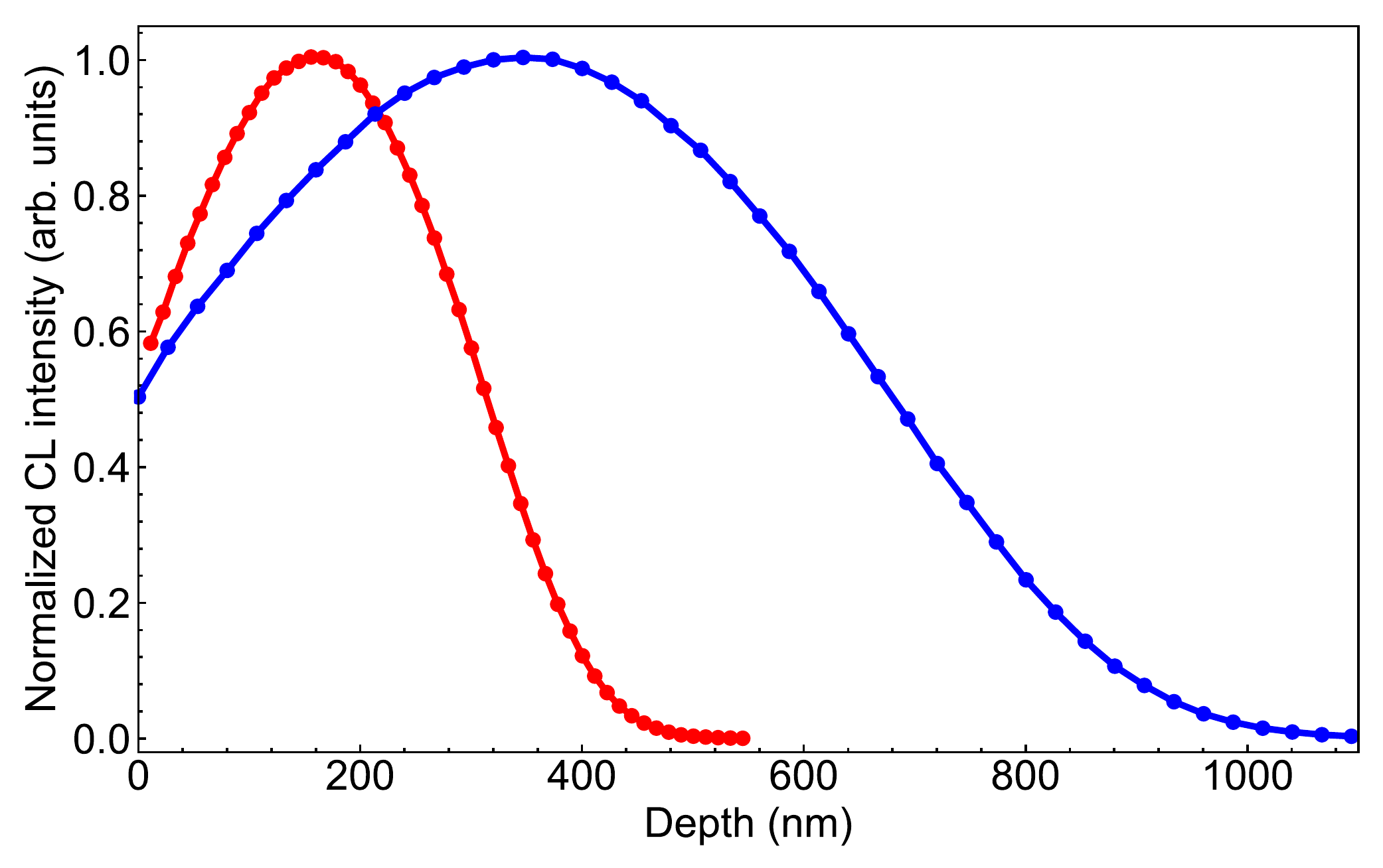}
\caption{Generated CL intensity versus sample depth for sample II (bare AlN) illuminated with a 7~keV electron beam (red) and sample III (6.7~nm of GaN on AlN) irradiated with an 11~keV electron beam (blue). The data was simulated using ‘monte CArlo SImulation of electroN trajectory in sOlids (CASINO)’ \citenum{drouinCASINOV2422007}.}
\label{Figure_S7}
\end{figure}

To ascertain that the cathodoluminescence (CL) spectra are dominated by the signal from the epitaxial layers and not the AlN substrates, the depth distribution of the CL intensity is estimated from Monte Carlo simulations of the deposited energy in Fig.~\ref{Figure_S7}. Note that these simulations do not take into account the temperature-dependence of the electron-phonon interaction during the relaxation of hot carriers to the band edges (hot carrier diffusion), which increases the depth of the generation volume\cite{jahnCarrierDiffusionGaN2022}. For our measurements at 10~K and beam energies of 7 and 11~keV, we estimate that the CL generation profiles should extend about 100--150~nm deeper into the sample. This picture does not include reabsorption of emitted photons, which differs between free and bound exciton emission.

To assess the \textbf{excitation density} at an acceleration voltage of $V_0=11$~keV, the generation rate is estimated according to Wu and Wittry \cite{wuInvestigationMinorityCarrier1978} with the parameters adapted from Ref.~\citenum{Jahn_jap_2003}:

$$G_0 = 6.25 \times 10^{21} V_0 I_b \frac{1- \delta\cdot \bar{V}/V_0}{\epsilon},$$

where $I_b = 24$~nA is the beam current, $\delta=0.1$ is the fraction of backscattered electrons, $\bar{V} = 0.65 V_0$ is the mean energy of the backscattered electrons, $\epsilon = 3E_g = 18$~eV is the mean energy required to generate an electron-hole pair (three times the bandgap energy).
The carrier density is then obtained as

$$\rho = \tau \frac{G_0}{V} = 4\times10^{16} \mathrm{cm}^{-3},$$
assuming a carrier lifetime of $\tau = 100$~ps and approximating the generation volume $V$ by a cylinder of diameter $d=600$~nm and depth $h=700$~nm, corresponding to a volume comprising 90\% of the CASINO energy loss distribution.

This calculation illustrates that despite a high beam current, the choice of a high acceleration voltage and thus a large excitation volume leads to a moderate excitation density. Naturally, the excitation density in the core of the volume will be higher, but overall the large volume should be dominating the CL signal.

\section{Phase diagram of the coupled exciton-carrier system in \texorpdfstring{A\MakeLowercase{l}N}{AlN}}

\begin{figure}[H]
\centerline{\includegraphics[width=12cm]{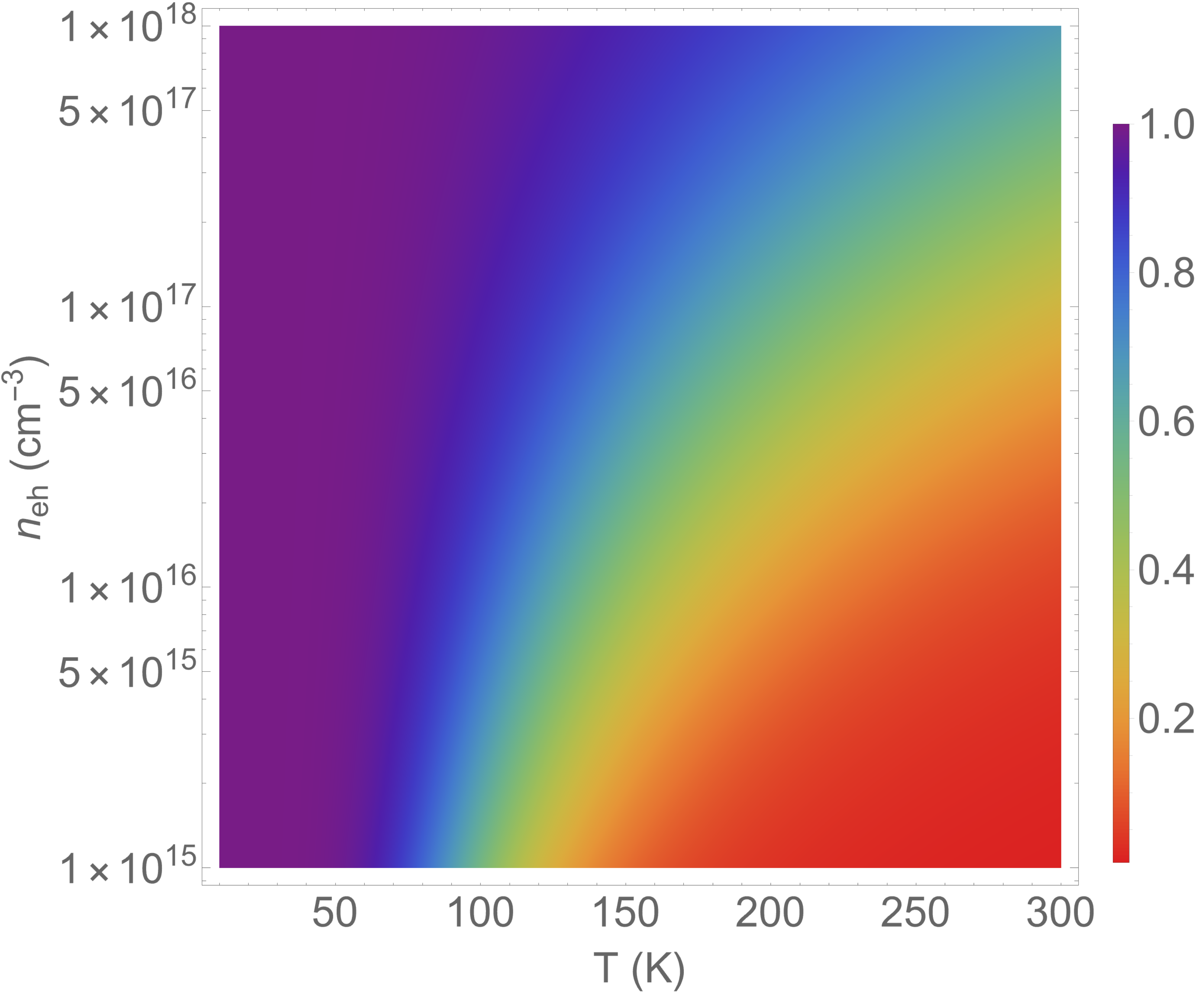}}
\caption{Phase diagram of the coupled exciton-carrier system in AlN with zero background carrier density. The color code represents the exciton fraction $f_x = n_x/(n_x + \Delta n)$ according to Saha's law, where $n_x$ and $\Delta n$ are the densities of excitons and free carriers, respectively, with the total cathodogenerated density $n_\text{eh} = n_x + \Delta n$. For the calculation, we assume a reduced density-of-states mass of 0.24 $m_0$ with the electron mass $m_0$, and an exciton binding energy of 57~meV.}
\label{Figure_S8}
\end{figure}
\newpage

\section{Extended table of near-bandedge transitions in \texorpdfstring{A\MakeLowercase{l}N}{AlN}}

\begin{table}[H]
\centering
\begin{tabular}{cccc}
\hline Transition & $E$ & $\Delta E(\Gamma^{n=1}_{5})$ & $\Delta E_\mathrm{lit}(\Gamma_{5}^{n=1})$\\
\text{} & $(\mathrm{eV})$ & $(\mathrm{meV})$ & $(\mathrm{meV})$ \\
\hline $\boldsymbol{\Gamma}^{\mathbf{n}=\mathbf{1}}_{\mathbf{5}}-\mathbf{2} \mathbf{L} \mathbf{O}$ & 5.815 & $-223$ & $-225$,\cite{leutePhotoluminescenceHighlyExcited2009} $-231$\cite{ishiiStimulatedEmissionMechanism2022} \\
$\boldsymbol{\Gamma}^{\mathbf{n} \geq \mathbf{2}}_{\mathbf{5}}-\mathbf{2} \mathbf{L} \mathbf{O}$ & $5.849$ & $-189$ & $-188$\cite{leutePhotoluminescenceHighlyExcited2009} \\
$\boldsymbol{\Gamma}^{\mathbf{n}=\mathbf{1}}_{\mathbf{5}}-\mathbf{L} \mathbf{O}$ & $5.926$ & $-112$ & $-109$,\cite{leutePhotoluminescenceHighlyExcited2009} $-117$ \cite{ishiiStimulatedEmissionMechanism2022} \\
TES($\mathrm{D^{0}X_{1}}$) & {} & {} & $-76.6$, \cite{neuschlDirectDeterminationSilicon2013} $-77$\cite{ishiiStimulatedEmissionMechanism2022} \\
$\boldsymbol{\Gamma}^{\mathbf{n} \geq \mathbf{2}}_{\mathbf{5}}-\mathbf{L O}$ & $5.965$ & $-73$ & $-73$\cite{leutePhotoluminescenceHighlyExcited2009} \\
$\mathrm{A^{0}X_{\alpha}}$ & {} & {} & $-50$\cite{nepalGrowthPhotoluminescenceStudies2006} \\
$\mathrm{A^{0}X_{\beta}}$ & {} & {} & $-40$,\cite{nepalAcceptorboundExcitonTransition2004} $-40$,\cite{nakarmiCorrelationOpticalElectrical2006} $-41$\cite{sedhainBerylliumAcceptorBinding2008} \\
$\mathrm{D^{0}X_{\alpha}}$ & {} & {} & $-37$\cite{bryanExcitonTransitionsOxygen2014} \\
$\mathrm{D^{0}X_{\zeta}(\Gamma_{1})}$ & {} & {} & $-34$,\cite{bryanExcitonTransitionsOxygen2014} $-34.2$,\cite{funatoHomoepitaxyPhotoluminescenceProperties2012} $-34.7$,\cite{chichibu2019} $-35.1$\cite{murotaniTemperatureDependenceExcitonic2009} \\
$\mathrm{A^{0}X_{\gamma}}$ & {} & {} & $-33$\cite{sedhainBerylliumAcceptorBinding2008} \\
$\mathrm{D^{0}X_{\eta}(\Gamma_{1})}$ & {} & {} & $-30$\cite{bryanExcitonTransitionsOxygen2014} \\
$\mathbf{D^{\mathbf{0}} \mathbf{X}_{\mathbf{1}} }$ & $6.010$ & $-28$ & $-28$, \cite{bryanExcitonTransitionsOxygen2014} $-28.2$,\cite{fenebergHighexcitationHighresolutionPhotoluminescence2010} $-28.2$,\cite{chichibu2019} $-28.5$,\cite{funatoHomoepitaxyPhotoluminescenceProperties2012} $-28.5$,\cite{neuschlOpticalIdentificationSilicon2012} $-28.7$,\cite{fenebergSharpBoundFree2011} $-28.7$\cite{ishiiLongrangeElectronholeExchange2020} \\
$\mathrm{D^{0}X_{\beta}}$ & {} & {} & $-25$\cite{bryanExcitonTransitionsOxygen2014} \\
$\mathrm{D^{0}X_{\gamma}}$ & {} & {} & $-21$,\cite{funatoHomoepitaxyPhotoluminescenceProperties2012} $-22.1$,\cite{fenebergHighexcitationHighresolutionPhotoluminescence2010} $-22.4$,\cite{fenebergSharpBoundFree2011} $-22.5$\cite{neuschlOpticalIdentificationSilicon2012} \\
$\mathrm{D^{0}X_{\delta}}$ & {} & {} &  $-19$,\cite{bryanExcitonTransitionsOxygen2014} $-19$,\cite{murotaniTemperatureDependenceExcitonic2009} $-19.0$ ,\cite{neuschlOpticalIdentificationSilicon2012} $-19.8$\cite{chichibu2019} \\
    $\mathbf{D}^{\mathbf{0}} \mathbf{X}_{\mathbf{2}}$ & $6.025$ & $-13$ & $-12.8$,\cite{fenebergHighexcitationHighresolutionPhotoluminescence2010} $-13$,\cite{leutePhotoluminescenceHighlyExcited2009} $-13$, \cite{bryanExcitonTransitionsOxygen2014} $-13.1$,\cite{chichibu2019} $-13.3$,\cite{neuschlOpticalIdentificationSilicon2012} $-13.3$,\cite{ishiiLongrangeElectronholeExchange2020} $-13.4$, \cite{fenebergSharpBoundFree2011}  $-13.6$ \cite{funatoHomoepitaxyPhotoluminescenceProperties2012} \\
$\mathrm{D^{0}X_{\epsilon}}$ & {} & {} & $-9.5$,\cite{funatoHomoepitaxyPhotoluminescenceProperties2012} $-9.5$\cite{neuschlOpticalIdentificationSilicon2012} \\
$\boldsymbol{\Gamma^{\mathbf{n}=\mathbf{1}}_{1}}$ & 6.031 & {$-7$} & $-4$ (w.r.t. $\Gamma^{n=1}_{5T}$),\cite{fenebergHighexcitationHighresolutionPhotoluminescence2010} $-8$,\cite{bryanExcitonTransitionsOxygen2014} $-9$ (w.r.t. $\Gamma^{n=1}_{5L}$),\cite{fenebergHighexcitationHighresolutionPhotoluminescence2010} $-9.0$,\cite{ishiiLongrangeElectronholeExchange2020} $-9.8$\cite{chichibu2019} \\
$\boldsymbol{\Gamma}^{\mathbf{n}=\mathbf{1}}_{\mathbf{5}}$ & $6.038$ & {0} & {0}\\
$\Gamma^{\mathrm{n=2}}_{1}$ & {} & {} & $+28.7$,\cite{chichibu2019} $+31$\cite{bryanExcitonTransitionsOxygen2014} \\
$\boldsymbol{\Gamma}^{\mathbf{n}=\mathbf{2}}_{\mathbf{5}}$ & $6.079$ & $+41$ & $+37$,\cite{leutePhotoluminescenceHighlyExcited2009} $+38$,\cite{chichibuExcitonicEmissionDynamics2013} $+38.1$,\cite{chichibu2019} $+39.3$,\cite{fenebergSharpBoundFree2011}
$+39.4$,\cite{funatoHomoepitaxyPhotoluminescenceProperties2012} $+39.4$,\cite{neuschlOpticalIdentificationSilicon2012} $+41$, \cite{bryanExcitonTransitionsOxygen2014} $+43$ \cite{murotaniTemperatureDependenceExcitonic2009} \\
$\boldsymbol{\Gamma}^{\mathbf{n}=\mathbf{3}}_{\mathbf{5}}$ & $6.089$ & $+51$ & \\
$\boldsymbol{\Gamma}^{\mathbf{n} \rightarrow \infty}_{\mathbf{5}}$ & $6.095$ & $+57$ & \\
\hline
\end{tabular}
\caption{Low-temperature (LHe) near-bandedge transitions in AlN (detailed version). Transitions resolved in the homoepitaxial, N-polar AlN layers measured in the present work are highlighted in \textbf{bold}. All of these values have an experimental uncertainty of $\pm 1$~meV. We also provide the shift with respect to the $\Delta E(\Gamma^{n=1}_{5})$ transitions and, where available, corresponding literature values.}
\label{table:1}
\end{table}

\bibliography{supplement}